\documentclass[11pt,a4paper]{article}
\pdfoutput=1
\usepackage{amssymb,amsmath,amsfonts, mathtools, mathrsfs}
\usepackage[dvipsnames]{xcolor}

\newif\ifnatbibsort\natbibsorttrue

\relax

\ifnatbibsort\RequirePackage[numbers,sort&compress]{natbib}\else\RequirePackage[numbers,compress]{natbib}\fi
\RequirePackage[colorlinks=true
,urlcolor=blue
,anchorcolor=blue
,citecolor=blue
,filecolor=blue
,linkcolor=blue
,menucolor=blue
,pagecolor=blue
,linktocpage=true
,pdfproducer=medialab
,pdfa=true
]{hyperref}

\usepackage{multicol}
\usepackage{tikz,pgf,pgfplots}
\usepackage{graphicx}
\usepackage{float}
\usepackage{subcaption}
\usepackage{caption}
\usepackage{enumerate}
\usepackage{dsfont}
\usepackage{verbatim}
\usepackage{amsfonts,euscript,amssymb,braket}
\usepackage{tikz}
\usetikzlibrary{arrows,decorations.markings,patterns}
\usepackage{slashed}

\def\clock{{\count0=\time
		\divide\count0 60
		\ifnum\count0<10 0\fi\the\count0
		\multiply\count0 -60 \advance\count0 \time
		:\ifnum\count0<10 0\fi \the\count0
}}
\newcommand{\timestamp}{{\small\vbox{\hbox{\tt\jobname.tex}
			\hbox{\the\day/\the\month/\the\year, \clock}}}}

\newcommand{\bea}{\begin{eqnarray}}
\newcommand{\eea}{\end{eqnarray}}

\newcommand{\be}{\begin{equation}}
\newcommand{\ee}{\end{equation}}

\makeatletter
\let\old@startsection=\@startsection
\let\oldl@section=\l@section
\renewcommand{\@startsection}[6]{\old@startsection{#1}{#2}{#3}{#4}{#5}{#6\mathversion{bold}}}
\renewcommand{\l@section}[2]{\oldl@section{\mathversion{bold}#1}{#2}}
\makeatother

\numberwithin{equation}{section}

\usepackage{color}

\begin{document}
	\renewcommand{\thefootnote}{\arabic{footnote}}

	\overfullrule=0pt
	\parskip=2pt
	\parindent=12pt
	\headheight=0in \headsep=0in \topmargin=0in \oddsidemargin=0in

	\vspace{ -3cm} \thispagestyle{empty} \vspace{-1cm}
	\begin{flushright} 
		\footnotesize
		\textcolor{red}{\phantom{print-report}}
	\end{flushright}

\begin{center}
	\vspace{.0cm}

	{\Large\bf \mathversion{bold}
	On the continuum limit of the entanglement Hamiltonian
	}
	\\
	\vspace{.25cm}
	\noindent
	{\Large\bf \mathversion{bold}
	of a sphere for the free massless scalar field}

	\vspace{0.8cm} {
		Nina Javerzat
		and Erik Tonni
	}
	\vskip  0.7cm
	
	\small
	{\em
		SISSA and INFN Sezione di Trieste, via Bonomea 265, 34136, Trieste, Italy 
	}
	\normalsize

\end{center}

\vspace{0.3cm}
\begin{abstract} 

We study the continuum limit of the entanglement Hamiltonian of a sphere 
for the massless scalar field in its ground state 
by employing the lattice model defined through the discretisation of the radial direction. 
In two and three spatial dimensions and 
for small values of the total angular momentum, 
we find numerical results in agreement with 
the corresponding ones derived from
the entanglement Hamiltonian predicted by conformal field theory.
When the mass parameter in the lattice model is large enough,
the dominant contributions come from
the on-site and the nearest-neighbour terms,
whose weight functions are straight lines.

\end{abstract}

\newpage

\tableofcontents

\section{Introduction}
\label{sec_intro}

The reduced density matrix of a subsystem is a central quantity to study 
in order to understand the entanglement properties of a quantum state 
for the spatial bipartition.
Denoting by $A$ a spatial subregion and by  $\bar{A}$ its complement, 
under the assumption that the Hilbert space of the system factorises 
as $\mathcal{H} = \mathcal{H}_A \otimes \mathcal{H}_{\bar{A}}$,
the reduced density matrix is $\rho_A \equiv \textrm{Tr}_{\mathcal{H}_{\bar{A}}} \rho$,
where $\rho$ is the density matrix of the entire system. 
Considering a system in its ground state,
the entanglement entropy 
$S_A \equiv - \,\textrm{Tr}_{\mathcal{H}_A} (\rho_A \log \rho_A)$
measures the bipartite entanglement in this pure state;
hence it has been widely explored during the past two decades
\cite{Bombelli:1986rw, Srednicki:1993im, Callan:1994py, Holzhey:1994we, Vidal:2002rm, Calabrese:2004eu}
(see \cite{Eisert:2008ur, EislerPeschel:2009review,Casini:2009sr} for reviews).
The reduced density matrix can always be written as 
$\rho_A = e^{-K_A}/\mathcal{Z}_A$,
where the constant $\mathcal{Z}_A = \textrm{Tr}_{\mathcal{H}_A} ( e^{-K_A} ) $
guarantees the normalisation condition $\textrm{Tr}_{\mathcal{H}_A} \rho_A = 1$
and the operator $K_A$ is the entanglement Hamiltonian 
(also known as modular Hamiltonian).

In certain relativistic quantum field theories (QFTs) 
and for some particular choices of states
and geometric bipartitions, the entanglement Hamiltonian is given by 
the integral over $A$ of the energy density multiplied by a suitable weight factor. 
The most important example
has been found by Bisognano and Wichmann 
\cite{Bisognano:1975ih, Bisognano:1976za}:
for a Lorentz invariant quantum field theory 
in the $d+1$ dimensional Minkowski spacetime in its ground state, 
when $A$ is the half space $x > 0$,
the entanglement Hamiltonian 
$K_A$ is the boost generator in the $x$-direction.

In conformal field theories (CFTs), this result and the conformal invariance 
have been employed to obtain other entanglement Hamiltonians. 
A seminal example is 
the entanglement Hamiltonian of a sphere $B$ of radius $\mathcal{R}$
for a $d+1$ dimensional CFT in Minkowski spacetime and in its ground state.
It reads
\cite{Hislop:1981uh,Casini:2011kv} (see also \cite{Haag:1992hx})
\be
\label{EH-intro}
K_B
\,=\,  
\mathcal{R} 
\int_{B} \beta(r) \, T_{tt}(\boldsymbol{x}) \, \textrm{d}^d\boldsymbol{x}
\ee
where the weight function $ \beta(r) $ is the following parabola
\be
\label{beta-def-intro}
 \beta(r) 
\, \equiv \,
 2\pi\left[ \,
 \frac{1}{2} \left( 1 - \frac{r^2}{ \mathcal{R}^2}\,\right)
 \right] .
\ee
When $d=1$, 
specific conformal mappings have been constructed to obtain
other entanglement Hamiltonians $K_A$ in the local form,
i.e. written as an the integral over $A$ 
of the energy density multiplied by the proper weight factor
\cite{Wong:2013gua, Cardy:2016fqc, Tonni:2017jom}.

It is important to understand how these QFT results  
can be obtained as the continuum limit 
of the corresponding entanglement Hamiltonians
in many-body quantum systems on the lattice. 
For free fermionic and bosonic systems, 
the Gaussian nature of the ground state allows to obtain 
explicit expressions for the entanglement Hamiltonian of a lattice subsystem 
for a generic number of spatial dimensions $d$
\cite{EislerPeschel:2009review,Peschel:2003rdm, Peschel_2004, Casini:2009sr, Banchi:2015aaa}.
These lattice operators are characterised 
by long-range and inhomogeneous couplings 
\cite{Arias:2016nip, Eisler:2017cqi, Eisler:2018ugn}.
In the special case of $d=1$,
the continuum limit of the entanglement Hamiltonian of 
a block made by consecutive sites
in a chain of free fermions in the ground state
has been studied analytically in \cite{Eisler:2019rnr}
by employing the results of \cite{Eisler:2017cqi},
finding the parabolic weight function (\ref{beta-def-intro}) expected from CFT.
In a massless harmonic chain, 
the corresponding analysis has been performed numerically in
\cite{Arias:2016nip, DiGiulio:2019cxv}.

The entanglement Hamiltonians of a block of consecutive sites 
have been studied numerically also for non-critical chains in their ground state,
finding also in these cases that the entanglement Hamiltonian matrices contain 
long-range and inhomogeneous couplings.
Far away from criticality, 
a triangular profile for the weight function has been observed \cite{Eisler:2020lyn},
which has been understood through the analytic expressions derived
for the entanglement Hamiltonian of the half infinite chain \cite{Peschel-Truong91, Peschel_1999}.

In this manuscript we consider the entanglement Hamiltonian 
of a sphere  $B \in \mathbb{R}^d$, 
mainly focussing on the CFT given by the massless scalar field 
in the $d+1$ dimensional Minkowski spacetime and in its ground state.
We study the continuum limit that leads to 
the entanglement Hamiltonian $K_B$ of the sphere 
given by (\ref{EH-intro}) and (\ref{beta-def-intro})
specialised to  this model. 
 Our analysis is mostly numerical and it is
based on the method developed in 
\cite{Arias:2016nip, Eisler:2017cqi, Eisler:2019rnr, DiGiulio:2019cxv} 
for $d=1$.
In the special case of $d=1$, 
we recover the results for the entanglement Hamiltonian 
of a segment at the beginning of the semi-infinite line
with Dirichlet boundary conditions obtained in \cite{DiGiulio:2019cxv}.
In the massive regime, 
we adapt to the higher dimensional case of the sphere
the analysis made in \cite{Eisler:2020lyn} 
for the entanglement Hamiltonian of the segment 
in the massive harmonic chain on the infinite line.

The layout of this paper is as follows.
In Sec.\,\ref{sec-ham-eh} we introduce the model of the massive scalar field,
the lattice regularisations of its Hamiltonian along the radial direction,
the CFT prediction for the entanglement Hamiltonian of the sphere
(\ref{EH-intro}) for this model
and the corresponding expressions on the lattice 
employed in our numerical analysis.
In Sec.\,\ref{sec_EH_ball_massless}
we focus on the massless case
and study numerically 
the continuum limit of the entanglement Hamiltonian.
In Sec.\,\ref{sec_massive} we discuss 
the regime where the mass parameter in the lattice model is large.
Conclusions are drawn in Sec.\,\ref{sec_conclusions}.
In the appendices\;\ref{app_corr_check} and \ref{app_kmax}
we report further results that clarify 
and support some discussions in the main text.

\section{Entanglement Hamiltonian of a sphere for the scalar field}
\label{sec-ham-eh}

In this section we introduce the main expressions 
employed in this manuscript to study 
the entanglement Hamiltonian of a sphere 
in the $d+1$ dimensional Minkowski spacetime
for the scalar field in its ground state.
In Sec.\,\ref{subsec-ham} we briefly review the Hamiltonian of the scalar field 
and its lattice regularisation along the radial direction,
as done by Srednicki \cite{Srednicki:1993im} 
to study the entanglement entropy of a sphere.
In Sec.\,\ref{subsec-eh} we combine the result of this analysis 
with the expression for the entanglement Hamiltonian of a generic region 
in harmonic lattices found by Casini and Huerta \cite{Casini:2009sr}.

\subsection{Hamiltonian and radial regularisations}
\label{subsec-ham}

The Hamiltonian of the massive real scalar field 
in the $d+1$ dimensional Minkowski spacetime
reads
\be
\label{hamiltonian}
H = \frac{1}{2} \int_{\mathbb{R}^d}
\Big(
\Pi^2 -\Phi\,\Delta \Phi + m^2\,\Phi^2 
\,\Big) 
\textrm{d}^d \boldsymbol{x}
\ee
where  
$\boldsymbol{x}\in \mathbb{R}^d$ is the vector identifying the spatial position,
$\Delta$ denotes the $d$-dimensional Laplacian,
$\Phi=\Phi(\boldsymbol{x})$ is the real scalar field
and $\Pi=\Pi(\boldsymbol{x})$ its canonically conjugate momentum field.
In order to study the entanglement Hamiltonian $K_B$ of a sphere $B$, 
it is convenient to employ the hyperspherical polar coordinates of $\mathbb{R}^d$ 
with origin in the center of the sphere.
These coordinates are given by the radial coordinate 
$r = | \boldsymbol{x} | = \sqrt{x_1^2 + \dots + x_d^2} \geqslant 0$
and by the vector $\boldsymbol{\Omega} = (\phi, \theta_1, \dots , \theta_{d-2})$
collecting all the angular coordinates\footnote{Denoting by $x_1$ 
the coordinate along a given vertical axes, the angular coordinates are given by 
\be
\phi = \arctan \frac{x_2}{x_1}
\;\qquad\;
\theta_1 = \arctan \frac{\sqrt{x_1^2 + x_2^2}}{x_3}
\;\qquad\;
\dots
\;\qquad\;
\theta_{d-2} = \arctan \frac{\sqrt{x_1^2 + \dots + x_{d-1}^2}}{x_{d}}
\ee
in terms of the Cartesian coordinates of $\mathbb{R}^d$, whose ranges are $\phi \in [0,2\pi)$ and $\theta_j \in [0,\pi]$.}  
of the 
$d-1$ dimensional unit sphere $\mathbb{S}$.

In these hyperspherical polar coordinates, the metric reads 
$\textrm{d}s^2 = - \,\textrm{d}t^2 + \textrm{d}r^2 + r^2\, \textrm{d}\boldsymbol{\Omega}^2$
and the corresponding volume element to employ in (\ref{hamiltonian}) is 
$\textrm{d}^d \boldsymbol{x} = r^{d-1} \textrm{d}r \, \textrm{d} \boldsymbol{\Omega}$,
where $\textrm{d} \boldsymbol{\Omega}$ is the volume element of $\mathbb{S}$.
The Laplacian operator $\Delta $ reads
$ \Delta  = \partial_r^2 + \tfrac{d-1}{r} \; \partial_r + \tfrac{1}{r^2}\, \tilde{\Delta}$,
where $\tilde{\Delta}$ is the spherical Laplace operator in $d-1$ dimensions, 
whose eigenfunctions are the real spherical harmonics in $d$ dimensions
$Y_{\boldsymbol{l}}(\boldsymbol{\Omega})$ of degree $l$
(see e.g. chapter XI in \cite{Bateman} or \cite{Abramowitz:1964:HMF,frye2012spherical})
\begin{equation}
\label{spherical-harmonics-def}
\tilde{\Delta} \,Y_{\boldsymbol{l}}(\boldsymbol{\Omega}) 
= 
\lambda_d(l) \,Y_{\boldsymbol{l}}(\boldsymbol{\Omega})
\;\;\;\qquad\;\;\;
\lambda_d(l) \equiv \, -\, l \big(l+d-2\big)
\end{equation}
where $\boldsymbol{l} = (l,i) $ with $l \geqslant 0$ 
and $1 \leqslant i \leqslant N_{d,l}$ labels 
the  linearly independent spherical harmonics of degree $l$ 
whose total number is
\be
N_{d,l} = 
\left\{\begin{array}{ll}
1 
& 
l=0
\\
\rule{0pt}{.8cm}
\displaystyle
\binom{\,l+d-1\,}{l} 
-
\binom{\,l+d-3\,}{l-2} 
\,=\,
\frac{2l+d-2}{l} \, \binom{\,l+d-3\,}{l-1} 
\hspace{1.2cm}
&
 l > 0
 \end{array}
 \right.
\ee
which can be obtained from the degeneracy of the $SO(d)$ representations.

The entanglement entropy of a sphere for the massless scalar field
has been first studied by Srednicki in  \cite{Srednicki:1993im}.
Following his analysis,
we decompose the fields in (\ref{hamiltonian}) as 
\be
\label{rescaled-fields-inv}
\Phi(r, \boldsymbol{\Omega})
\,=\,
r^{-\frac{d-1}{2}} 
\sum_{\boldsymbol{l}}
\Phi_{\boldsymbol{l}}(r) \;
Y_{\boldsymbol{l}}(\boldsymbol{\Omega})
\;\;\;\;\qquad\;\;\;\;
\Pi(r, \boldsymbol{\Omega})
\,=\,
r^{-\frac{d-1}{2}} 
\sum_{\boldsymbol{l}}
\Pi_{\boldsymbol{l}}(r) \;
Y_{\boldsymbol{l}}(\boldsymbol{\Omega})
\ee
where the sums can be written as
$\sum_{\boldsymbol{l}}(\dots)  = \sum_{l=0}^{\infty} \sum_{i=1}^{N_{d,l}} (\dots)$.

In the special case of $d=1$, the scalar field is on the half-line $r \geqslant 0$ 
and the angular coordinates $\boldsymbol{\Omega} $ do not occur;
hence the decomposition in (\ref{rescaled-fields-inv}) becomes trivial. 
Instead, when either $d=2$ or $d=3$,  
where we have respectively $N_{2,l} = 2$ and $N_{3,l} = 2l+1$,
the decomposition of $\Phi$ in (\ref{rescaled-fields-inv}) reads respectively 
\be
\Phi(r,\theta) = \Phi_0(r) 
+ \sum_{l>0}\left( \Phi_{l}(r) \, e^{il\theta}+\Phi_{-l}(r) \,e^{-il\theta} \right)
\qquad
\Phi(r,\theta_1,\theta_2) = \sum_{l\geqslant 0}\sum_{m=-l}^l \Phi_{l,m}(r) \,Y_{l,m}(\theta,\phi)\,.
\ee

By employing the orthonormality condition 
$\int_{\mathbb{S}} 
Y_{\boldsymbol{l}_1}(\boldsymbol{\Omega}) \,
Y_{\boldsymbol{l}_2}(\boldsymbol{\Omega}) 
\, \textrm{d}\boldsymbol{\Omega}\, 
=
\delta_{\boldsymbol{l}_1,\boldsymbol{l}_2}\,$
for the spherical harmonics,
one finds that the partial wave components 
$\Phi_{\boldsymbol{l}}(r)  $ and $\Pi_{\boldsymbol{l}}(r)$ 
in (\ref{rescaled-fields-inv}) are 
\be
\label{rescaled-fields-def}
\Phi_{\boldsymbol{l}}(r) 
\,=\,
r^{\frac{d-1}{2}} \!
\int_{\mathbb{S}}
\Phi(r, \boldsymbol{\Omega})\;
Y_{\boldsymbol{l}}(\boldsymbol{\Omega})
\,\textrm{d}\boldsymbol{\Omega}
\;\;\;\;\qquad\;\;\;\;
\Pi_{\boldsymbol{l}}(r) 
\,=\,
r^{\frac{d-1}{2}} \!
\int_{\mathbb{S}}
\Pi(r, \boldsymbol{\Omega})\;
Y_{\boldsymbol{l}}(\boldsymbol{\Omega})
\,\textrm{d}\boldsymbol{\Omega}\,.
\ee
These fields obey the canonical commutation relations
\be
\big[ \Phi_{\boldsymbol{l}_1}(r_1)  , \Pi_{\boldsymbol{l}_2}(r_2) \big] 
=  \textrm{i} \,\delta_{\boldsymbol{l}_1 , \boldsymbol{l}_2}\, \delta(r_1 - r_2)\,.
\ee
We remark that, for any $\boldsymbol{l}$,
the fields (\ref{rescaled-fields-def}) satisfy Dirichlet boundary conditions at $r=0$.
This is a crucial feature in our analysis. 

By employing (\ref{spherical-harmonics-def})
and the decompositions (\ref{rescaled-fields-inv}) in the Hamiltonian (\ref{hamiltonian}),
it becomes
\be
\label{phys-H-decomposed}
H = \sum_{\boldsymbol{l}} H_{\boldsymbol{l}}
\;\;\;\qquad\;\;\;
H_{\boldsymbol{l}} 
\,\equiv
\int_0^{\infty} 
\!  \widetilde{T}_{tt}^{(\boldsymbol{l})} 
\,\textrm{d}r
\ee
where 
\be
\label{T00-scalar-l-mode}
\widetilde{T}_{tt}^{(\boldsymbol{l})} 
\equiv
\frac{1}{2} \,\bigg\{\,
\Pi_{\boldsymbol{l}}^2
-
\Phi_{\boldsymbol{l}}\, \partial_r^2 \Phi_{\boldsymbol{l}}
+ 
\left(
\frac{\mu_d(l)}{r^2}
+ 
m^2 
\right) \Phi_{\boldsymbol{l}}^2
\,\bigg\}
\ee
with $\mu_d(l)$ defined in terms of $\lambda_d(l)$ introduced in (\ref{spherical-harmonics-def}) 
as follows
\be
\label{mu-d-def}
\mu_d(l) \,\equiv \,
\frac{(d-3)(d-1)}{4} - \lambda_d(l)\,.
\ee
From  (\ref{phys-H-decomposed}) and the fact that the fields (\ref{rescaled-fields-def}) 
corresponding to different $\boldsymbol{l}$'s commute,
one concludes that the ground state of $H$ is the direct product of the ground states of $H_{\boldsymbol{l}}$ for different $\boldsymbol{l}$'s 
\cite{Srednicki:1993im}.
When $d=1$, only $l=0$ is allowed; 
hence the coefficient (\ref{mu-d-def}) vanishes
and the sum in (\ref{phys-H-decomposed}) contains a single term.
In this case (\ref{T00-scalar-l-mode})  
becomes  the energy density for the scalar field on the half-line 
satisfying Dirichlet boundary conditions at the beginning of the half-line  \cite{Liguori:1997vd}.

Following the regularisation procedure discussed in
\cite{Srednicki:1993im} (see also \cite{Bombelli:1986rw}),
the ultraviolet divergences are regularised
by introducing a discretisation of the radial direction (with lattice spacing $a$)
and the infrared ones by confining the system into a finite volume;
hence we consider a large but finite number $R_{\textrm{\tiny tot}}$ of sites 
along the discretised radial direction.
At the $j$-th site of the discretised radial 
direction\footnote{In the notation adopted throughout this manuscript, hatted operators are defined on the lattice.},
the position operator $\hat{q}_{\boldsymbol{l},j} $ 
and the momentum operator $\hat{p}_{\boldsymbol{l},j} $
are dimensionless, Hermitian and satisfy the canonical commutation relations
$ \big[ \hat{q}_{\boldsymbol{l}_1,i}  \, , \, \hat{p}_{\boldsymbol{l}_2,j} \big] 
= \textrm{i} \,\delta_{\boldsymbol{l}_1 , \boldsymbol{l}_2}\, \delta_{i,j} $.

In the continuum limit $a \to 0$ and $R_{\textrm{\tiny tot}} \to \infty$.
Since we keep $R_{\textrm{\tiny tot}} a = \mathcal{R}_{\textrm{\tiny tot}}$ finite in this limit, 
the system in the continuum is enclosed within 
the finite sphere  of radius $\mathcal{R}_{\textrm{\tiny tot}}$.
The radial position corresponds to $r = j a$, with $0 \leqslant r \leqslant  \mathcal{R}_{\textrm{\tiny tot}}$,
and the fields $\Phi_{\boldsymbol{l}}(r)$ and $\Pi_{\boldsymbol{l}}(r)$ in the continuum limit
(which vanish identically for $r> \mathcal{R}_{\textrm{\tiny tot}}$ \cite{Srednicki:1993im})
can be introduced through the position and momentum operators 
in the standard way
\be 
\label{qp-field-replacement}
\hat{q}_{\boldsymbol{l},j} \, \longrightarrow \,  \Phi_{\boldsymbol{l}}(r)
\;\; \qquad \;\;
\hat{p}_{\boldsymbol{l},j} \, \longrightarrow \, a \, \Pi_{\boldsymbol{l}}(r)\,.
\ee
The lattice regularisation defined above
leads to $(\hat{q}_{\boldsymbol{l},j+1} + \hat{q}_{\boldsymbol{l},j-1} - 2\hat{q}_{\boldsymbol{l},j})/a^2$
for the regularisation of $ \partial_r^2 \Phi_{\boldsymbol{l}}$.
Thus, for  the lattice regularisation $\widehat{H}_{\boldsymbol{l}}$ 
of the operator $H_{\boldsymbol{l}}$ 
in (\ref{phys-H-decomposed}) we find
\be
\label{Ham-fixed-mode}
\widehat{H}_{\boldsymbol{l}}
\,=\, 
\frac{1}{2a}\, 
\Big(  
\boldsymbol{\hat{p}}^{\textrm t}_{\boldsymbol{l}} \,
\boldsymbol{I}_{\textrm{\tiny tot}}
\, \boldsymbol{\hat{p}}_{\boldsymbol{l}}
 +
\boldsymbol{\hat{q}}^{\textrm t}_{\boldsymbol{l}} \,
M_{l} 
\,\boldsymbol{\hat{q}}_{\boldsymbol{l}}
\Big)
\ee
where $\boldsymbol{\hat{p}}_{\boldsymbol{l}}$ and $\boldsymbol{\hat{q}}_{\boldsymbol{l}}$ are the vectors whose $j$-th element is  given 
by $\hat{p}_{\boldsymbol{l},j}$ and $\hat{q}_{\boldsymbol{l},j}$ respectively.
The matrix $\boldsymbol{I}_{\textrm{\tiny tot}}$ is the $R_{\textrm{\tiny tot}} \times R_{\textrm{\tiny tot}}$ identity matrix
and $M_{l} $
is the $R_{\textrm{\tiny tot}} \times R_{\textrm{\tiny tot}}$ 
real, symmetric, positive definite and tri-diagonal matrix 
whose non vanishing entries are 
\be
\label{matrix-M}
\big( M_{l} \big)_{j,j}
= \,
2+\frac{\mu_d(l)}{j^2}+\omega^2
\;\;\;\qquad\;\;\;
\big( M_{l}\big)_{j,j+1} = \, -1
\ee
where $\mu_d(l)$ has been defined in (\ref{mu-d-def}),
the addition term $2$ in $( M_{l} )_{j,j}$
comes from the discretisation of $\Phi_{\boldsymbol{l}}\, \partial_r^2 \Phi_{\boldsymbol{l}}$
and $\omega$ corresponds to the mass parameter in the lattice along the radial direction,
which is related to the mass parameter $m$ occurring in the Hamiltonian (\ref{hamiltonian})
in the continuum limit as $\omega / a  \,\longrightarrow \,  m$.
Since $\mu_d(l)$ is quadratic in the parameters $d$ and $l$
and the coefficients of $d^2$ and $l^2$ are both positive,
the large $l$ regime at given $d$ is similar
to the regime given by large $d$ with $l$ fixed.

The ground state correlators 
$\langle \hat{q}_{\boldsymbol{l},i} \,\hat{q}_{\boldsymbol{l},j}\rangle $ and 
$\langle \hat{p}_{\boldsymbol{l},i} \,\hat{p}_{\boldsymbol{l},j}\rangle $ 
provide the elements of the $R_{\textrm{\tiny tot}} \times R_{\textrm{\tiny tot}}$ 
correlation matrices $Q_{l}$ and $P_{l}$ respectively,
which can be obtained from (\ref{matrix-M}) as 
\cite{Bombelli:1986rw, Srednicki:1993im, Audenaert_2002, Cramer:2005mx, Cramer_2006b}
\be
\label{corr-matrices-mode}
Q_{l} \, \equiv \, \frac{1}{2}\,  M_{l}^{-1/2}
\;\;\;\qquad\;\;
P_{l}\,  \equiv \, \frac{1}{2}\, M_{l}^{1/2}\,.
\ee
These correlation matrices are well defined when $\omega = 0$, for any $d \geqslant 1$
and $l \geqslant 0$.
In particular, since translation invariance does not occur in the lattice model along the radial direction,
we do not have to deal with the zero mode.
We checked numerically  that,
when $\mu_d(l)=0$ in (\ref{matrix-M}) and for large values of $R_{\textrm{\tiny tot}}$,
the elements of the correlation matrices (\ref{corr-matrices-mode}) 
agree with the analytical expressions for the correlators on the half line 
with Dirichlet boundary condition at the origin of the half-line
given in \cite{Calabrese:2012nk} for $\omega =0$ and in \cite{DiGiulio:2019cxv} for a generic $\omega \geqslant 0$
(see \cite{Botero04} for the corresponding correlators for the harmonic chain on the line, where the zero mode occurs).
In the appendix\;\ref{app_corr_check} 
we compare (\ref{corr-matrices-mode}) when $\omega = 0$
with the corresponding expressions obtained in the continuum 
through quantum field theory methods \cite{Frolov:1999an,Saharian:2000mw}.

Various discretisations of the Hamiltonian of the real scalar field
have been introduced to study the entanglement entropy of the sphere 
\cite{Srednicki:1993im, Riera_2006, Lohmayer:2009sq,Huerta:2011qi, Liu:2012eea, Safdi:2012sn, Klebanov:2012va, Casini:2015dsg, Cotler:2016acd,Nishioka:2018khk}.
We have employed the one characterised by (\ref{matrix-M})
(see e.g. also \cite{Herzog:2014fra}).
In the appendix\;\ref{app_corr_check},
the correlators (\ref{corr-matrices-mode}) corresponding to the matrix $M_l$
obtained through a different discretisation are also discussed 
\cite{Srednicki:1993im, Riera_2006, Huerta:2011qi}.


\subsection{Entanglement Hamiltonian of a sphere}
\label{subsec-eh}

We consider the spatial  bipartition of $\mathbb{R}^d$ 
given by  a sphere $B$ of radius $\mathcal{R}$ and its complement 
in the $d+1$ dimensional Minkowski spacetime at any given time slice. 
When the system is a CFT in its ground state,
the entanglement Hamiltonian $K_B $ is (\ref{EH-intro}),
which has been obtained in \cite{Hislop:1981uh,Casini:2011kv}
through a conformal transformation 
of the entanglement Hamiltonian found
by Bisognano and Wichmann  \cite{Bisognano:1975ih, Bisognano:1976za}.

In the special case of the free massless scalar field, 
$K_B $ is (\ref{EH-intro}) with $T_{tt}$ given by 
the integrand of (\ref{hamiltonian}) when $m=0$.
In the resulting $K_B$, we can straightforwardly adapt 
the steps that have led to write (\ref{hamiltonian}) as (\ref{phys-H-decomposed}),
finding that
\be
\label{KB-mode-sum}
K_B = \sum_{\boldsymbol{l}} K_{B,\boldsymbol{l}}
\ee
where
\be
\label{EH-ball-massless-scalar}
K_{B,\boldsymbol{l}}
\,\equiv\,
\mathcal{R}
\int_0^{\mathcal{R}} 
\!\! \beta(r)\,
\widetilde{T}_{tt}^{(\boldsymbol{l})} 
\,\textrm{d}r
\ee
with $\beta(r)$ being defined as the parabola (\ref{beta-def-intro})
and the operator $\widetilde{T}_{tt}^{(\boldsymbol{l})} $ by (\ref{T00-scalar-l-mode}) with $m=0$.


In the regularised model obtained by discretising the radial direction 
and described in Sec.\,\ref{subsec-ham}),
the sphere $B$ has radius $R < R_{\textrm{\tiny tot}}$.
In the continuum limit, 
$a \to 0$ and $R \to +\infty$ with $ Ra = \mathcal{R}$ fixed.
The reduced correlation matrices $Q_{l,B}$ and $P_{l,B}$ 
at a given value of the angular momentum parameter $l$
are the $R \times R$ symmetric and positive definite matrices 
obtained by restricting the corresponding correlation matrices in 
(\ref{corr-matrices-mode}) 
to the sites identifying the sphere $B$ along the radial direction, 
i.e. 
$ (Q_{l,B})_{i,j} =(Q_{l})_{i,j}$ 
and $ (P_{l,B})_{i,j} =(P_{l})_{i,j}$ 
with $1 \leqslant i \leqslant R$ and $1 \leqslant j \leqslant R$.

The reduced correlation matrices $Q_{l,B}$ and $P_{l,B}$ at given $l$ 
provide the reduced covariance matrix $Q_{l,B} \oplus P_{l,B}$ at fixed $l$,
whose symplectic spectrum 
$\{ \sigma_{l,k} \,; 1 \leqslant k \leqslant R\}$
is given by the eigenvalues of $\sqrt{Q_{l,B} \,P_{l,B}}$,
which are the same ones of its transpose $\sqrt{P_{l,B} \,Q_{l,B}}$.
The uncertainty principle implies that $\sigma_{l,k}>1/2$.
Since the ground state of this model is Gaussian
and the operators associated to different  $\boldsymbol{l}$'s commute,
the symplectic spectra of the covariance matrices $Q_{l,B} \oplus P_{l,B}$ at fixed $l$
corresponding to different values of $l$
give the R\'enyi entropies and the entanglement entropy
(see (\ref{EE-mode-sum}) and the corresponding discussion).

The Gaussian nature of the ground state, 
combined with the fact that operators corresponding to different $\boldsymbol{l}$'s commute,
leads to the entanglement Hamiltonian $\widehat{K}_B = \sum_{\boldsymbol{l}} \widehat{K}_{B,\boldsymbol{l}}$ in the regularised model,
where $\widehat{K}_{B,\boldsymbol{l}}$ can be obtained through the results 
obtained in \cite{Peschel:2003rdm, EislerPeschel:2009review, Casini:2009sr, Banchi:2015aaa}.
In particular, the contribution to $\widehat{K}_{B}$ of the degrees of freedom associated to the sector
labelled by $\boldsymbol{l}$ is given by the following quadratic operator \cite{Casini:2009sr}
\be
\label{ent-ham HC}
\widehat{K}_{\boldsymbol{l},B}
= \frac{1}{2}\, \boldsymbol{\hat{u}}_{\boldsymbol{l},B}^{\textrm t} \,H_{l,B}\, \boldsymbol{\hat{u}}_{\boldsymbol{l},B}
\hspace{.5cm} \qquad \hspace{.5cm} 
\boldsymbol{\hat{u}}_{\boldsymbol{l},B}
\equiv
\bigg(
\begin{array}{c}
\boldsymbol{\hat q}_{\boldsymbol{l}} \\  \boldsymbol{\hat p}_{\boldsymbol{l}}
\end{array}  \bigg)
\bigg|_{B}
\ee
where $H_{l,B}$ is the symmetric, positive definite and block diagonal matrix defined 
in terms of the reduced correlation matrices $Q_{l,B}$ and $P_{l,B}$ 
at fixed $l$ as follows
\bea
\label{eh-block-ch-version}
H_{l,B}
\,\equiv\,
V^{(l)} \oplus T^{(l)} 
&\equiv &
\Big( h\big(\sqrt{P_{l,B} \,Q_{l,B}}\,\big)   
\oplus 
h\big(\sqrt{Q_{l,B} \,P_{l,B}}\,\big) \Big) 
 \big( P_{l,B} \oplus Q_{l,B} \big)
 \\
&=&
 \big( P_{l,B} \oplus Q_{l,B} \big)
 \Big( h\big(\sqrt{Q_{l,B} \,P_{l,B}}\,\big)  
  \oplus 
  h\big(\sqrt{P_{l,B} \,Q_{l,B}}\,\big) \Big) 
 \nonumber
\eea
with 
\be
\label{h-function def}
h(y)  \equiv  \frac{1}{y} \, \log \!\left(\frac{y +1/2}{y - 1/2} \right).
\ee

The symplectic spectrum of $H_{l,B}$ in (\ref{eh-block-ch-version})
gives the single particle entanglement energies $\varepsilon_{l,k} $ at given $l$
(hence $\varepsilon^2_{l,k}$ are the eigenvalues of $V^{(l)} \,T^{(l)}$),
which are related to the symplectic eigenvalues of $Q_{l,B} \oplus P_{l,B}$
as $\varepsilon_{l,k} = 2\,\textrm{arccoth}(2\sigma_{l,k} ) = \log[(\sigma_{l,k}+1/2)/(\sigma_{l,k}-1/2)]$,
whose inverse is $\sigma_{l,k} = \tfrac{1}{2}\coth(\varepsilon_{l,k}/2)$.
Notice that both $\sigma_{l,k} $ and  $\varepsilon_{l,k}$ depend also on 
$d$ through (\ref{matrix-M}).

\section{Continuum limit of the entanglement Hamiltonian at $\omega=0$}
\label{sec_EH_ball_massless}

In this section we study the entanglement Hamiltonian $K_{B} $  
of a sphere $B \in \mathbb{R}^d$ 
for the free massless scalar field by taking the continuum limit 
of the corresponding entanglement Hamiltonian $\widehat{K}_{B} $ 
in the regularised model, 
whose contribution for a given $\boldsymbol{l}$ is (\ref{ent-ham HC}) at $\omega  = 0$.
The expected result is the CFT expression (\ref{EH-intro}) 
specialised to the free massless scalar field,
which can be written as in (\ref{KB-mode-sum}),
in terms of the operator (\ref{T00-scalar-l-mode}) with $m=0$.
In the resulting CFT expression
a non trivial $r$ dependent term occurs
whose coefficient (\ref{mu-d-def}) depends both on $d$ and $l$.
This coefficient vanishes identically when $d=1$.
In our numerical analysis
we follow the procedure employed in 
\cite{Arias:2016nip,Eisler:2017cqi,Eisler:2019rnr,DiGiulio:2019cxv} 
for free bosonic and fermionic chains (i.e. for $d=1$) 
at criticality and in their ground states
to study the entanglement Hamiltonians of an interval
either on the infinite line \cite{Arias:2016nip,Eisler:2017cqi,Eisler:2019rnr,DiGiulio:2019cxv} 
or at the beginning of the half-line with Dirichlet boundary conditions at the origin \cite{DiGiulio:2019cxv}.


Plugging the matrix (\ref{eh-block-ch-version}) into  (\ref{ent-ham HC}), 
one finds that the operator $\widehat{K}_{\boldsymbol{l},B} $ 
can be written as
\be
\label{KA/2}
\widehat{K}_{\boldsymbol{l},B} 
= 
\frac{1}{2}
\Big(
\widehat{H}_{T,\boldsymbol{l}} + \widehat{H}_{V,\boldsymbol{l}}
\Big)
\ee
where the operators $\widehat{H}_{T,\boldsymbol{l}} $ and $\widehat{H}_{V,\boldsymbol{l}} $ are defined 
through the symmetric matrices $T^{(l)}$ and $V^{(l)}$ as follows
\be
\label{H_T and H_V operators}
\widehat{H}_{T,\boldsymbol{l}} 
\equiv
\sum_{i,j =1}^R T^{(l)}_{i,j} \,\hat{p}_{\boldsymbol{l},i}  \, \hat{p}_{\boldsymbol{l},j} 
\;\;\;\;\qquad\;\;\;\;
\widehat{H}_{V,\boldsymbol{l}} 
\equiv
\sum_{i,j =1}^R V^{(l)}_{i,j} \,\hat{q}_{\boldsymbol{l},i}  \, \hat{q}_{\boldsymbol{l},j} \,.
\ee

These sums can be organised in various  ways,
as discussed in \cite{DiGiulio:2019cxv} for $d=1$.
We find it convenient to write the sums in (\ref{H_T and H_V operators})
by decomposing the contribution coming from the $i$-th row of the matrices 
$V^{(l)}$ and $T^{(l)}$, 
as also done in \cite{Arias:2016nip,DiGiulio:2019cxv} in the case of $d=1$.
This choice leads to 
\bea
\label{H_T Casini}
\widehat{H}_{T,\boldsymbol{l}} 
&=&
R
\sum_{i=1}^R
\left( \frac{T^{(l)}_{i,i}}{R} \,\hat{p}_{\boldsymbol{l},i} ^2
+
 \sum_{k =1}^{R-i} \frac{T^{(l)}_{i,i+k}}{R} \;
 \hat{p}_{\boldsymbol{l},i}  \, \hat{p}_{\boldsymbol{l},i+k} 
 +
\sum_{k =1}^{i-1} \frac{T^{(l)}_{i,i-k}}{R} \;
 \hat{p}_{\boldsymbol{l},i}  \, \hat{p}_{\boldsymbol{l},i-k} 
\right)
\\
\rule{0pt}{1cm}
\label{H_V Casini}
\widehat{H}_{V,\boldsymbol{l}} 
&=&
R
\sum_{i=1}^R
\left( \frac{V^{(l)}_{i,i}}{R} \,\hat{q}_{\boldsymbol{l},i} ^2
+
 \sum_{k =1}^{R-i} \frac{V^{(l)}_{i,i+k}}{R} \;
  \hat{q}_{\boldsymbol{l},i}  \, \hat{q}_{\boldsymbol{l},i+k} 
 +
\sum_{k =1}^{i-1} \frac{V^{(l)}_{i,i-k}}{R} \;
 \hat{q}_{\boldsymbol{l},i}  \, \hat{q}_{\boldsymbol{l},i+k} 
\right) .
\eea

Following the analysis made in  \cite{Eisler:2017cqi, DiGiulio:2019cxv} for $d=1$,
we conjecture the existence of the limits 
\be
\label{munuk def}
\lim_{R \to \infty} \frac{T^{(l)}_{i,i + \eta  k}}{R}  
\,\equiv\, 
\tau_{l,\eta k}(r_{\eta k})
\;\;\;\qquad\;\;\;
\lim_{R \to \infty}  \frac{V^{(l)}_{i,i + \eta k}}{R}  
\,\equiv\, 
\nu_{l,\eta k}(r_{\eta k})
\ee
where $k \geqslant 0$ and $\eta = \pm 1$ are discrete parameters and 
\be
\label{r_k def}
r_{\tilde{k}}
\equiv \frac{1}{R} \bigg( i + \frac{\tilde{k}}{2}\,\bigg)\,.
\ee
Notice that $i \pm k/2$ in (\ref{munuk def})
corresponds to the midpoint between the $i$-th and the $(i\pm k)$-th site 
along the discretised radial direction. 
Since $T^{(l)}$ and $V^{(l)}$ are symmetric matrices,
their diagonals labelled by $+k$ and $-k$ in  (\ref{munuk def}) contain the same elements.

In our numerical analyses we considered 
lattices along the radial direction made by 
$R_{\textrm{\tiny tot}} \in \{ 200,\,400,\,600,\,800 \}$ 
and kept $R / R_{\textrm{\tiny tot}}  \ll 1 $ fixed
(we choose $R / R_{\textrm{\tiny tot}} = 1/10$).
This choice is imposed by the fact that 
the continuum result that we are studying
holds in the limit where the volume of the space is infinite.
%
%
As already remarked in previous studies on entanglement Hamiltonians in free chains
\cite{Arias:2016nip,Eisler:2017cqi,DiGiulio:2019cxv}, 
high numerical precision is usually needed.
In our case, they are required to evaluate (\ref{eh-block-ch-version})
because many symplectic eigenvalues of $Q_{l,B}  \oplus P_{l,B}$
are very close to $1/2$ and they must be distinguished from $1/2$ 
in order to be employed in (\ref{h-function def}).
Higher precision is needed as $l$ or $d$ or $\omega$ increases.
Thus, for $d=2$ and $d=3$ we have restricted our numerical analysis to $l\leqslant 10$ and $\omega\leqslant 5$,
working up to a precision of 2000 digits for the highest values of these parameters.

\begin{figure}[H]
\vspace{-.8cm}
\centering
\begin{subfigure}{.4\textwidth}
\hspace{-1.2cm}
\vspace{.3cm}
\includegraphics[scale=.34]{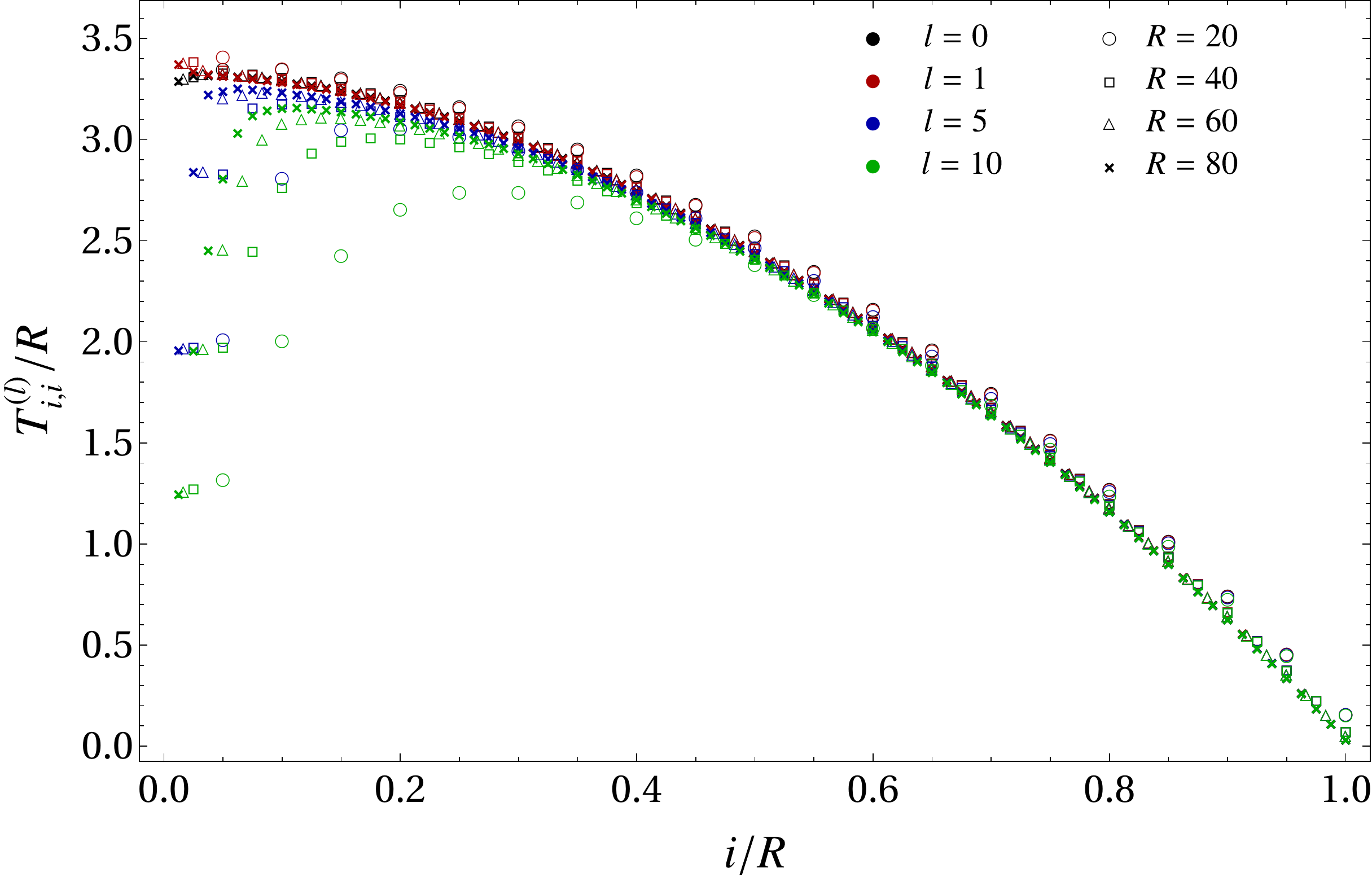}
\end{subfigure}
\hfill
\begin{subfigure}{.4\textwidth}
\hspace{-1.26cm}
\vspace{.3cm}
\includegraphics[scale=.342]{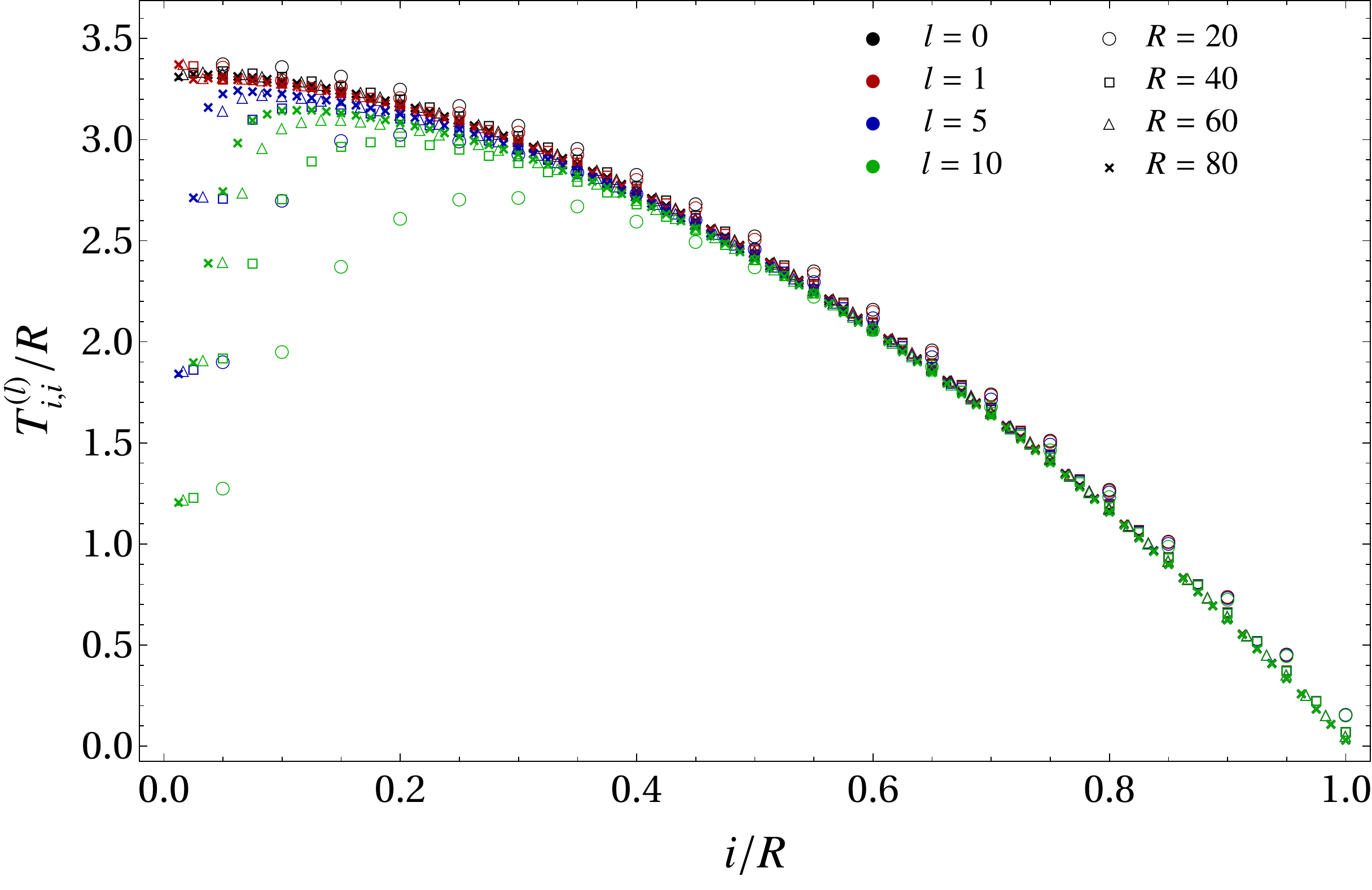}
\end{subfigure}

\begin{subfigure}{.4\textwidth}
\hspace{-1.5cm}
\vspace{.3cm}
\includegraphics[scale=.35]{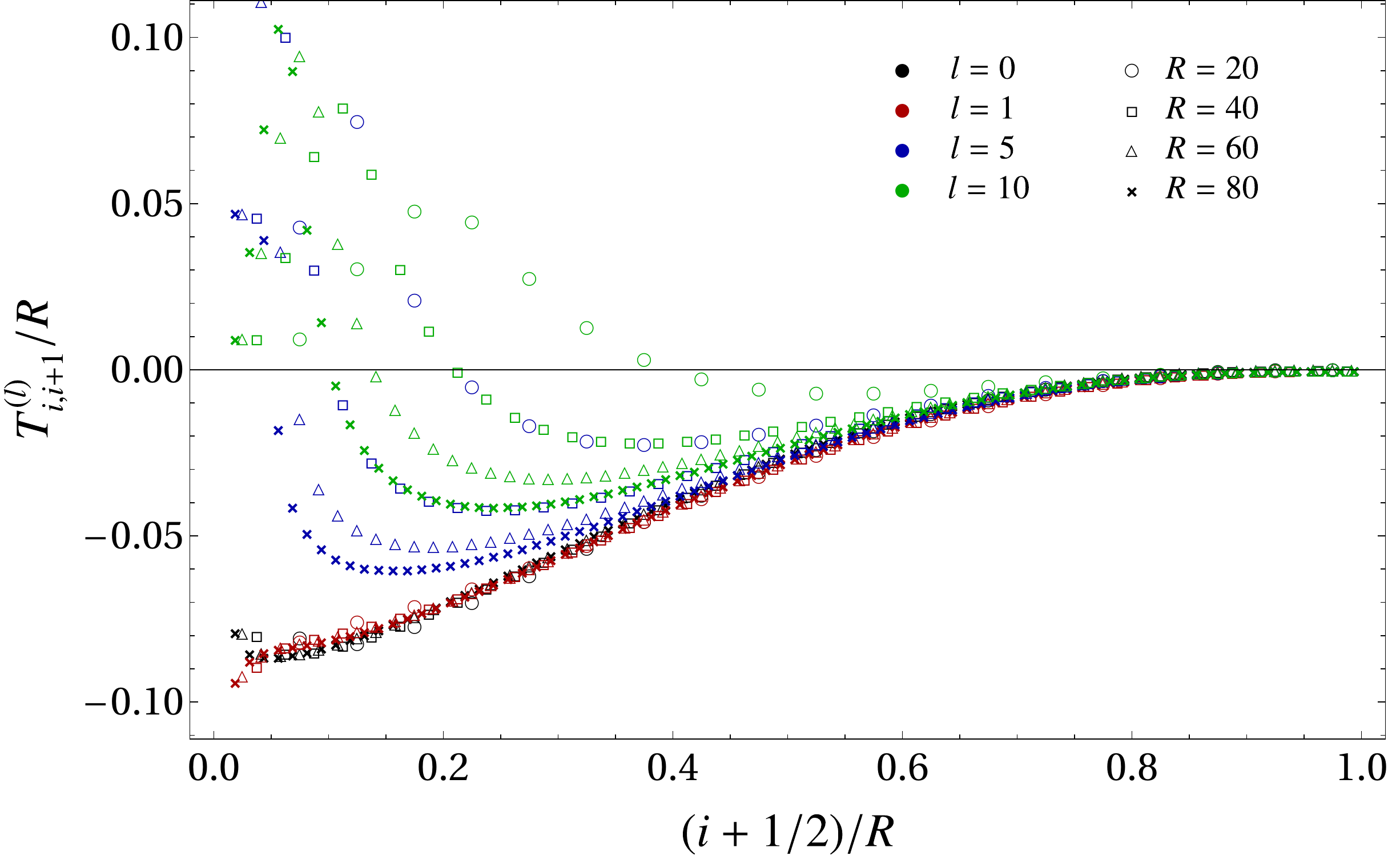}
\end{subfigure}
\hfill
\begin{subfigure}{.4\textwidth}
\hspace{-1.5cm}
\vspace{.3cm}
\includegraphics[scale=.35]{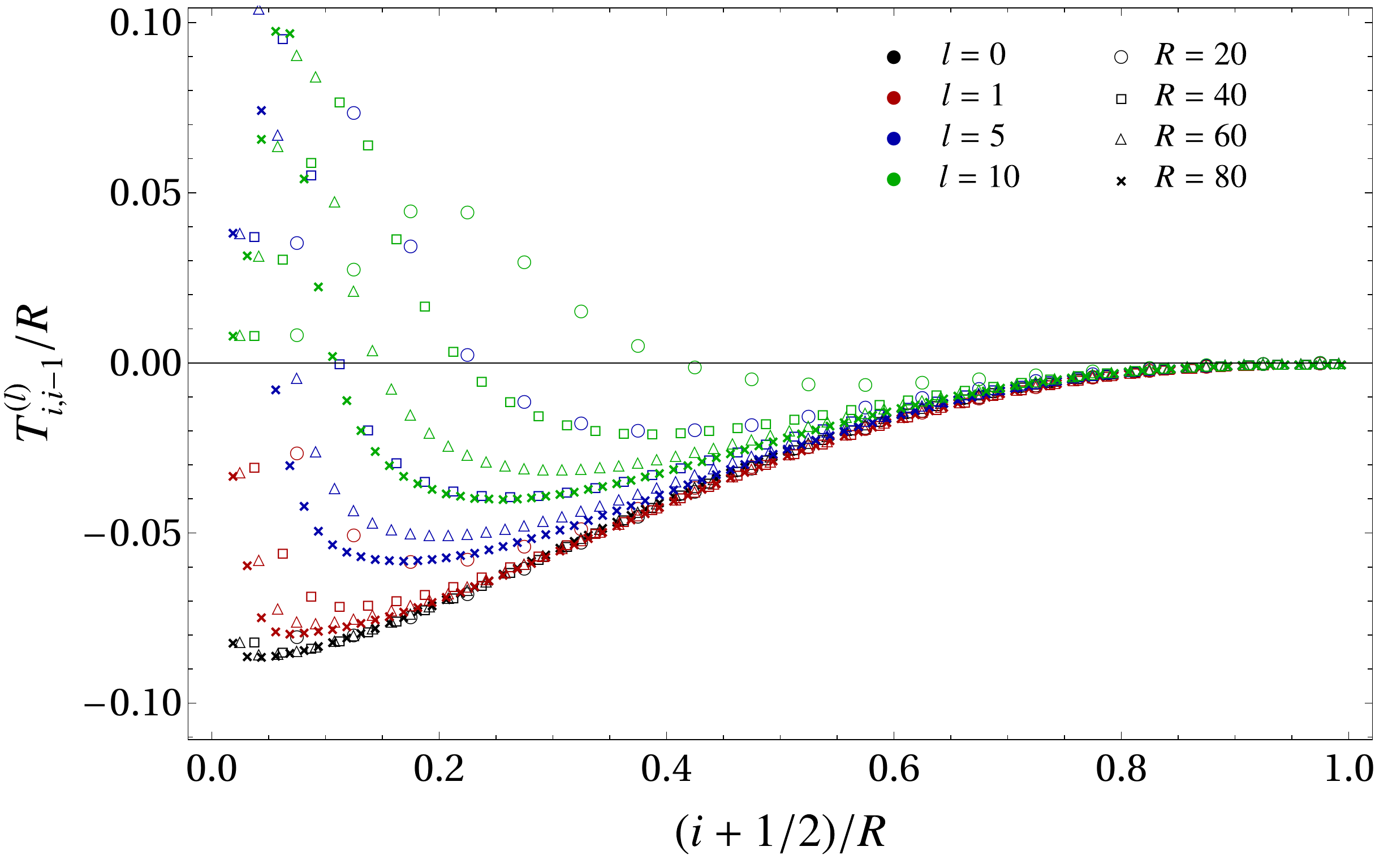}
\end{subfigure}

\begin{subfigure}{.4\textwidth}
\hspace{-1.5cm}
\vspace{.3cm}
\includegraphics[scale=.35]{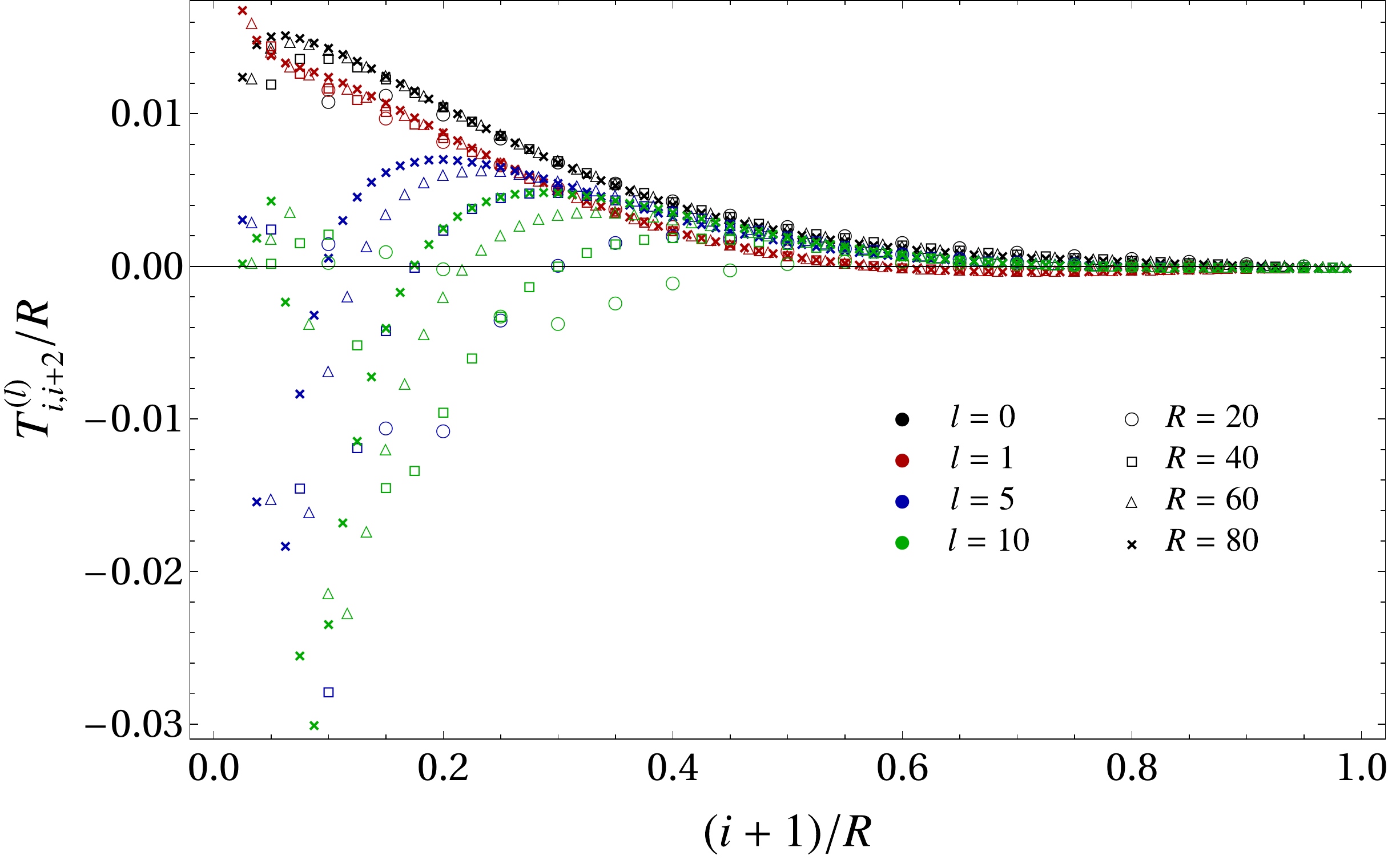}
\end{subfigure}
\hfill
\begin{subfigure}{.4\textwidth}
\hspace{-1.5cm}
\vspace{.3cm}
\includegraphics[scale=.35]{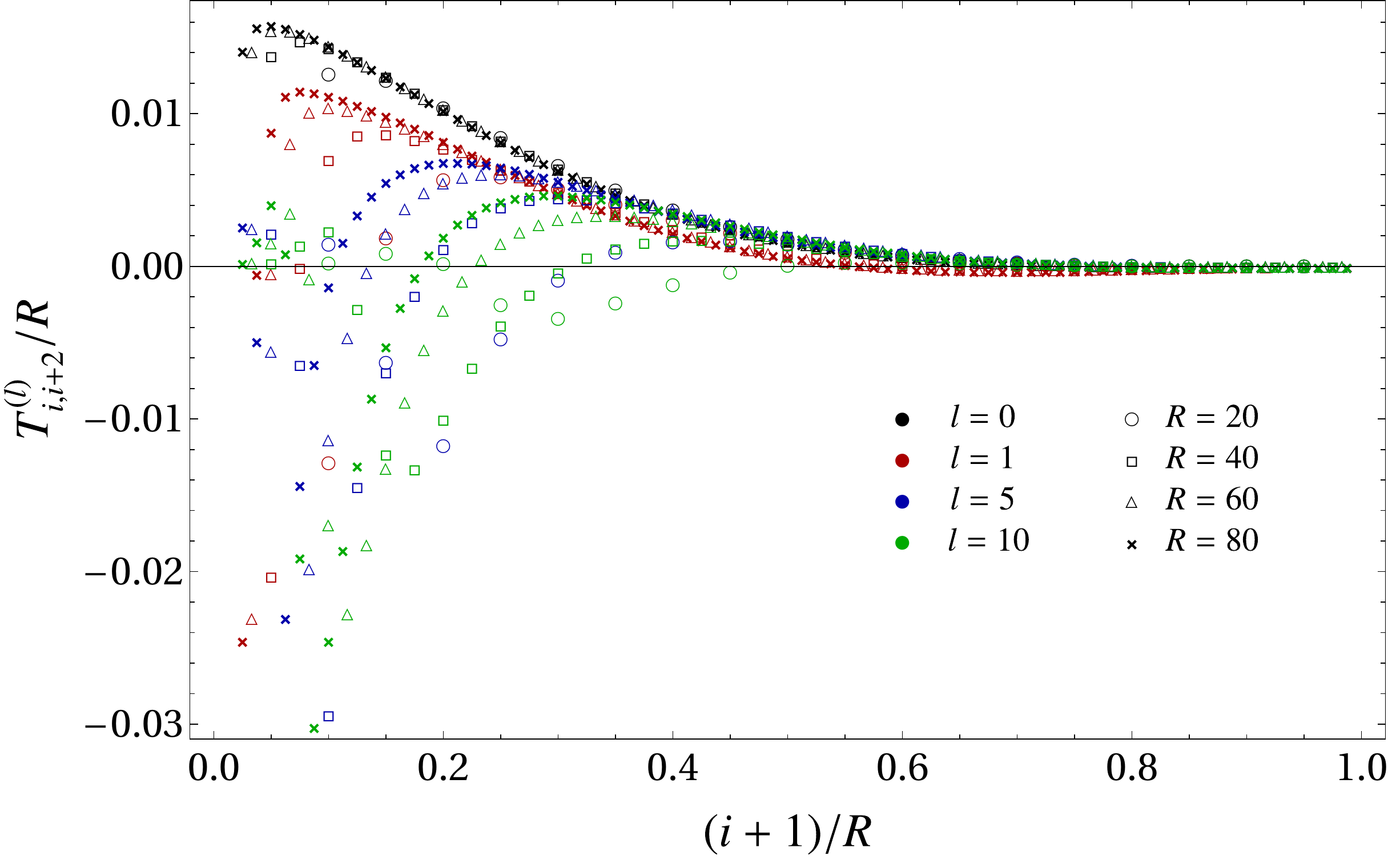}
\end{subfigure}

\begin{subfigure}{.4\textwidth}
\hspace{-1.6cm}
\vspace{.3cm}
\includegraphics[scale=.355]{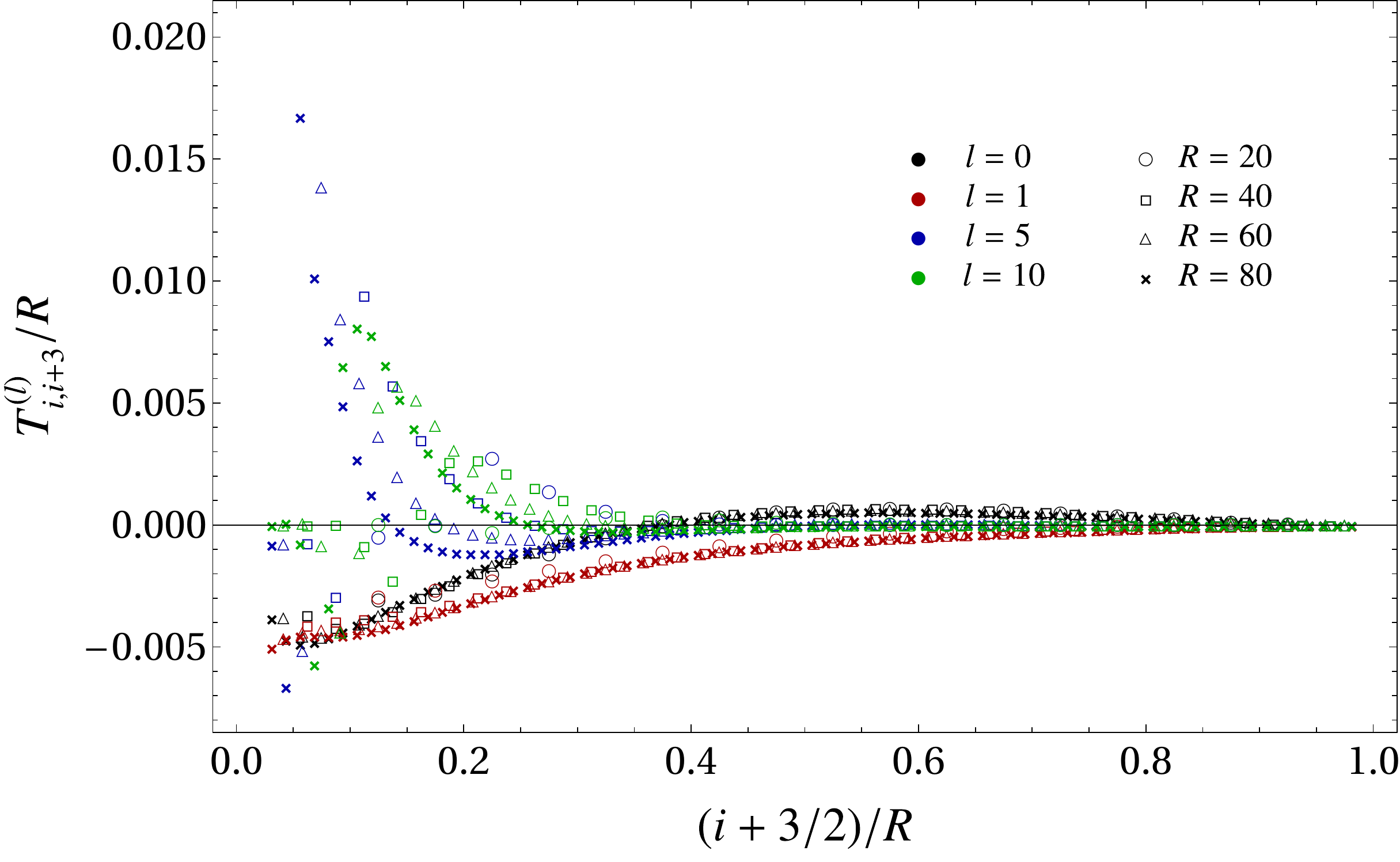}
\end{subfigure}
\hfill
\begin{subfigure}{.4\textwidth}
\hspace{-1.5cm}
\vspace{.3cm}
\includegraphics[scale=.355]{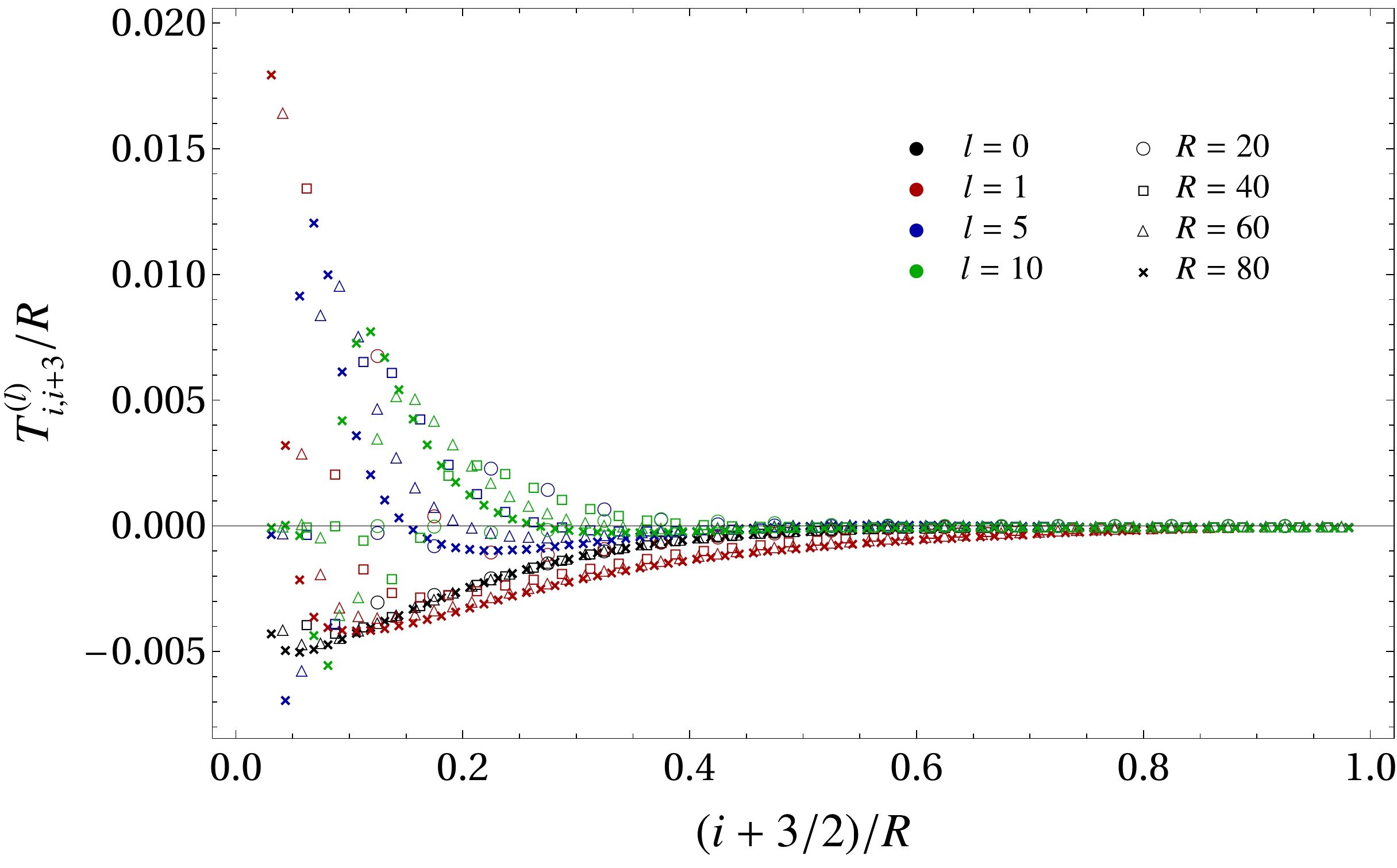}
\end{subfigure}
\vspace{-.3cm}
\caption{
Short-range couplings in the matrix $T^{(l)}/R$ 
(see (\ref{H_T and H_V operators}) and (\ref{munuk def})) 
when $\omega = 0$ and
either $d=2$ (left panels) or $d=3$ (right panels),
for some small $l$'s and different sizes $R$.
}
\label{fig-diagonals-T}
\end{figure}

\begin{figure}[H]
\vspace{-.8cm}
\centering
\begin{subfigure}{.4\textwidth}
\hspace{-1.3cm}
\vspace{.3cm}
\includegraphics[scale=.34]{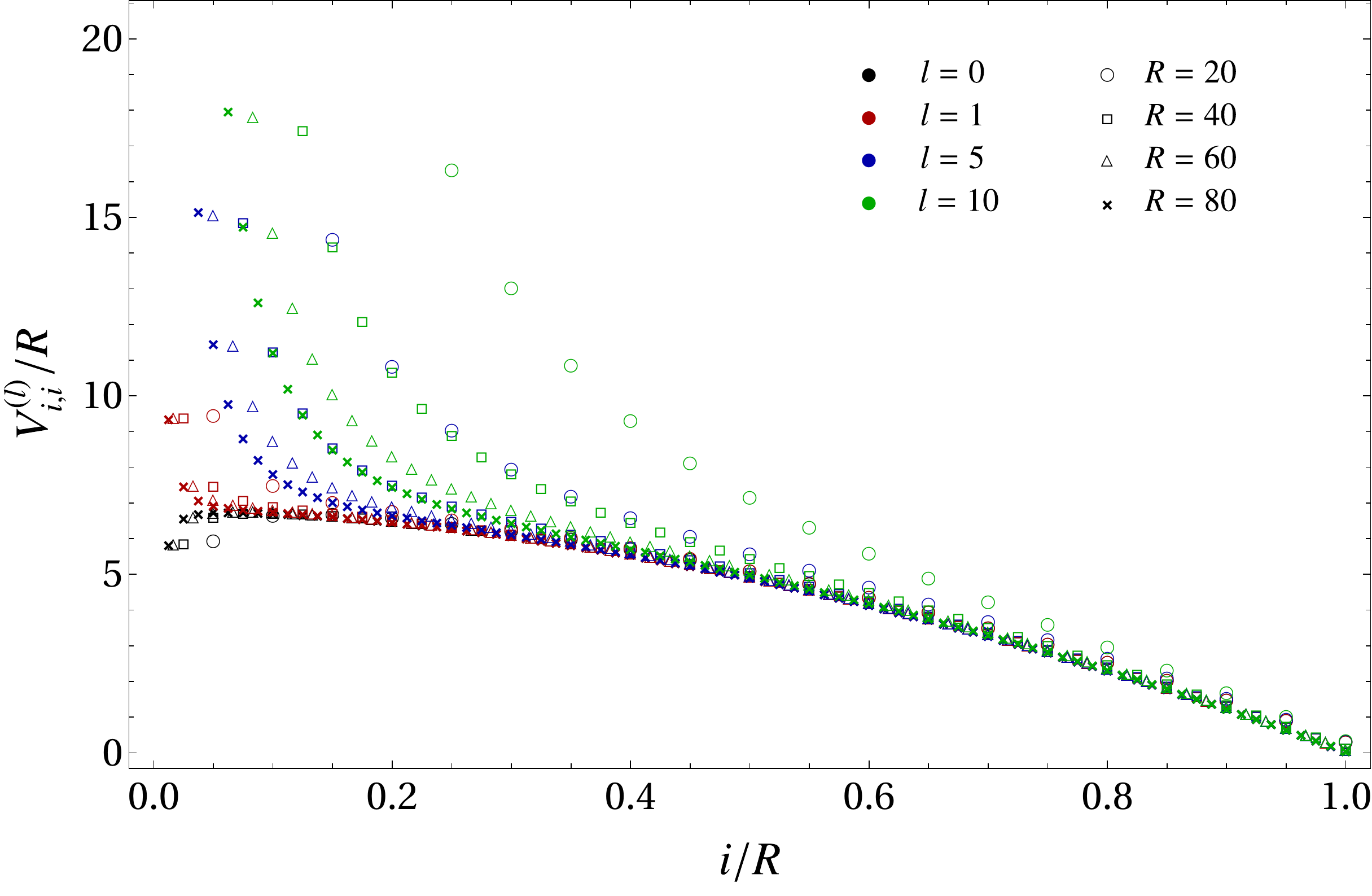}
\end{subfigure}
\hfill
\begin{subfigure}{.4\textwidth}
\hspace{-1.3cm}
\vspace{.3cm}
\includegraphics[scale=.34]{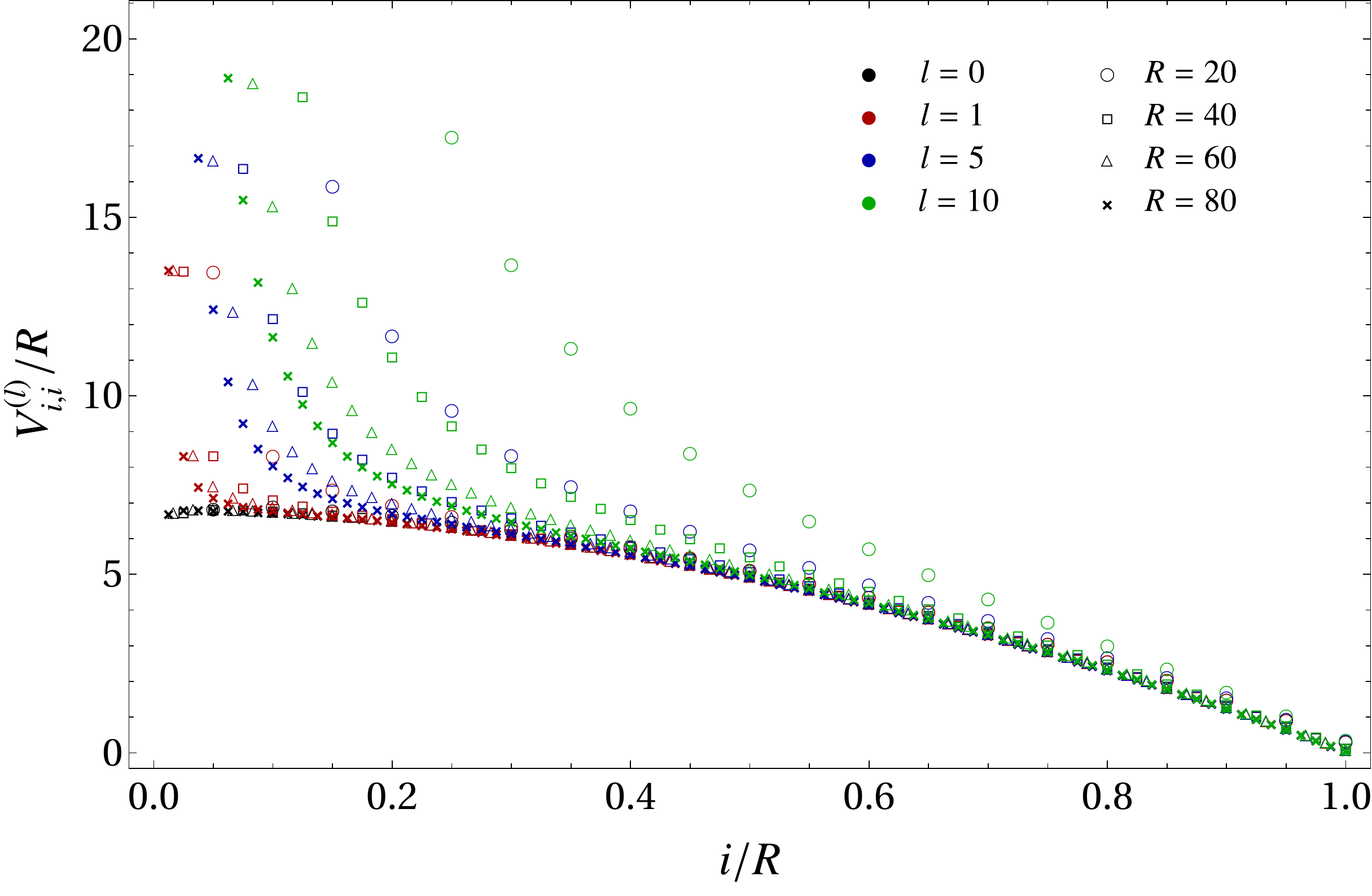}
\end{subfigure}

\begin{subfigure}{.4\textwidth}
\hspace{-1.5cm}
\vspace{.3cm}
\includegraphics[scale=.35]{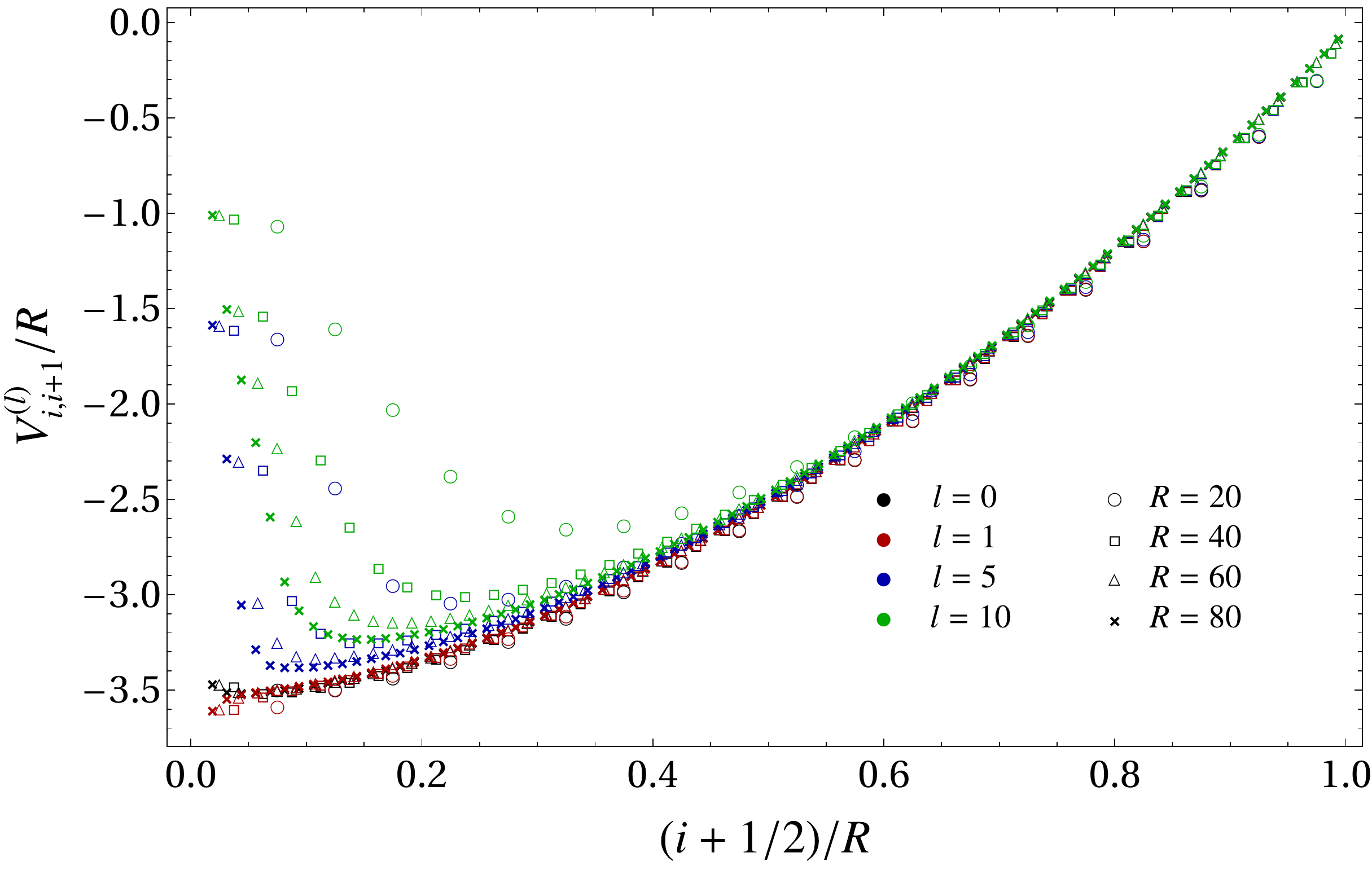}
\end{subfigure}
\hfill
\begin{subfigure}{.4\textwidth}
\hspace{-1.5cm}
\vspace{.3cm}
\includegraphics[scale=.35]{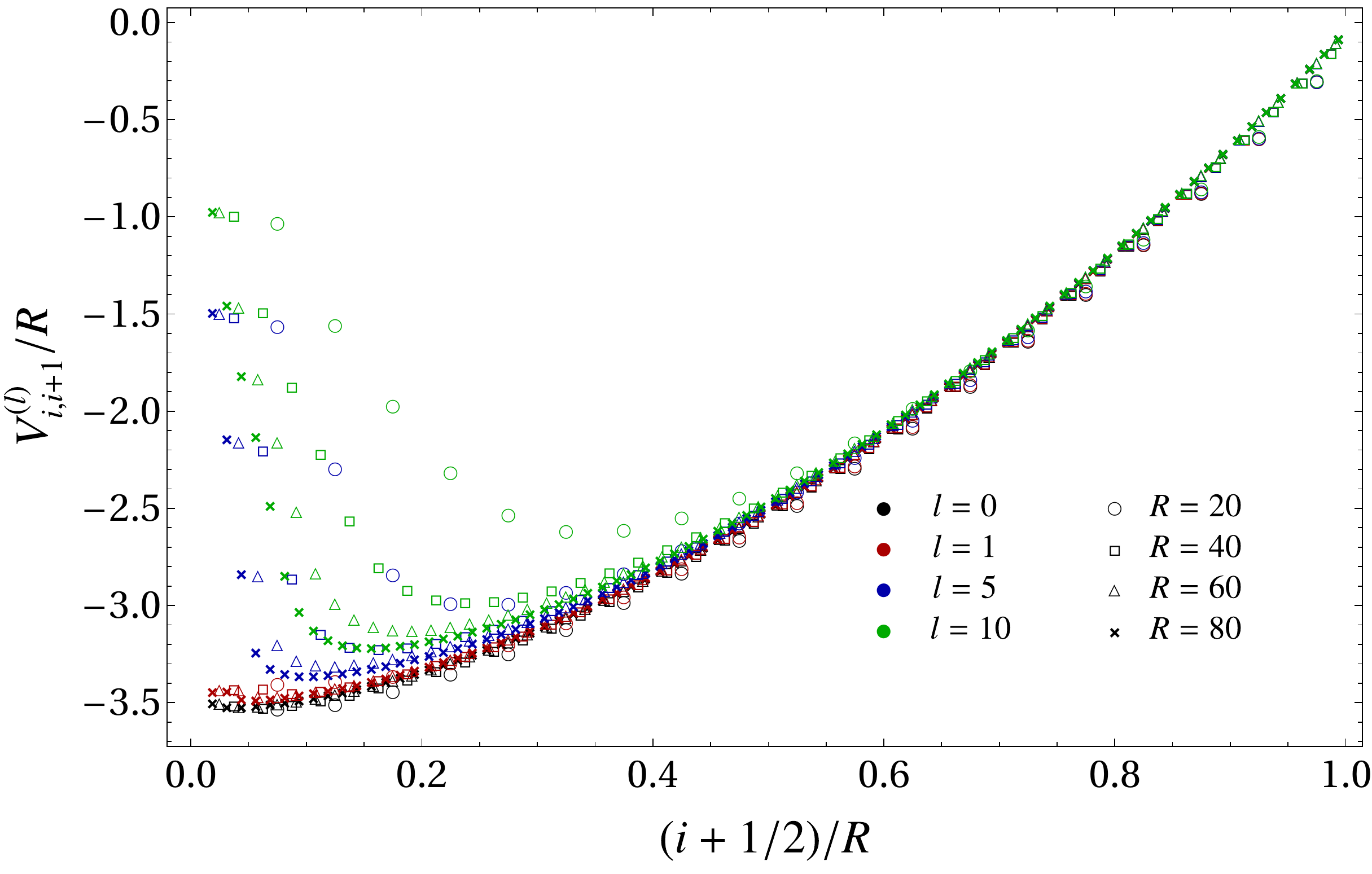}
\end{subfigure}

\begin{subfigure}{.4\textwidth}
\hspace{-1.5cm}
\vspace{.3cm}
\includegraphics[scale=.35]{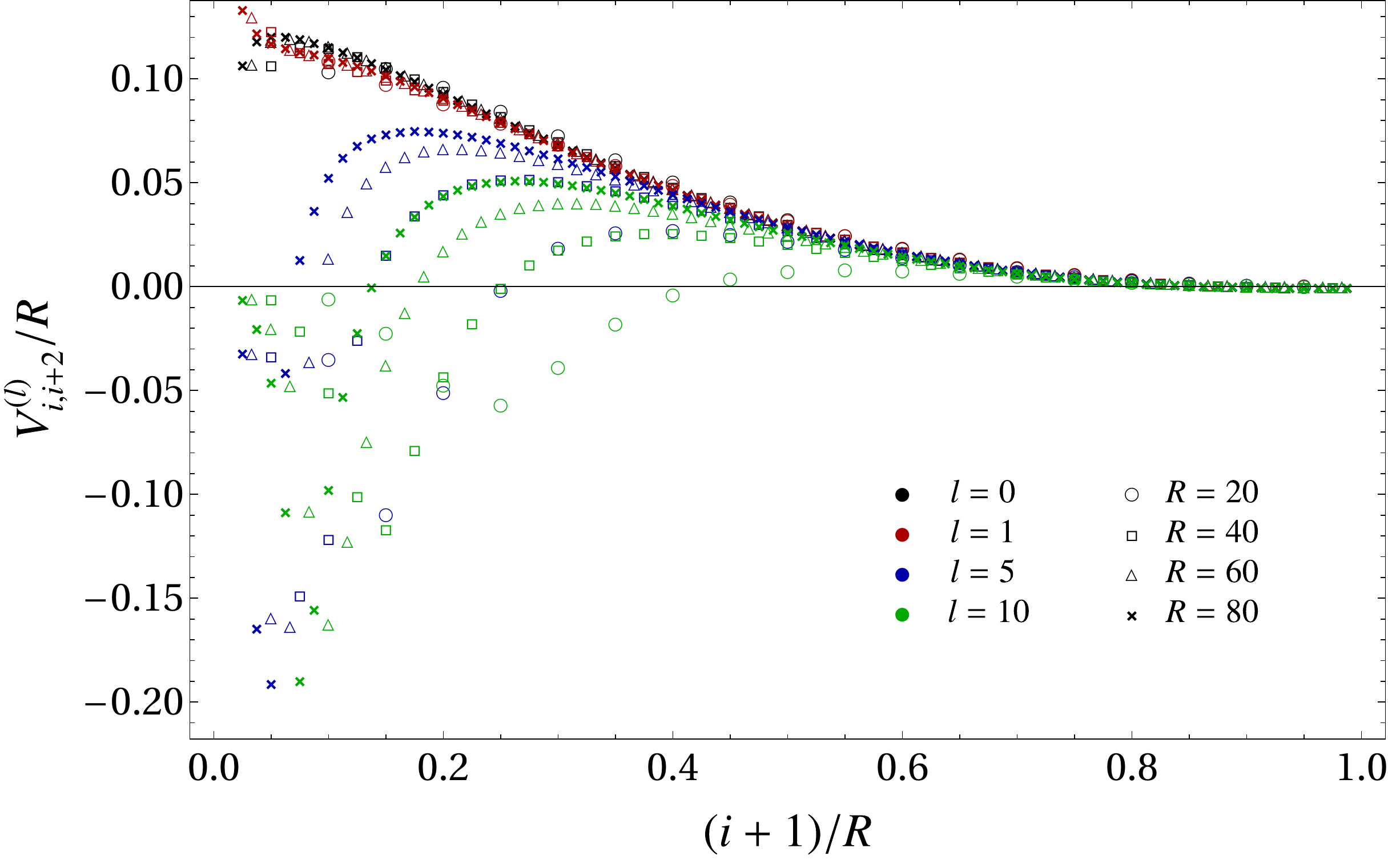}
\end{subfigure}
\hfill
\begin{subfigure}{.4\textwidth}
\hspace{-1.5cm}
\vspace{.3cm}
\includegraphics[scale=.35]{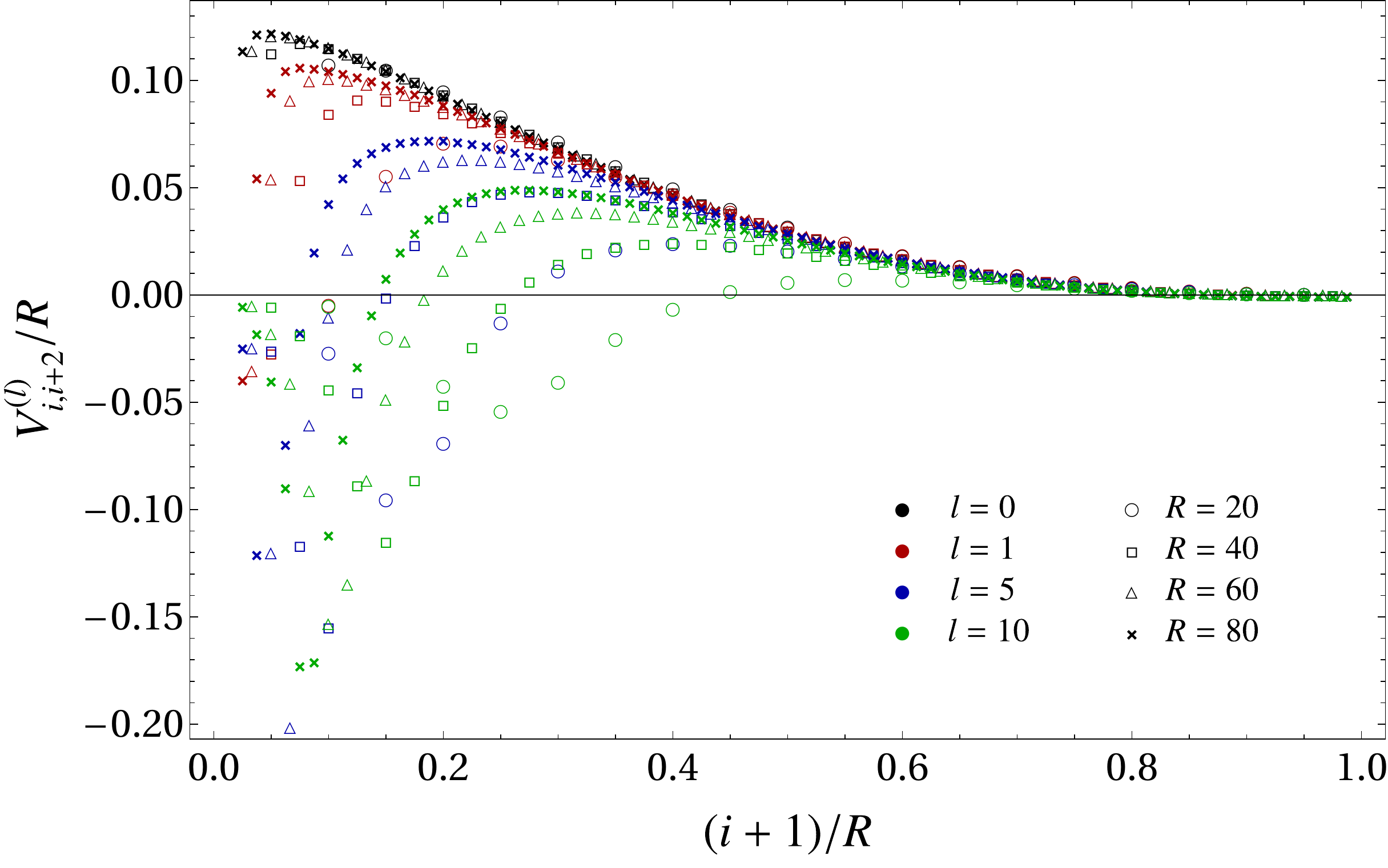}
\end{subfigure}

\begin{subfigure}{.4\textwidth}
\hspace{-1.5cm}
\vspace{.3cm}
\includegraphics[scale=.35]{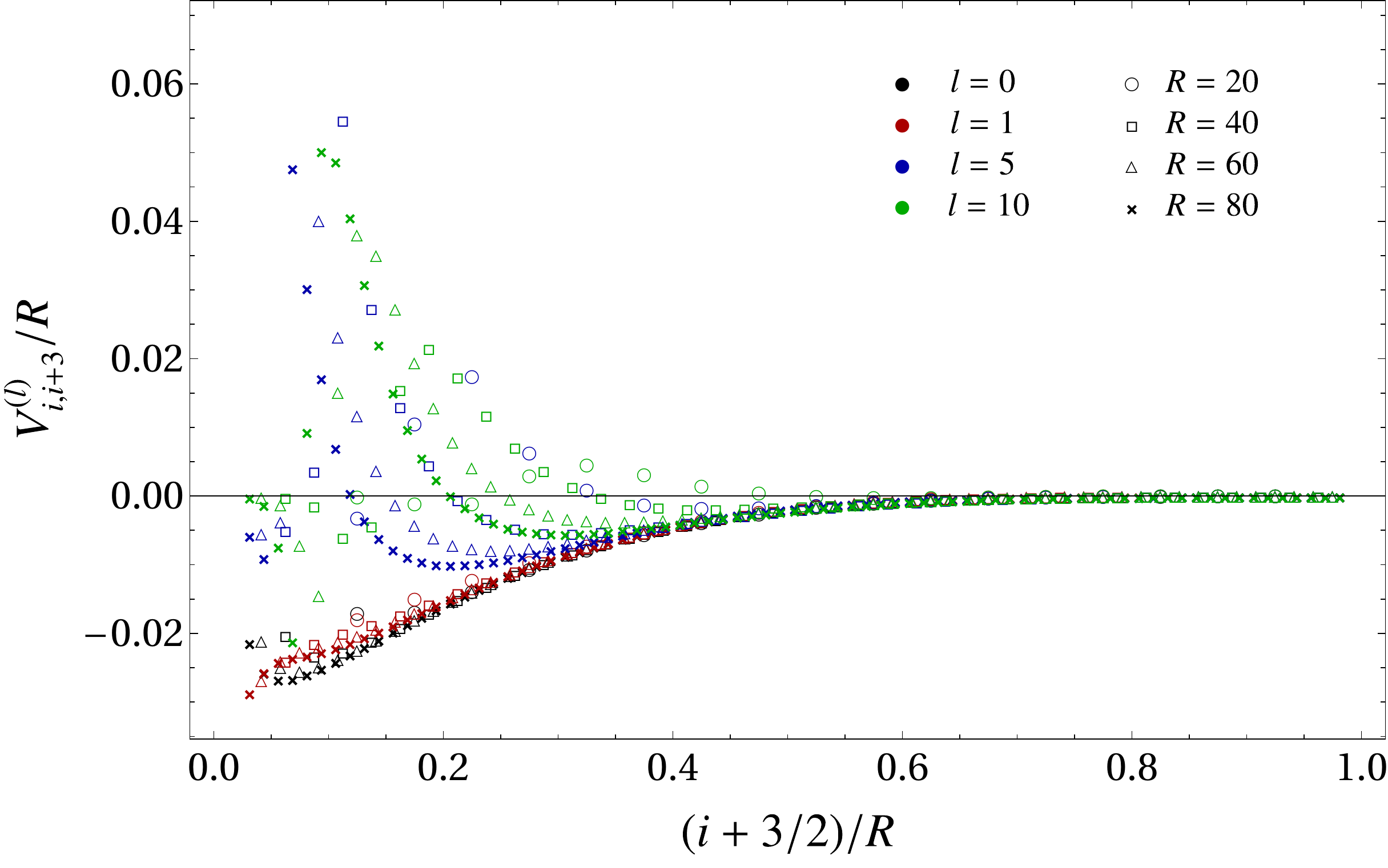}
\end{subfigure}
\hfill
\begin{subfigure}{.4\textwidth}
\hspace{-1.35cm}
\vspace{.3cm}
\includegraphics[scale=.35]{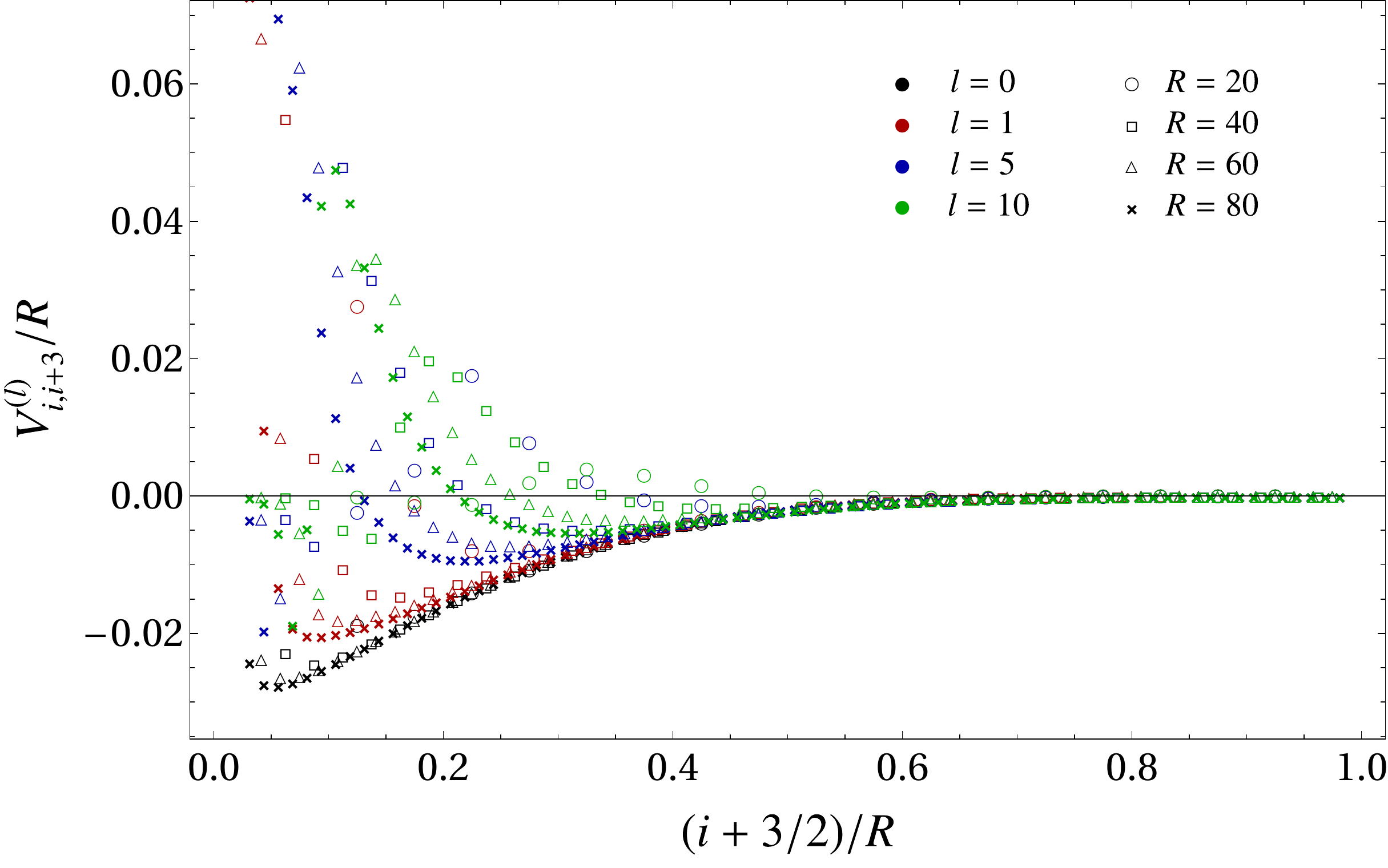}
\end{subfigure}
\vspace{-.3cm}
\caption{
Short-range couplings in the matrix $V^{(l)}/R$ 
(see (\ref{H_T and H_V operators}) and (\ref{munuk def})) 
when $\omega = 0$ and
either $d=2$ (left panels) or $d=3$ (right panels),
for some small $l$'s and different sizes $R$.
}
\label{fig-diagonals-V}
\end{figure}

In Fig.\,\ref{fig-diagonals-T} and Fig.\,\ref{fig-diagonals-V} 
we show some numerical results for the elements in and near the main diagonals of
the matrices $T^{(l)}$ and $V^{(l)}$ when $\omega = 0$.
From previous analyses for $d=1$ \cite{DiGiulio:2019cxv,Eisler:2020lyn},
they are expected to be extensive;
hence we consider $T^{(l)}_{i,i+k}/R$ and $V^{(l)}_{i,i+k}/R$ 
(only the data corresponding to $k\in \{0,1,2,3\}$ are reported).
The data points in Fig.\,\ref{fig-diagonals-T} and Fig.\,\ref{fig-diagonals-V}
display good collapse close to the boundary of the sphere $B$ as $R$ increases.
For small values of $l$, this agreement is observed almost all over the range of the spatial index
except near the center of the sphere.
These data collapses provide some numerical support to the conjecture (\ref{munuk def})
for small values of $l$.
%
Comparing the left and  the right panels in Fig.\,\ref{fig-diagonals-T} and Fig.\,\ref{fig-diagonals-V},
we do not find significant differences between $d=2$ and $d=3$.
As $l$ increases, we expect that larger values of $R$ are needed to observe 
good collapses of the data points. 
This lack of convergence for high values of $l$ 
and our limited capability to treat numerically large systems
forces us to focus only on small values of $l$.

We consider the continuum limit given by
$a \to 0$, $R \to \infty$ and $R_{\textrm{\tiny tot}} \to \infty$,
while $R a \equiv \mathcal{R}$ and $R_{\textrm{\tiny tot}} a \equiv \mathcal{R}_{\textrm{\tiny tot}}$
are kept fixed (hence $R/R_{\textrm{\tiny tot}}$ is fixed as well).
The parameters $\mathcal{R}$ and $\mathcal{R}_{\textrm{\tiny tot}}$
are the radii respectively of the sphere $B$ and of the entire system, 
which is a larger concentric sphere. 

In the continuum,
the radial position is labelled by $r = i a$ with $0< r < \mathcal{R}$.
This leads to write (\ref{r_k def}) also as 
\be
r_{\tilde{k}}
= \frac{r}{\mathcal{R}} +\frac{\tilde{k} a/2}{\mathcal{R}}
\ee
which suggests to expand 
$\tau_{l,\eta k}$ and $\nu_{l,\eta k}$ in (\ref{munuk def})
as Taylor series when $a \to 0$.
In order to study the continuum limit of the operators in (\ref{H_T Casini}) and (\ref{H_V Casini}),
first we write these sums
as $ R \sum_{i=1}^R (\dots )  = \frac{(R a)}{a^2}\sum_{i=1}^R (\dots ) a$ 
and then employ that 
$\sum_{i=1}^R (\dots)a  \rightarrow  \int_0^{\mathcal{R}}  (\dots)  \textrm{d}r$.
From (\ref{qp-field-replacement}),
for the operators $\hat{q}_{\boldsymbol{l},i+\eta k} $ and $\hat{p}_{\boldsymbol{l},i+\eta k} $ 
in the continuum limit we have
\bea
\label{q-der-field-replacement}
\hat{q}_{\boldsymbol{l},i+\eta k}
& \longrightarrow &
\Phi_{\boldsymbol{l}}(r+ \eta k a)
=
 \sum_{p \geqslant 0} \frac{(\eta ka)^p}{p!} \, \partial_r^p \Phi_{\boldsymbol{l}}(r)
\\
\rule{0pt}{.6cm}
\label{p-der-field-replacement}
 \hat{p}_{\boldsymbol{l},i+\eta k} 
& \longrightarrow &
 a\,\Pi_{\boldsymbol{l}}(r+ \eta k a)
=\,
a \sum_{p \geqslant 0} \frac{(\eta ka)^p}{p!} \, \partial_r^p \Pi_{\boldsymbol{l}}(r)
\eea
where the Taylor expansions of the fields as $a \to 0$ have been used. 
Combining (\ref{qp-field-replacement}), (\ref{q-der-field-replacement}) and (\ref{p-der-field-replacement}),
for the operators (\ref{H_T Casini}) and (\ref{H_V Casini}) in the continuum limit
one obtains 
$\widehat{H}_{T,\boldsymbol{l}} \longrightarrow H_{T,\boldsymbol{l}} $ 
and 
$\widehat{H}_{V,\boldsymbol{l}} \longrightarrow H_{V,\boldsymbol{l}} $ 
respectively, with
\bea
\label{H_T def}
&& \hspace{-.5cm}
H_{T,\boldsymbol{l}} 
=
\mathcal{R}
\int_0^{\mathcal{R}}
\!\bigg( \tau_{l,0}(r) \, \Pi_{\boldsymbol{l}}(r)^2
+
\sum_{\eta} 
\sum_{k =1}^{k_{\textrm{\tiny max}}} 
\tau_{l,k,\eta}(r + \eta ka/2)
\, \Pi_{\boldsymbol{l}}(r)\, \Pi_{\boldsymbol{l}}(r + \eta k a)
\bigg)
\textrm{d}r
\\
\label{H_V def}
&& \hspace{-.5cm}
H_{V,\boldsymbol{l}} 
=
\frac{\mathcal{R}}{a^2}
\int_0^{\mathcal{R}}
\!\bigg( \nu_{l,0}(r) \, \Phi_{\boldsymbol{l}}(r)^2
+ \sum_{\eta} 
 \sum_{k =1}^{k_{\textrm{\tiny max}}} 
\nu_{l,k,\eta }(r+\eta ka/2)
\, \Phi_{\boldsymbol{l}}(r)\, \Phi_{\boldsymbol{l}}(r+\eta k a)
\bigg)
\textrm{d}r
\eea
where $k_{\textrm{\tiny max}}$ parameterises the number of diagonals included in the sums.
The parameter $k_{\textrm{\tiny max}}$
plays an important role throughout our numerical analysis. 
In the continuum limit, also $k_{\textrm{\tiny max}}$ is infinite 
and therefore all the diagonals should be taken into account.
However, in order to be consistent with (\ref{munuk def}),
where  $R \to \infty$ for any finite $0 \leqslant k \leqslant k_{\textrm{\tiny max}}$,
in our numerical analysis
(where both $R$ and $k_{\textrm{\tiny max}}$ are finite)
we have to consider $k_{\textrm{\tiny max}} \ll R$.
%
In the appendix\;\ref{app_kmax} the role of $k_{\textrm{\tiny max}} $ is further discussed. 

\begin{figure}[t!]
\vspace{-.5cm}
\begin{subfigure}{.4\textwidth}
\hspace{-1.4cm}
\vspace{.4cm}
\includegraphics[scale=.35]{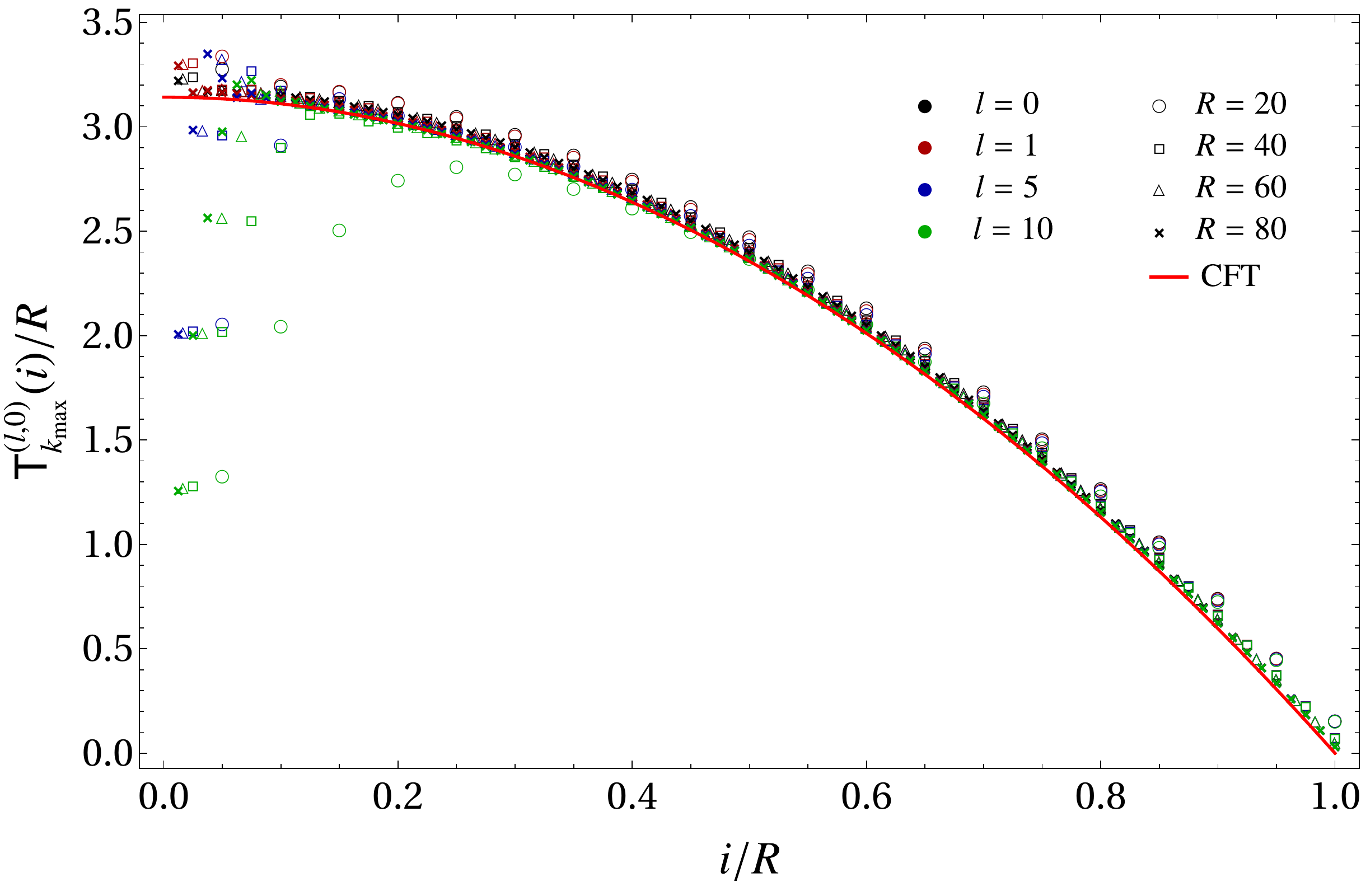}
\end{subfigure}\hfill
\begin{subfigure}{.4\textwidth}
\hspace{-1.5cm}
\vspace{.4cm}
\includegraphics[scale=.35]{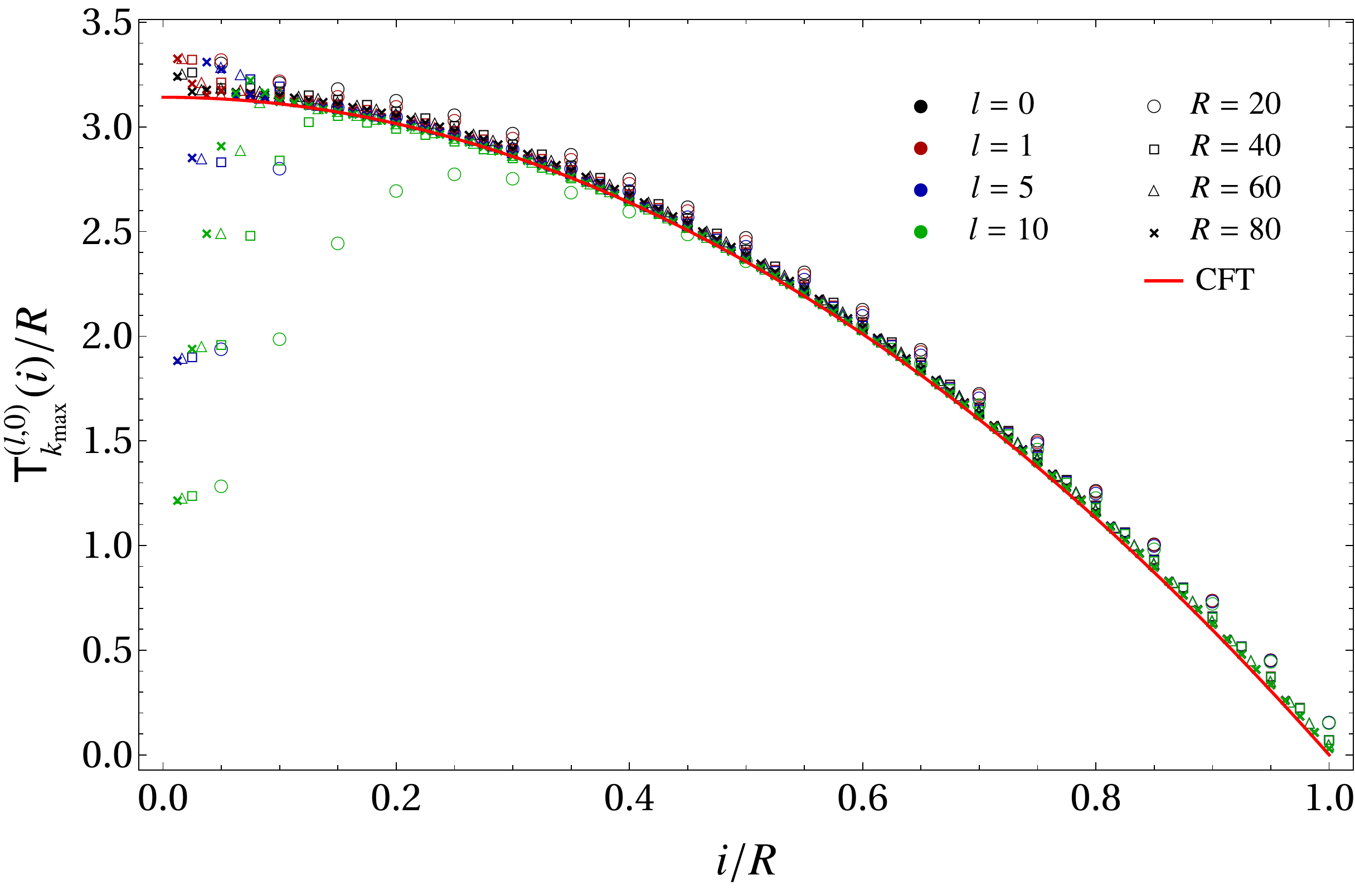}
\end{subfigure}
\begin{subfigure}{.4\textwidth}
\hspace{-1.4cm}
\includegraphics[scale=.35]{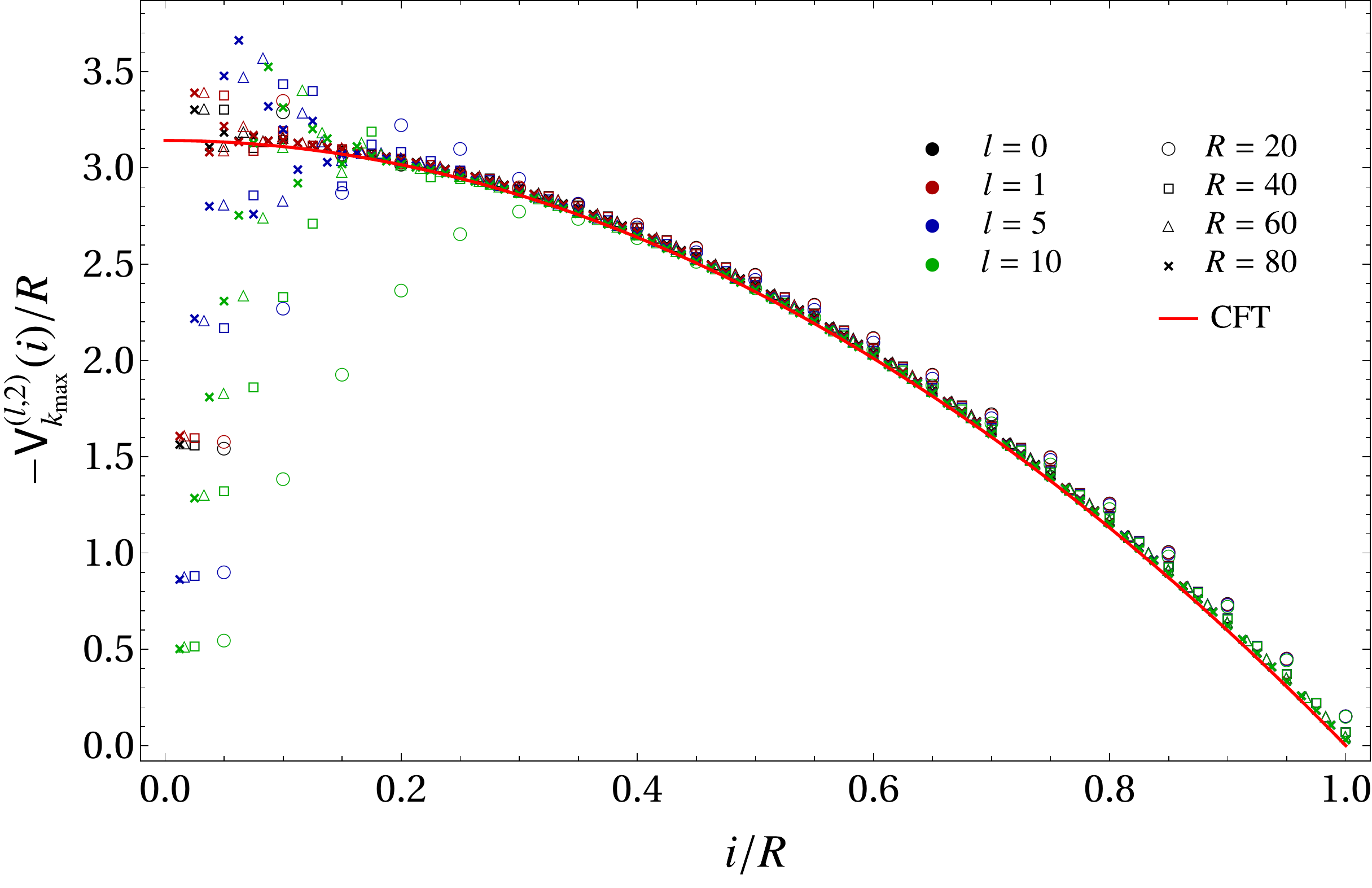}
\end{subfigure}\hfill
\begin{subfigure}{.4\textwidth}
\hspace{-1.15cm}
\includegraphics[scale=.3425]{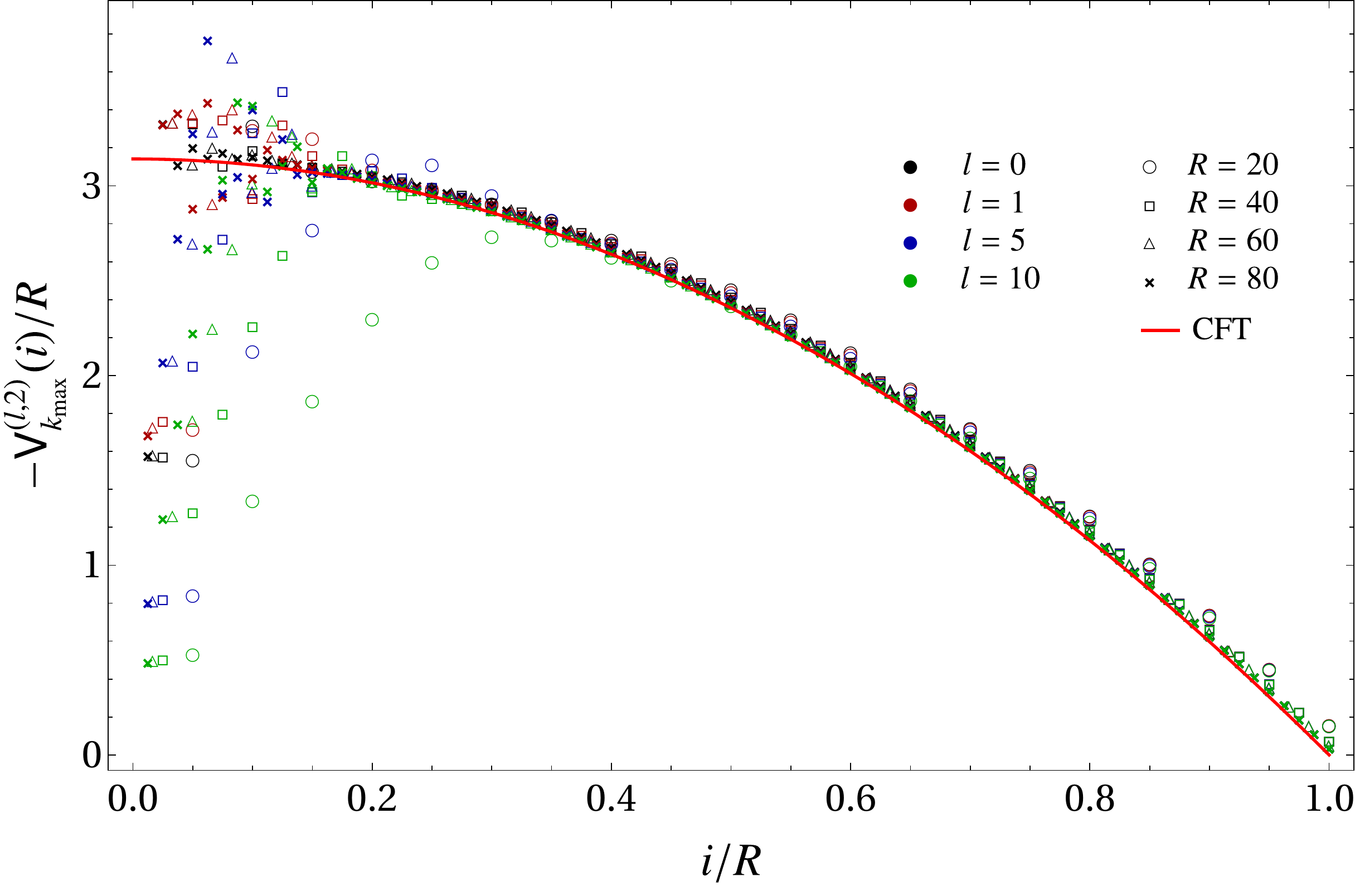}
\end{subfigure}
\vspace{.1cm}
\caption{ 
\label{beta-TV-parabola}
$\mathsf{T}_{k_\textrm{\tiny max}}^{(l,0)}/R$ from (\ref{T0-def}) (top panels) 
and $-\,\mathsf{V}_{k_\textrm{\tiny max}}^{(l,2)}/R$  from (\ref{V2-def}) (bottom panels), 
when $\omega = 0$ and
either  $d = 2$ (left panels) or $d = 3$ (right panels), for small values of $l$ 
and $k_{\textrm{\tiny max}} = R/10$ for every $l$
(other choices for $k_{\textrm{\tiny max}}$ have been considered in Fig.\,\ref{fig:T0andV2app}).
The red curve corresponds to the CFT prediction (\ref{beta-def-intro}).
}
\end{figure}

Since $a \to 0$ in the continuum limit,
we expand the integrands in (\ref{H_T def}) and (\ref{H_V def}),
keeping only the terms that could lead to a non vanishing contribution after the limit. 
The expansion of (\ref{H_T def}) gives
\be
\label{Noperator expansion1}
H_{T,\boldsymbol{l}} 
\,=\,
\mathcal{R}
\int_0^{\mathcal{R}}
\mathcal{T}_{k_\textrm{\tiny max}}^{(l,0)}(r) \,  \Pi_{\boldsymbol{l}}(r)^2 \,\textrm{d}r 
+
O(a)
\ee
where  we have introduced
\be
\label{N-summation k^0 continuum}
\mathcal{T}_{k_\textrm{\tiny max}}^{(l,0)}(r)
 \equiv
 \lim_{R \to \infty} \frac{ \mathsf{T}_{k_\textrm{\tiny max}}^{(l,0)}(i)}{R}
\ee
with 
\be
\label{T0-def}
 \mathsf{T}_{k_\textrm{\tiny max}}^{(l,0)}(i)
 \,\equiv\,
T^{(l)}_{i,i} 
+ 
\sum_{\eta}
\sum_{k=1}^{k_\textrm{\tiny max}^{(\eta)}} 
T^{(l)}_{i,i+\eta k}
\ee
being $k_\textrm{\tiny max}^{(\pm)}$ defined as follows
\be
\label{kmax-pm-def}
k_\textrm{\tiny max}^{(+)}
\equiv
\mathrm{min} \big(k_\textrm{\tiny max}\, ,R-i\big)
\;\;\;\;\;\qquad\;\;\;\;\;
k_\textrm{\tiny max}^{(-)}
\equiv
\mathrm{min}\big(k_\textrm{\tiny max}\, ,i-1\big)\,.
\ee
In the top panels of Fig.\,\ref{beta-TV-parabola}
we show numerical results supporting the evidence that the limit (\ref{N-summation k^0 continuum}) 
leads to a well defined finite function (see (\ref{conj-infty-beta}))
when $k_\textrm{\tiny max}$ is large enough,
at least for the small values of $l$ explored.

\begin{figure}[t!]
\vspace{.5cm}
\begin{subfigure}{.4\textwidth}
\hspace{-1.4cm}
\includegraphics[scale=.355]{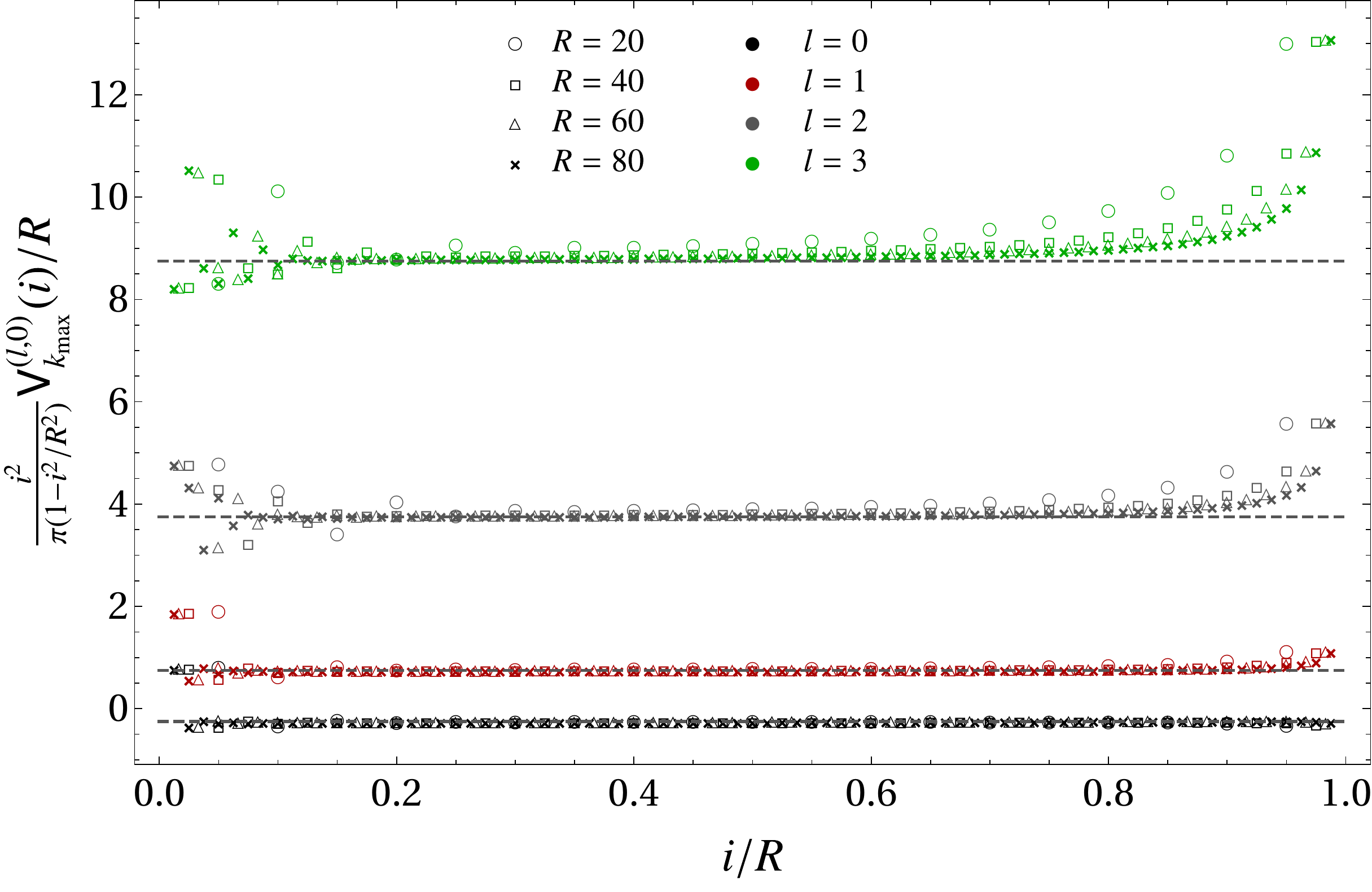}
\end{subfigure}\hfill
\begin{subfigure}{.4\textwidth}
\hspace{-1.4cm}
\includegraphics[scale=.355]{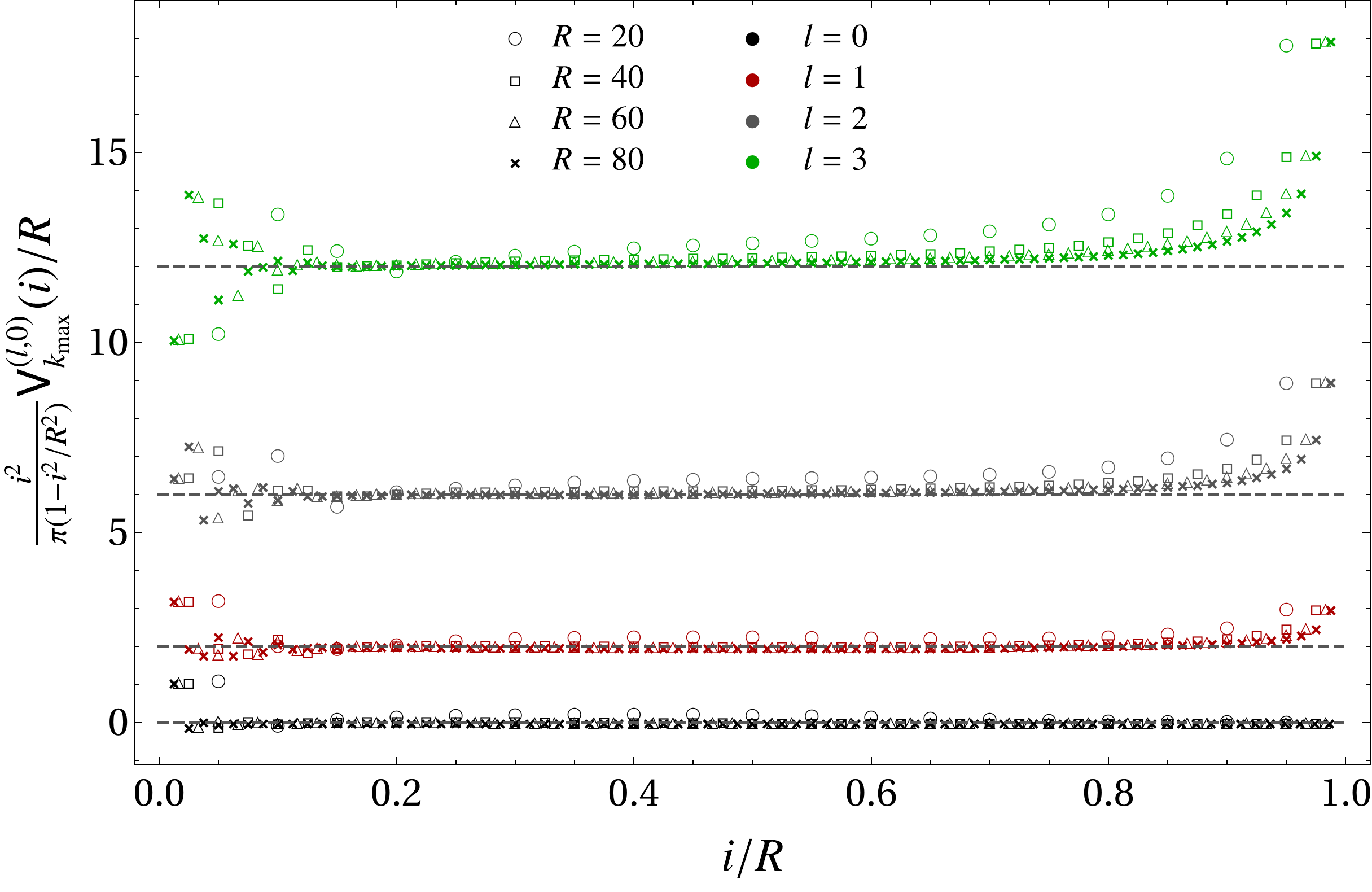}
\end{subfigure}
\\
\vspace{-.0cm}
\rule{0pt}{3.2cm}
\begin{subfigure}{.4\textwidth}
\hspace{-1.75cm}
\includegraphics[scale=.36]{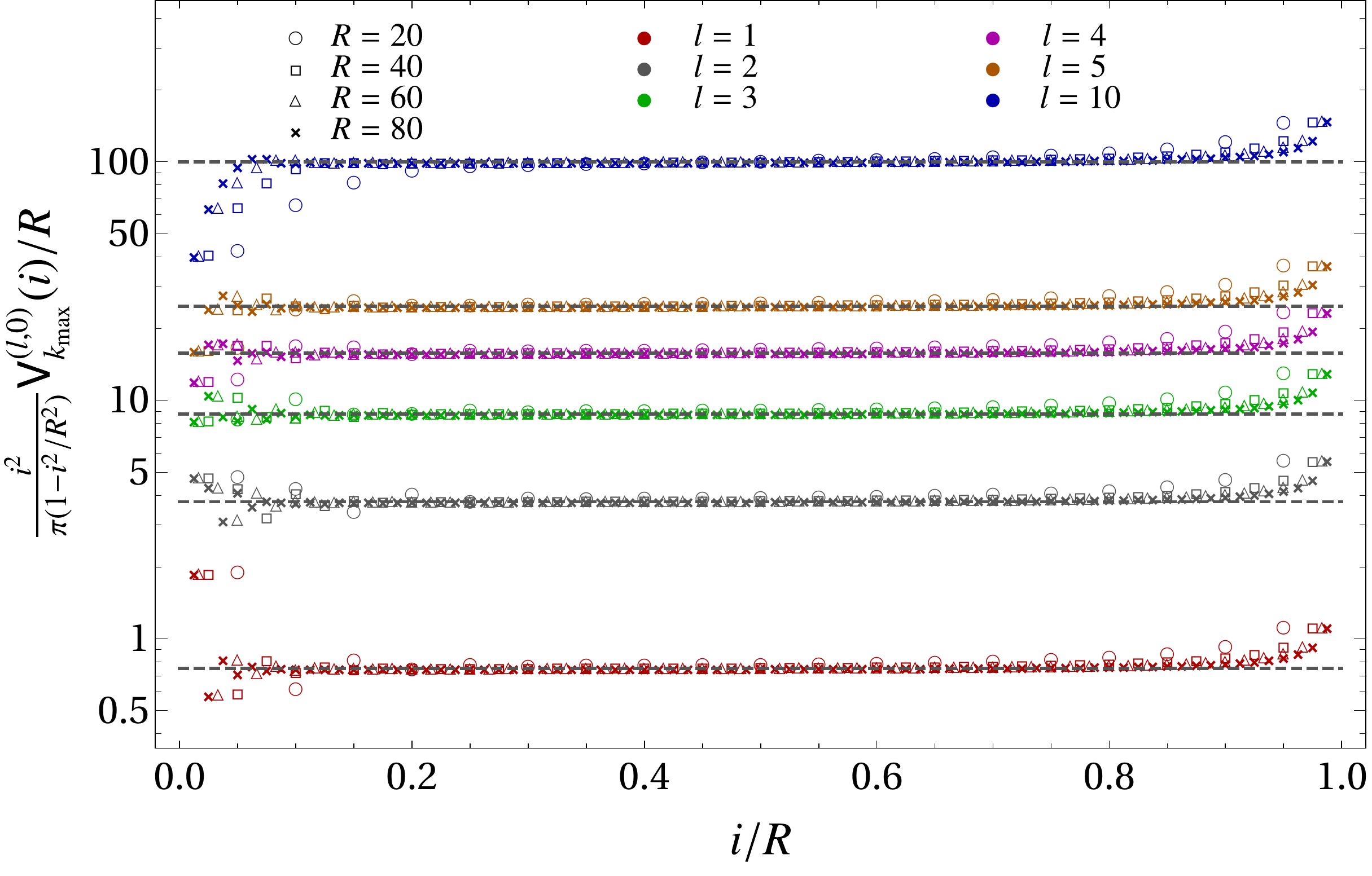}
\end{subfigure}\hfill
\begin{subfigure}{.4\textwidth}
\hspace{-1.4cm}
\includegraphics[scale=.36]{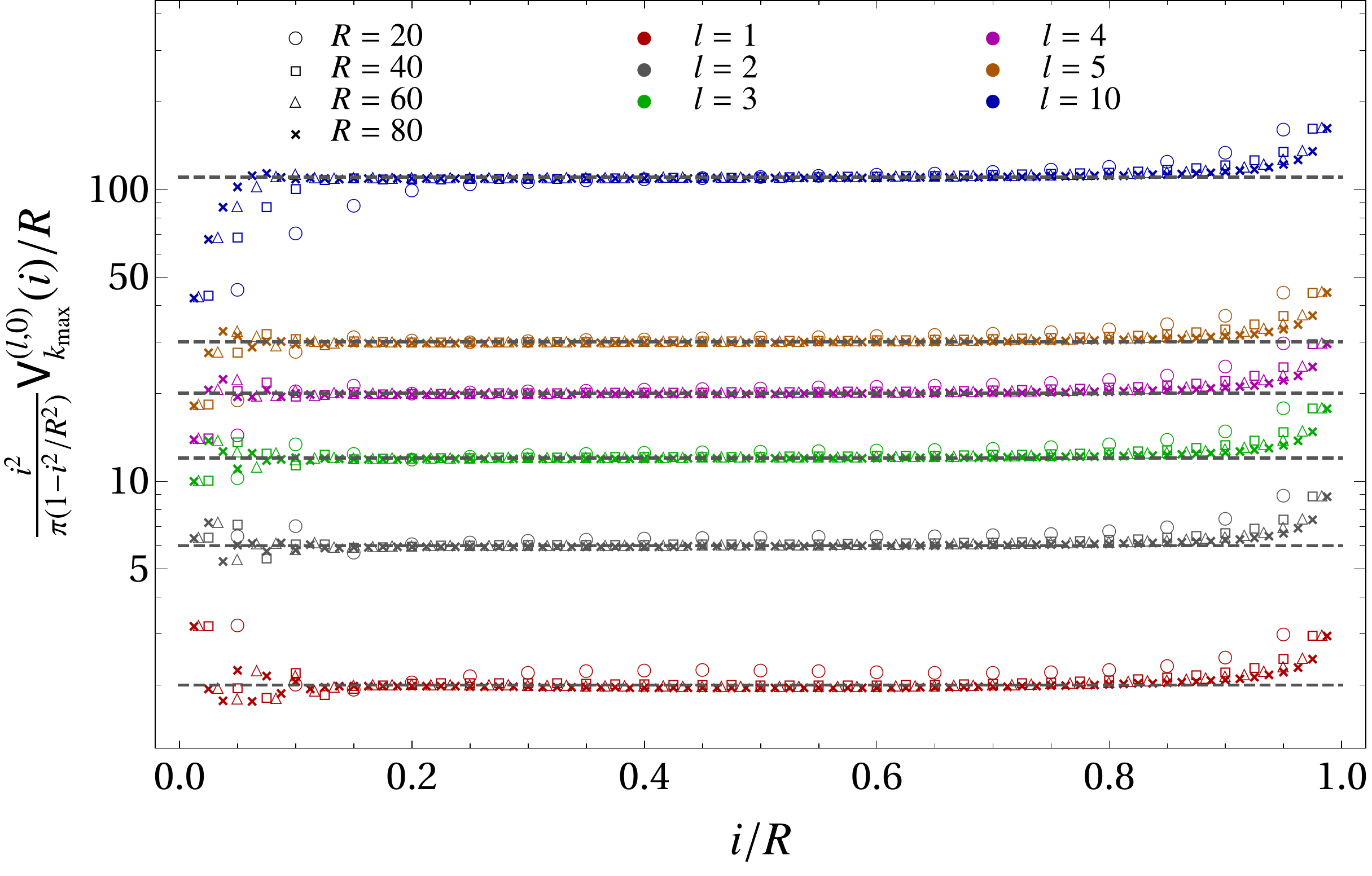}
\end{subfigure}
\vspace{.1cm}
\caption{ 
\label{beta-V0-horizontal-lines-mode}
$\mathsf{V}_{k_\textrm{\tiny max}}^{(l,0)}/R$ from (\ref{V0-def})
when $\omega = 0$ and
either  $d = 2$ (left panels) or $d = 3$ (right panels), for small values of $l$ and 
$k_{\textrm{\tiny max}} = R/10$  for every $l$
(other choices for $k_{\textrm{\tiny max}} $ have been considered in Fig.\,\ref{fig:V0app}).
The horizontal dashed lines correspond to (\ref{mu-d-def}), 
i.e. to $l^2-1/4$ in the left panels and to $l(l+1)$ in the right panels.
In the bottom panels the data corresponding to $1 \leqslant l \leqslant 10$ are shown
by adopting the logarithmic scale for the vertical axes. 
}
\end{figure}

The expansion of (\ref{H_V def}) gives
\bea
\label{Moperator expansion1}
& &\hspace{-.3cm}
H_{V,\boldsymbol{l}} 
=
\frac{\mathcal{R}}{a^2}
\int_0^{\mathcal{R}}
\Bigg\{  
\mathcal{V}_{k_\textrm{\tiny max}}^{(l,0)}(r) 
\, \Phi_{\boldsymbol{l}}(r)^2
 +
a 
\sum_{\eta} \sum_{k =1}^{k_{\textrm{\tiny max}}} \eta k
\bigg[ \,\frac{1}{2}\,
\nu_{l,k,\eta}'(r) \, \Phi_{\boldsymbol{l}}(r) 
+ \nu_{l,k,\eta}(r) \, \Phi_{\boldsymbol{l}}(r)' \,\bigg] \Phi_{\boldsymbol{l}}(r)
\nonumber
\\
& & \hspace{4cm}
+ \;a^2\,
\sum_{\eta} 
 \sum_{k =1}^{k_{\textrm{\tiny max}}} 
 \frac{k^2}{2}
\left[ \, \frac{1}{4} \,\nu_{l,k,\eta}''(r) \, \Phi_{\boldsymbol{l}}(r) 
+ \partial_r \Big( \nu_{l,k,\eta} (r) \, \Phi_{\boldsymbol{l}}'(r) \Big)  \,\right] \Phi_{\boldsymbol{l}}(r)
\Bigg\} \,\textrm{d}r
\nonumber
\\
& &
\eea
where $O(a)$ terms have been neglected
and  
\be
\label{M-summation k^0 continuum}
\mathcal{V}_{k_\textrm{\tiny max}}^{(l,0)}(r)
 \equiv
 \lim_{R \to \infty} \frac{ \mathsf{V}_{k_\textrm{\tiny max}}^{(l,0)}(i)}{R}
\ee
with $ \mathsf{V}_{k_\textrm{\tiny max}}^{(l,0)}(i)$ being defined as 
\be
\label{V0-def}
 \mathsf{V}_{k_\textrm{\tiny max}}^{(l,0)}(i)
 \,\equiv\,
V^{(l)}_{i,i} 
+ 
\sum_{\eta}
\sum_{k=1}^{k_\textrm{\tiny max}^{(\eta)}} 
V^{(l)}_{i,i+ \eta k}\,.
\ee
The numerical results displayed in Fig.\,\ref{beta-V0-horizontal-lines-mode}
indicate that the limit (\ref{M-summation k^0 continuum}) 
provides a well defined finite function when $k_\textrm{\tiny max}$ is large enough
(see (\ref{V-infty-beta-new-term})),
at least for small the values of $l$ that we have considered.

\begin{figure}[t!]
\vspace{-.5cm}
\begin{subfigure}{.4\textwidth}
\hspace{-1.4cm}
\vspace{.4cm}
\includegraphics[scale=.35]{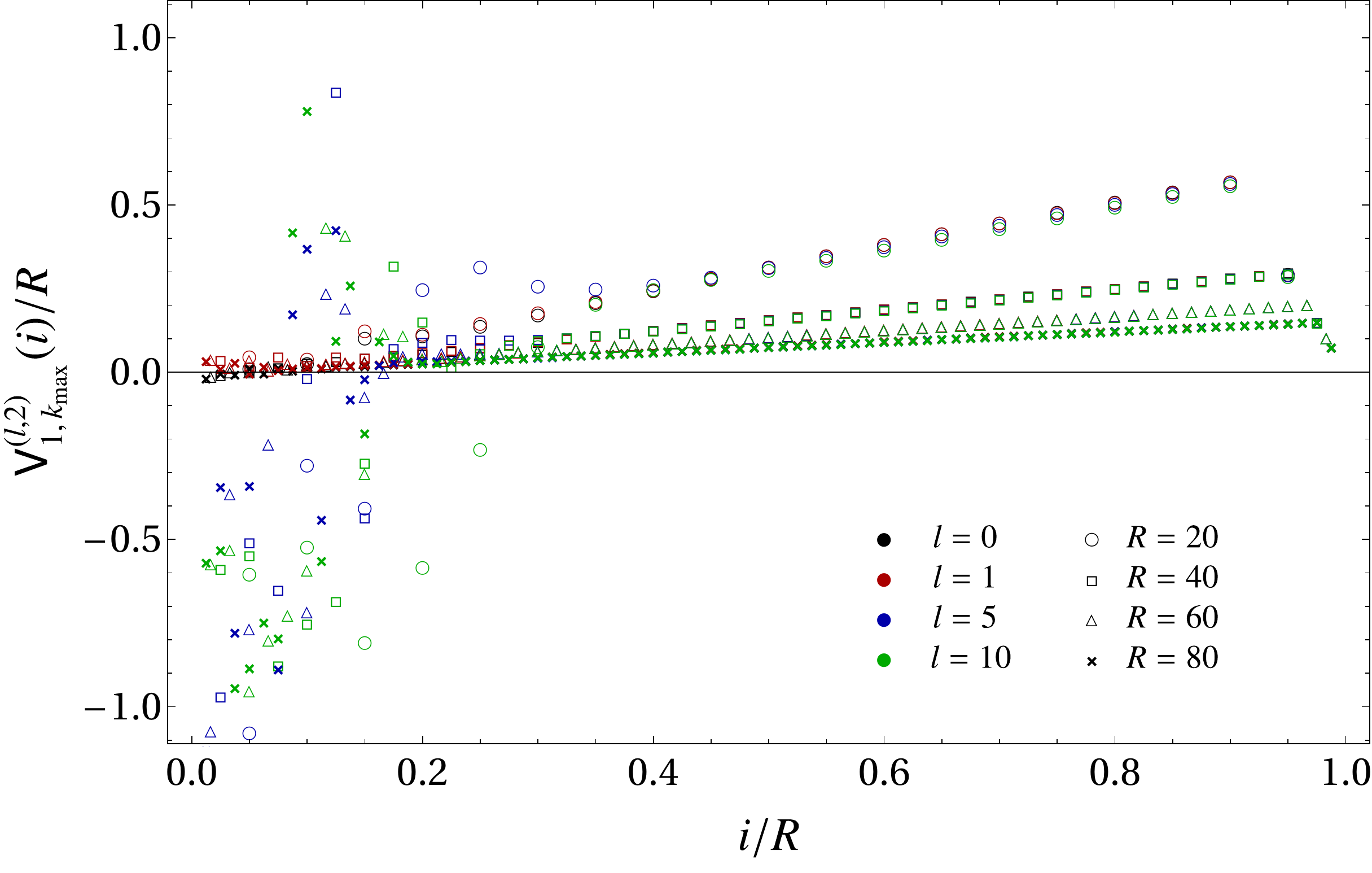}
\end{subfigure}\hfill
\begin{subfigure}{.4\textwidth}
\hspace{-1.6cm}
\vspace{.4cm}
\includegraphics[scale=.35]{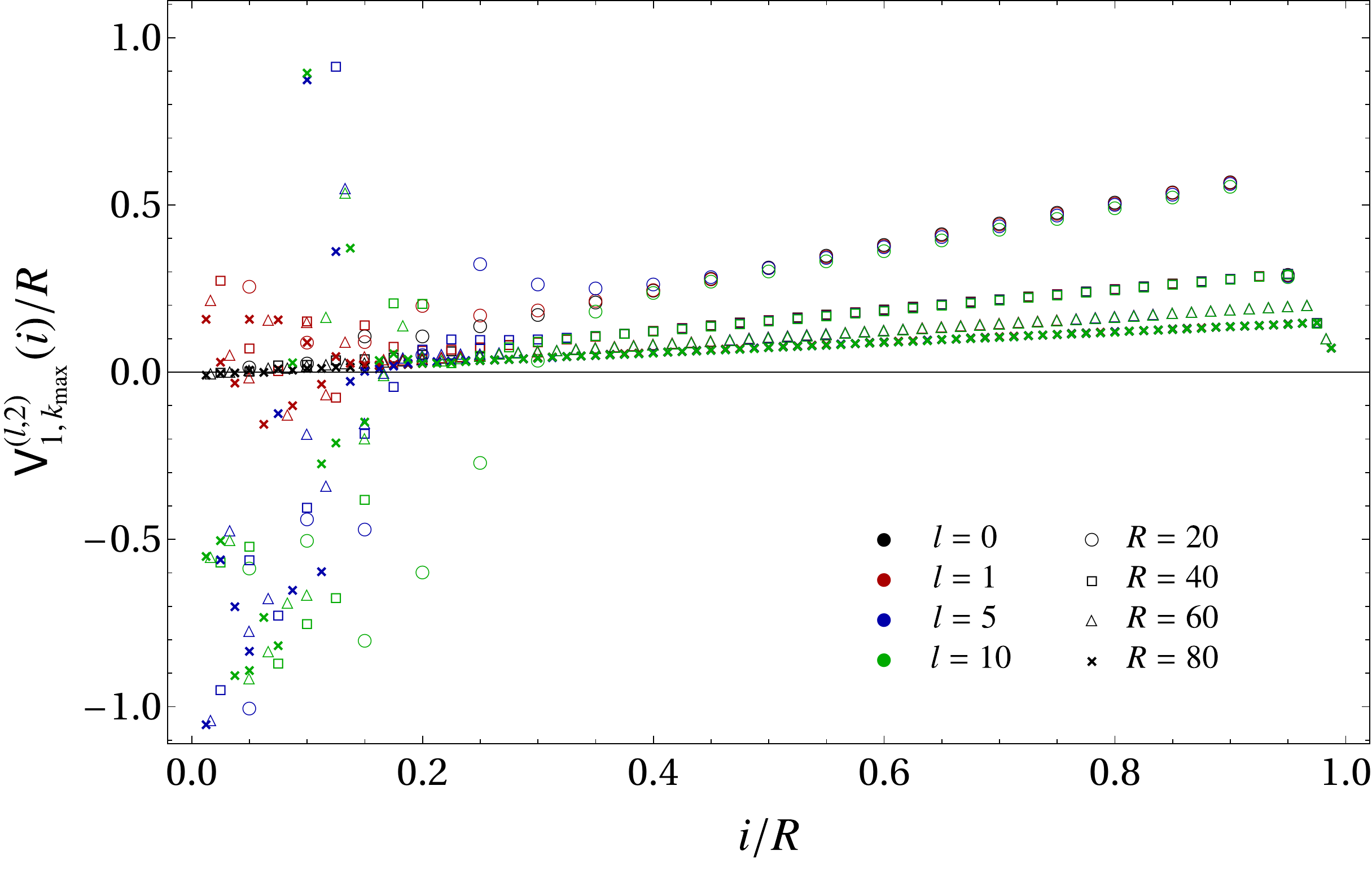}
\end{subfigure}
\begin{subfigure}{.4\textwidth}
\hspace{-1.2cm}
\includegraphics[scale=.34]{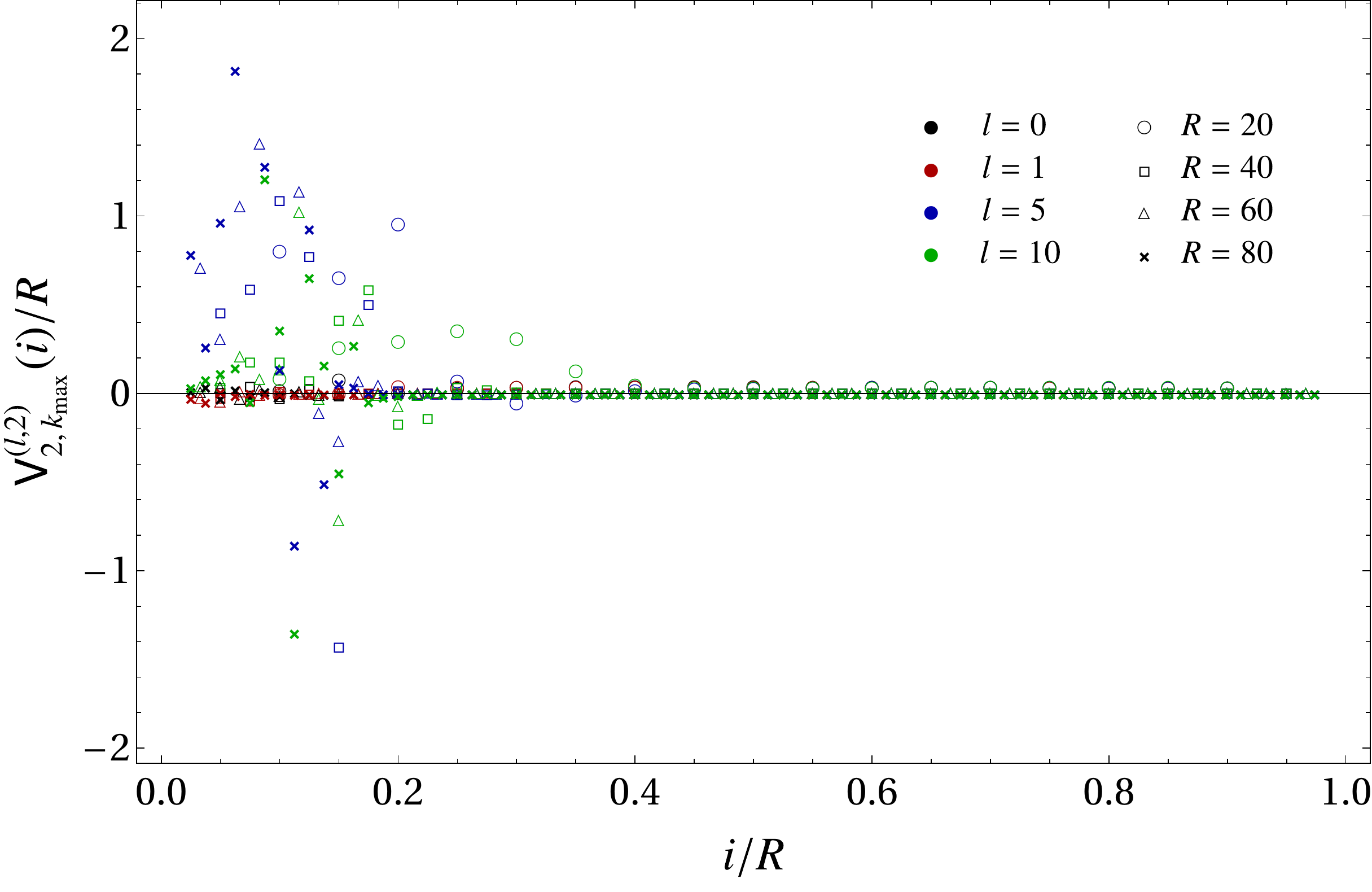}
\end{subfigure}\hfill
\begin{subfigure}{.4\textwidth}
\hspace{-1.2cm}
\includegraphics[scale=.34]{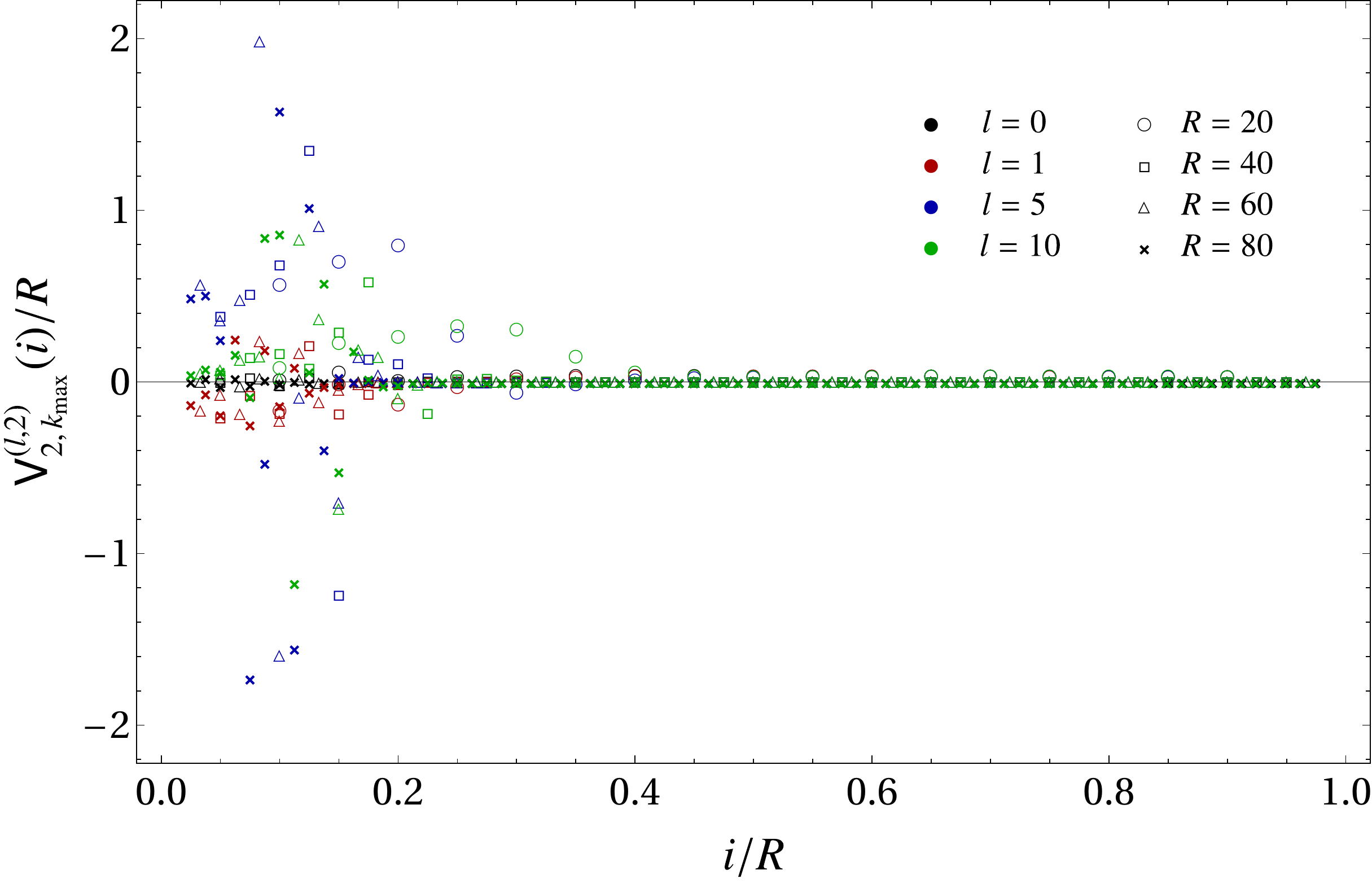}
\end{subfigure}
\vspace{.1cm}
\caption{ 
\label{beta-Vd1-zero}
$\mathsf{V}_{1,k_\textrm{\tiny max}}^{(l,2)}/R$ (top panels) 
and $\mathsf{V}_{2,k_\textrm{\tiny max}}^{(l,2)}/R$ (bottom panels),
from (\ref{V12-def})  and (\ref{V22-def}) respectively,
when $\omega = 0$ and
either  $d = 2$ (left panels) or $d = 3$ (right panels), for various values of $l$ and $R$.
Here $k_{\textrm{\tiny max}} = R/10$  for every $l$.
}
\end{figure}

Since the integrand of the $O(1/a)$ term in (\ref{Moperator expansion1})
is proportional to the total derivative $\partial_r [\nu_{l,k,\eta}(r)  \, \Phi(r)^2] $,
the corresponding integral gives the boundary terms 
$[\nu_{l,k,\eta}(r)  \, \Phi(r)^2 ] |^{r=\mathcal{R}}_{r=0}\,$.
One of these boundary terms 
vanishes because $ \nu_{l,k,\eta}(\mathcal{R})  =0$,
while the other one does not contribute because of
the Dirichlet boundary condition $\Phi_{\boldsymbol{l}}(0) = 0$ imposed at the origin. 
Thus, (\ref{Moperator expansion1}) simplifies to
\bea
\label{Moperator expansion1 bdy}
& &\hspace{-.9cm}
H_{V,\boldsymbol{l}} 
=
\frac{\mathcal{R}}{a^2}
\int_0^{\mathcal{R}}
\mathcal{V}_{k_\textrm{\tiny max}}^{(l,0)}(r) 
\, \Phi_{\boldsymbol{l}}(r)^2
\,\textrm{d}r
\\
& & \hspace{.3cm}
+ \;\mathcal{R}
\int_0^{\mathcal{R}} \;
 \sum_{\eta} \sum_{k =1}^{k_{\textrm{\tiny max}}} 
 \frac{k^2}{2}
\left[ \, \frac{1}{4} \,\nu_{l,k,\eta}''(r) \, \Phi_{\boldsymbol{l}}(r) 
+  \nu_{l,k,\eta}' (r) \, \Phi_{\boldsymbol{l}}'(r)  
+ \nu_{l,k,\eta} (r) \, \Phi_{\boldsymbol{l}}''(r)   \,\right] 
\Phi_{\boldsymbol{l}}(r) \,\textrm{d}r\,.
\nonumber
\eea
The last term of this expression, 
whose integrand is $\nu_{l,k,\eta}(r) \,  \Phi''_{\boldsymbol{l}}(r)\, \Phi_{\boldsymbol{l}}(r)$,
can be studied by employing (\ref{munuk def}) and introducing
\be
\label{M-summation k^2 der0 continuum}
\mathcal{V}_{k_\textrm{\tiny max}}^{(l,2)}(r)
 \equiv
 \lim_{R \to \infty} \frac{ \mathsf{V}_{k_\textrm{\tiny max}}^{(l,2)}(i)}{R}
\ee
where
\be
\label{V2-def}
 \mathsf{V}_{k_\textrm{\tiny max}}^{(l,2)}(i)
 \equiv
\sum_{\eta}
\sum_{k=1}^{k_\textrm{\tiny max}^{(\eta)}} 
\frac{k^2}{2}\,V^{(l)}_{i,i+\eta k}\,.
\ee
In the bottom panels of  Fig.\,\ref{beta-TV-parabola} we display numerical results 
indicating that the limit (\ref{M-summation k^2 der0 continuum}) 
gives a well defined finite function when $k_\textrm{\tiny max}$ is large enough 
(see (\ref{conj-infty-beta})).

As for the term in (\ref{Moperator expansion1 bdy}) whose integrand is 
$\nu_{l,k,\eta}' (r) \, \Phi_{\boldsymbol{l}}'(r)\, \Phi_{\boldsymbol{l}}(r)$,
since  the analytic expressions for $\nu_{l,k,\eta} (r)$ are not known,
we approximate $\nu_{l,k,\eta}' (r)$ through finite differences, 
i.e. by replacing this function with 
$[\nu_{l,k,\eta}(r+a) - \nu_{l,k,\eta}(r)]/a$.
This approximation, combined with (\ref{munuk def}) and (\ref{Moperator expansion1 bdy}),
leads to introduce
\be
\label{M-summation k^2 der1 continuum}
\mathcal{V}_{1,k_\textrm{\tiny max}}^{(l,2)}(r)
 \equiv
 \lim_{R \to \infty} \frac{ \mathsf{V}_{1,k_\textrm{\tiny max}}^{(l,2)}(i)}{R}
\ee
with $ \mathsf{V}_{1,k_\textrm{\tiny max}}^{(l,2)}(i)$ being defined as follows
\be
\label{V12-def}
 \mathsf{V}_{1,k_\textrm{\tiny max}}^{(l,2)}(i)
 \equiv
\sum_{\eta}
\sum_{k=1}^{k_\textrm{\tiny max}^{(\eta)}} 
\frac{k^2}{2}
\Big(V^{(l)}_{i+1,i+\eta k+1}-V^{(l)}_{i,i+\eta k}\Big)\,.
\ee

Similarly, for the term in (\ref{Moperator expansion1 bdy})
whose integrand is $\nu''_{l,k,\eta}(r) \,  \Phi^2_{\boldsymbol{l}}(r)$
we approximate $\nu_{l,k,\eta}''(r)$ through finite differences.
This leads to define
\be
\label{M-summation k^2 der2 continuum}
\mathcal{V}_{2,k_\textrm{\tiny max}}^{(l,2)}(x)
 \equiv
 \lim_{R \to \infty} \frac{ \mathsf{V}_{2,k_\textrm{\tiny max}}^{(l,2)}(i)}{R}
\ee
with
\bea
\label{V22-def}
 \mathsf{V}_{2,k_\textrm{\tiny max}}^{(l,2)}(i)
\,\equiv\,
\sum_\eta
\sum_{k=1}^{k_\textrm{\tiny max}^{(\eta)}} 
\frac{k^2}{2} 
\Big(
V^{(l)}_{i+1,i+\eta k+1}
-2\,V^{(l)}_{i,i+\eta k}
+V^{(l)}_{i-1,i+\eta k-1}
\Big)\,.
\eea
The subindices $1$ and $2$ in the l.h.s.'s of 
(\ref{M-summation k^2 der1 continuum})-(\ref{V12-def})
and  of (\ref{M-summation k^2 der2 continuum})-(\ref{V22-def})
respectively
indicate that the corresponding quantities are related to 
$\nu_{l,k,\eta}' (r)$ and $\nu_{l,k,\eta}'' (r)$.
In the top and bottom panels of  Fig.\,\ref{beta-Vd1-zero} we show some numerical results 
telling us that the limits in (\ref{M-summation k^2 der1 continuum}) and (\ref{M-summation k^2 der2 continuum}) respectively
give the function that vanishes identically (see (\ref{conj-infty-zero})).

Finally, 
by employing the expressions 
(\ref{N-summation k^0 continuum}), (\ref{M-summation k^2 der0 continuum}), 
(\ref{M-summation k^2 der1 continuum}) and (\ref{M-summation k^2 der2 continuum})
as discussed above,
we take the limit $k_{\textrm{\tiny max}} \to \infty$ 
in (\ref{Noperator expansion1}) and (\ref{Moperator expansion1 bdy}), 
finding for the non vanishing contributions the following expression
\bea
\label{MplusNoperator expansion bdy}
\frac{H_{T,\boldsymbol{l}}  + H_{V,\boldsymbol{l}}}{2}
&=&
\frac{\mathcal{R}}{a^2}
\int_0^{\mathcal{R}}
\frac{1}{2}
\left[ \,\mathcal{V}_{\infty}^{(l,0)}(r) 
+  \frac{1}{4}\, \mathcal{V}_{2,\infty}^{(l,2)}(r) \right] 
\Phi_{\boldsymbol{l}}(r)^2\, \textrm{d}r\,
\\
\rule{0pt}{.8cm}
& & 
+\, \mathcal{R}
\int_0^{\mathcal{R}}
\frac{1}{2}\,
\Big[ \, 
\mathcal{T}_{\infty}^{(l,0)}(r) \,  \Pi_{\boldsymbol{l}}(r)^2 
+ \mathcal{V}_{1,\infty}^{(l,2)}(r)\, \Phi_{\boldsymbol{l}}'(r) \, \Phi_{\boldsymbol{l}}(r)
+ \mathcal{V}_{\infty}^{(l,2)}(r) \, \Phi_{\boldsymbol{l}}''(r) \, \Phi_{\boldsymbol{l}}(r)
\,\Big] \textrm{d}r
\nonumber
\eea
where it is assumed that the weight functions are well defined.
Since the CFT expression (\ref{EH-intro}) is valid when $\mathcal{R}_{\textrm{\tiny tot}} \to \infty$,
we must consider $R \ll R_{\textrm{\tiny tot}}$ 
in order to compare our numerical results with this CFT formula specialised to the massless scalar field.
The main outcomes of our numerical analysis 
are shown in Fig.\,\ref{beta-TV-parabola}, Fig.\,\ref{beta-V0-horizontal-lines-mode} and Fig.\,\ref{beta-Vd1-zero},
where the data reported  in the left panels and in the right panels 
correspond to $d=2$ and $d=3$ respectively. 
These results provide some numerical evidence supporting 
the conjecture that (\ref{MplusNoperator expansion bdy}) 
provides the CFT prediction (\ref{EH-intro})
for the massless scalar field.

In Fig.\,\ref{beta-TV-parabola}
the combinations of diagonals defined in (\ref{T0-def}) and (\ref{V2-def}) are considered.
The numerical data shown in this figure lead us to conjecture that
\be
\label{conj-infty-beta}
\mathcal{T}_{\infty}^{(l,0)}(r)
\,=\,
\beta(r)
\;\;\;\qquad\;\;\;
-\mathcal{V}_{\infty}^{(l,2)}(r)
\,=\,
 \beta(r)
\ee
where $\beta(r)$ is the parabola (\ref{beta-def-intro}) restricted to $0 \leqslant r \leqslant \mathcal{R}$,
which is independent both of the dimensionality parameter $d$ and of the mode parameter $l$.
%

In Fig.\,\ref{beta-V0-horizontal-lines-mode} we report
numerical data points for the combination of diagonals (\ref{M-summation k^0 continuum}) 
which support the following conjecture
\be
\label{V-infty-beta-new-term}
\frac{\mathcal{V}_{k_\textrm{\tiny max}}^{(l,0)}(r)}{a^2}
\;\longrightarrow\;
\mu_d(l)\,
\frac{\beta(r)}{r^2}\,.
\ee
The horizontal dashed lines in both the panels of Fig.\,\ref{beta-V0-horizontal-lines-mode}
correspond to the coefficient $\mu_d(l)$ defined in (\ref{mu-d-def}).
We find it worth highlighting that, although 
the diagonals shown in Fig.\,\ref{fig-diagonals-V} for $d=2$ and $d=3$ seem identical,
their combination (\ref{V0-def}) displays the peculiar dependence on $d$
given by (\ref{mu-d-def}),
as shown in Fig.\,\ref{beta-V0-horizontal-lines-mode}.
This is a characteristic feature of the fact that 
we are considering an entanglement Hamiltonian in a 
Minkowski spacetime with a number of spatial dimensions strictly larger than one;
indeed, the term corresponding to  (\ref{V-infty-beta-new-term}) 
gives the vanishing function when $d=1$.
Near the boundary of the sphere, i.e. where $i/r \sim 1^-$,
the discrepancy between the data points 
(which are obtained as a ratio of two quantities that are both vanishing at the boundary of the sphere)
and (\ref{V-infty-beta-new-term}) increases with $l$.
These discrepancies becomes very similar when the logarithmic scale is adopted
(see the bottom panels of Fig.\,\ref{beta-V0-horizontal-lines-mode}).

In Fig.\,\ref{beta-Vd1-zero} 
we show numerical data for the combinations of diagonals introduced in (\ref{V12-def})  and (\ref{V22-def}). 
In this case the curves for different sizes $R$ do not collapse and tend to zero as $R$ increases;
hence it is natural to conjecture that
\be
\label{conj-infty-zero}
\mathcal{V}_{1,\infty}^{(l,2)}(r) \,=\, 0
\;\;\;\qquad\;\;\;
\mathcal{V}_{2,\infty}^{(l,2)}(r) \,=\, 0\,.
\ee

We emphasise that large values of $k_{\textrm{\tiny max}}$ are needed 
to obtain the numerical results described above.
In appendix\;\ref{app_kmax} we show that the expected CFT results are not obtained
when $k_{\textrm{\tiny max}}$ is not large enough.
This crucial message can be appreciated 
e.g. for (\ref{conj-infty-beta}) and (\ref{V-infty-beta-new-term})
by comparing 
the right panels of 
Fig.\,\ref{beta-TV-parabola} and Fig.\,\ref{beta-V0-horizontal-lines-mode} 
with Fig.\,\ref{fig:T0andV2app} and Fig.\,\ref{fig:V0app} respectively.
The conjecture (\ref{V-infty-beta-new-term}) naturally leads to ask 
how $k_{\textrm{\tiny max}} \to \infty$ should be taken in the continuum limit. 
In Fig.\,\ref{beta-TV-parabola}, Fig.\,\ref{beta-V0-horizontal-lines-mode} 
and Fig.\,\ref{beta-Vd1-zero}
the ratio $k_{\textrm{\tiny max}}/R$ is kept fixed 
(and equal to $1/10$) as $R$ increases.
In the appendix\;\ref{app_kmax} we report and discuss also numerical results 
obtained by keeping $k_{\textrm{\tiny max}}$ fixed as $R$ increases 
(see Fig.\,\ref{fig:T0andV2app} and Fig.\,\ref{fig:V0app}). 
We find that these two different limiting procedures give the same results
for the quantities that we are considering.
However, this question deserves further investigations.
Notice also that
all the data reported in 
Fig.\,\ref{beta-TV-parabola}, Fig.\,\ref{beta-V0-horizontal-lines-mode} 
and Fig.\,\ref{beta-Vd1-zero}
do not display good collapses around the center of the sphere
(i.e. where $i/R \sim 0^+$):
we expect that larger systems are needed to observe them. 

The conjectures 
(\ref{conj-infty-beta}),  (\ref{V-infty-beta-new-term}) and (\ref{conj-infty-zero})
have been formulated for any value of $l$.
However, the data points 
in Fig.\,\ref{beta-TV-parabola}, Fig.\,\ref{beta-V0-horizontal-lines-mode} and Fig.\,\ref{beta-Vd1-zero}
provide numerical support to these conjectures only for the small values of $l$
that we have been able to explore. 
The validity of (\ref{conj-infty-beta}),  (\ref{V-infty-beta-new-term}) and (\ref{conj-infty-zero})
for any $l$ is a strong assumption and we find it 
worth improving the numerical analysis described above in order to check it
also for higher values of $l$.

Finally, by inserting 
(\ref{conj-infty-beta}), (\ref{V-infty-beta-new-term}) and (\ref{conj-infty-zero}) 
into (\ref{MplusNoperator expansion bdy}),
one concludes that (\ref{EH-ball-massless-scalar}) 
is the continuum limit of the operator  (\ref{ent-ham HC}).
Then, the summation (\ref{KB-mode-sum}) provides the final 
CFT result for the entanglement Hamiltonian of the sphere for the massless scalar field.

\begin{figure}[t!]
\centering
\begin{subfigure}{.4\textwidth}
\hspace{-1.5cm}
\includegraphics[scale=.3]{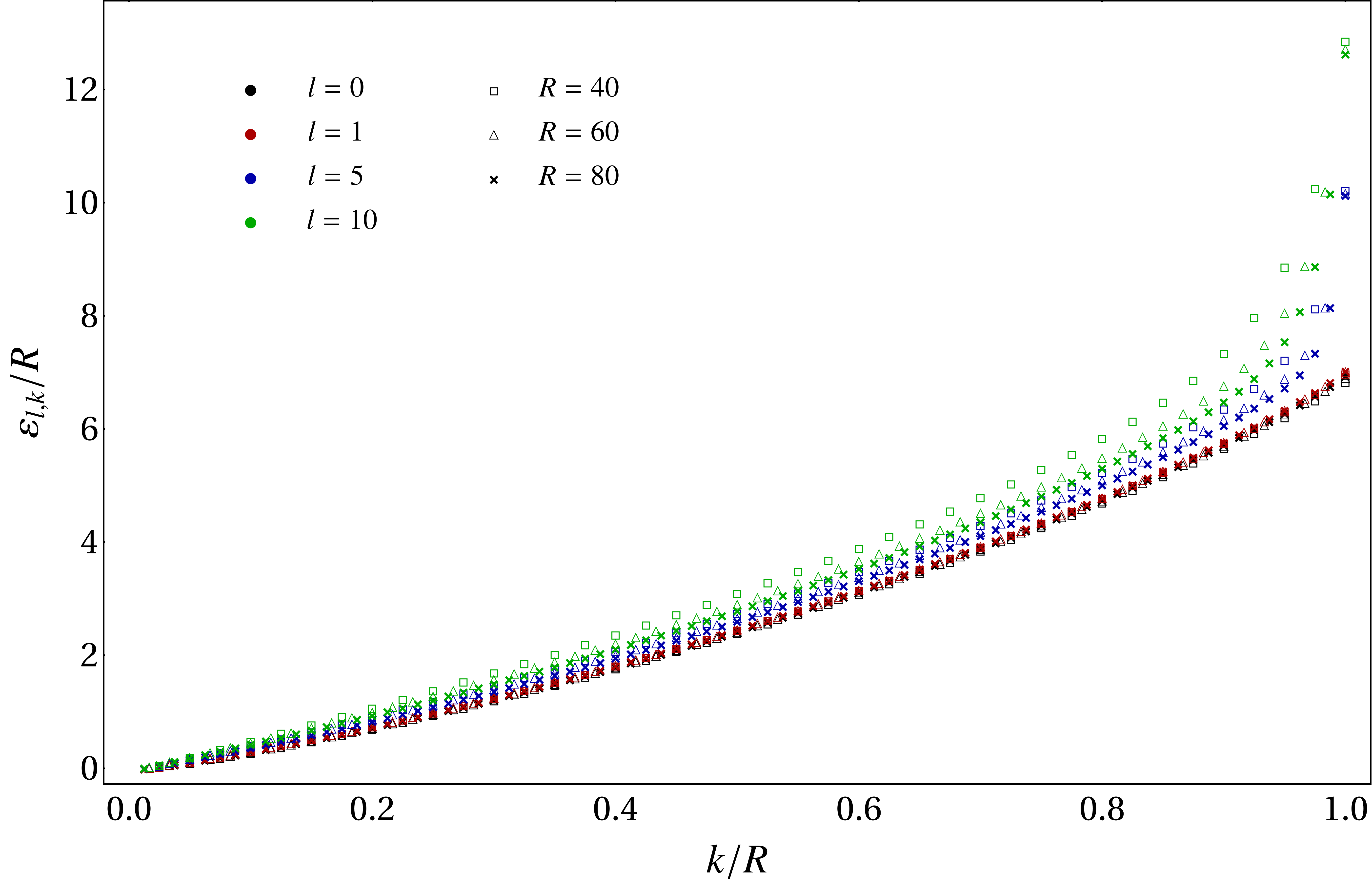}
\end{subfigure}
\hfill
\begin{subfigure}{.4\textwidth}
\hspace{-1.5cm}
\includegraphics[scale=.3]{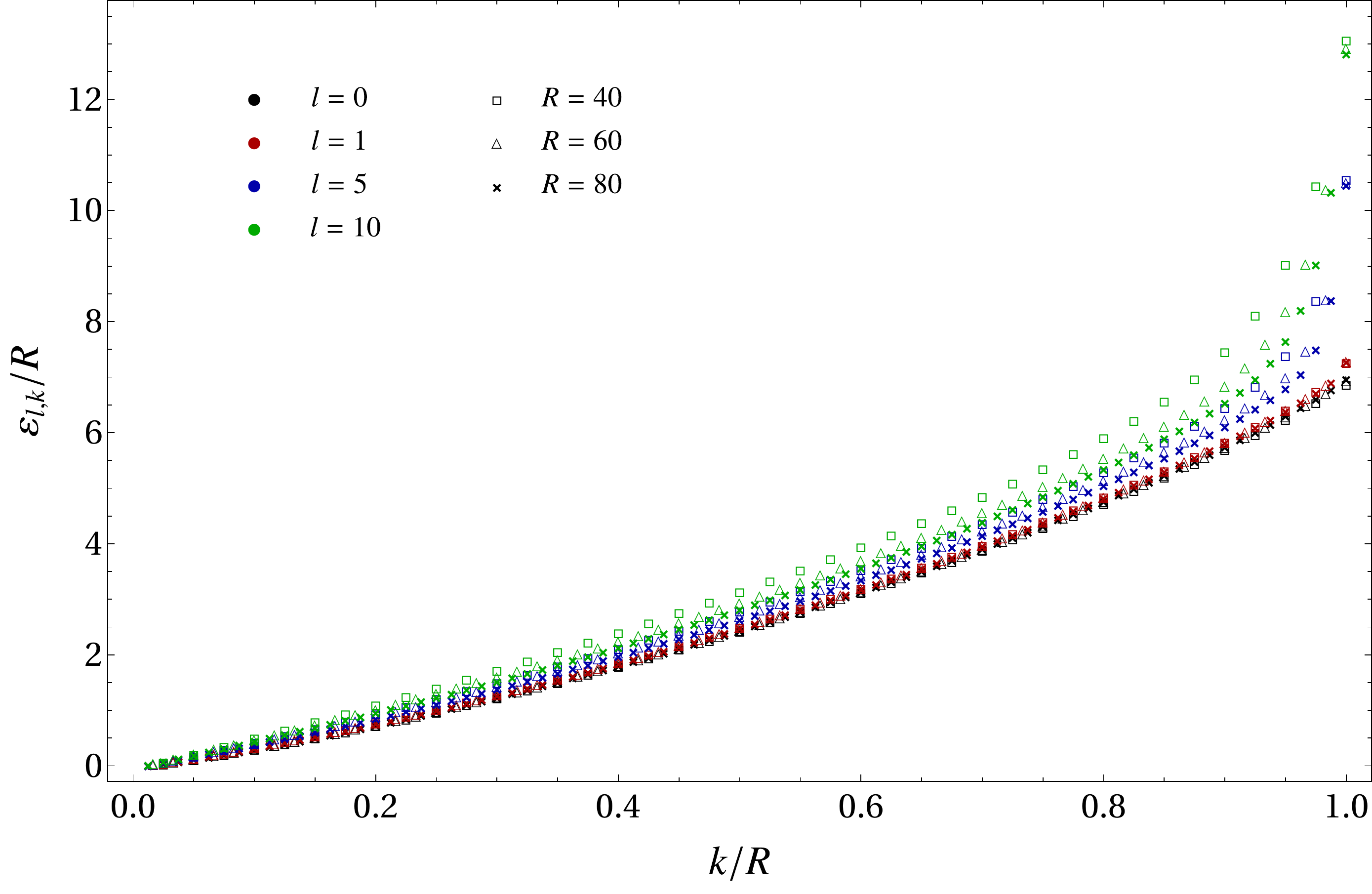}
\end{subfigure}
\hfill
\caption{Single particle entanglement energies at given $l$ in the case of $\omega =0$
when $d=2$ (left panel) and $d=3$ (right panel),
for various values of $l$ and $R$.
}
\label{fig:eps0}
\end{figure}

We find worth discussing also some numerical results for the entanglement entropy. 
Since the ground state is the direct product of the ground states 
corresponding to $\widehat{H}_{\boldsymbol{l}}$,
the entanglement entropy of the sphere $S_B$ is obtained
by summing the contributions corresponding to all the different 
values of $\boldsymbol{l}$ \cite{Srednicki:1993im}.
These contributions can be evaluated through the symplectic eigenvalues 
of the reduced covariance matrix $Q_{l,B} \oplus P_{l,B}$ at given $l$
or, equivalently, through the single particle entanglement energies at given $l$
\cite{Botero04,Audenaert_2002,Plenio:2004he,Cramer:2005mx,Schuch_2006, Weedbrook12b},
as briefly anticipated in Sec.\,\ref{subsec-eh}.
Since $Q_{l,B} \oplus P_{l,B}$ depends only on $l$,
the entanglement entropy of the sphere $B$ is computed as follows \cite{Srednicki:1993im}
\be
\label{EE-mode-sum}
S_B = 
\sum_{l=0}^\infty 
N_{d,l} \,S_{B,l}
\;\;\;\;\qquad\;\;\;\;
S_{B,l}
\equiv
\sum_{k=1}^R
\left(\,
\frac{\varepsilon_{l,k}}{e^{\varepsilon_{l,k}}\ - 1} - \log\! \big( 1 - e^{- \varepsilon_{l,k}}\big)
\right)
\ee
which has been explored numerically in various studies
\cite{Srednicki:1993im, Casini:2009sr, Lohmayer:2009sq, Huerta:2011qi, Klebanov:2012va, Liu:2012eea}.

In Fig.\,\ref{fig:eps0} we show the single particle entanglement energies 
$\varepsilon_{l,k}/R$ at given $l$ in terms of $k/R$ when $\omega =0$,
for $d=2$ (left panel) and $d=3$ (right panel).
The summand in $S_{B,l}$ diverges as $\varepsilon_{l,k} \to 0^+$;
hence the low-lying part of the single particle entanglement spectrum at fixed $l$ 
provides the largest contribution to $S_{B,l}$.
Clear differences between $\varepsilon_{l,k}/R$ for $d=2$ and $d=3$ are not visible, 
despite the fact that $Q_{l,B} \oplus P_{l,B}$ depends explicitly on $d$ (see (\ref{matrix-M})).

\begin{figure}[t!]
\centering
\begin{subfigure}{.4\textwidth}
\hspace{-1.5cm}
\includegraphics[scale=.3]{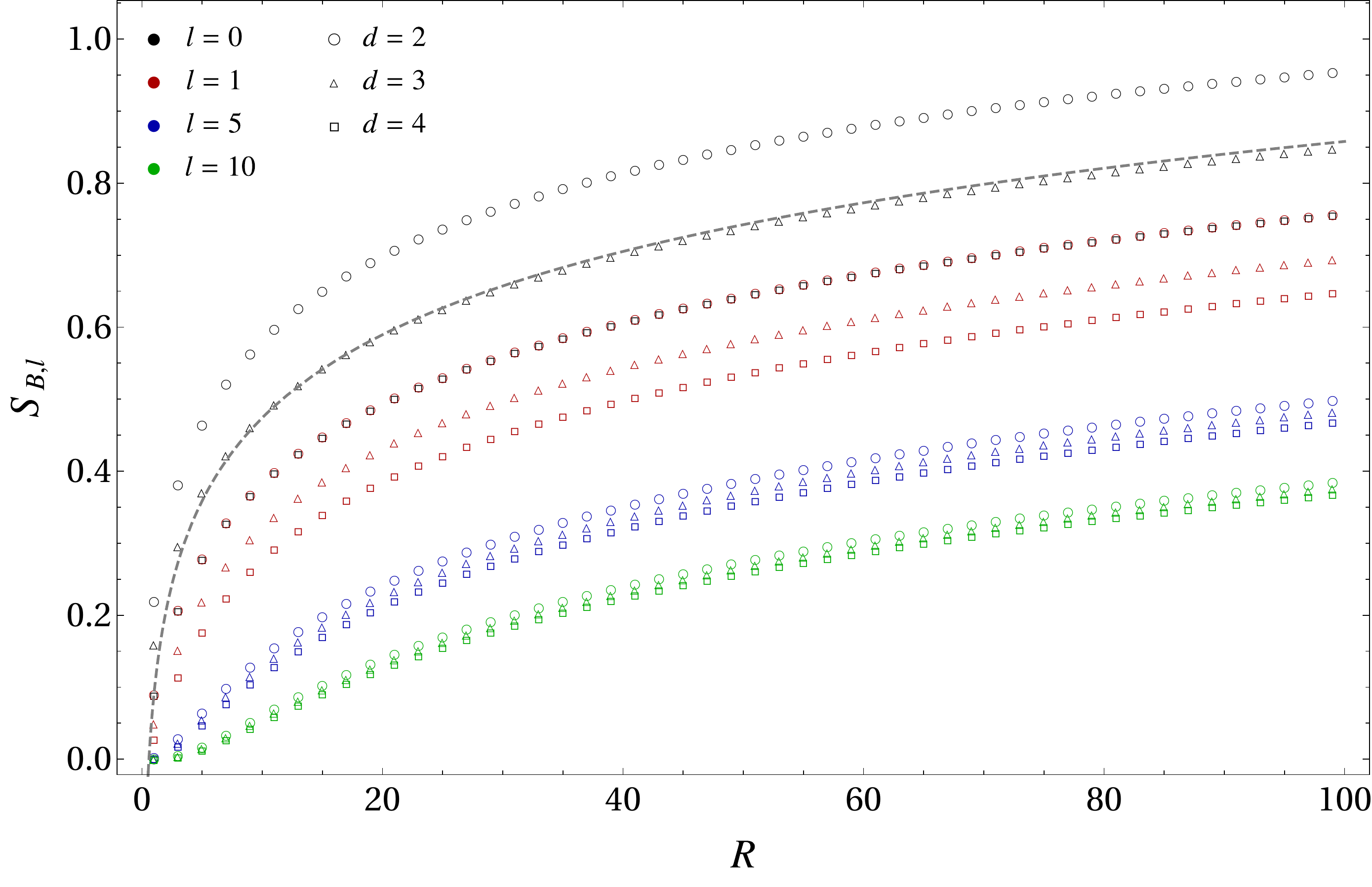}
\end{subfigure}
\hfill
\begin{subfigure}{.4\textwidth}
\hspace{-1.5cm}
\includegraphics[scale=.35]{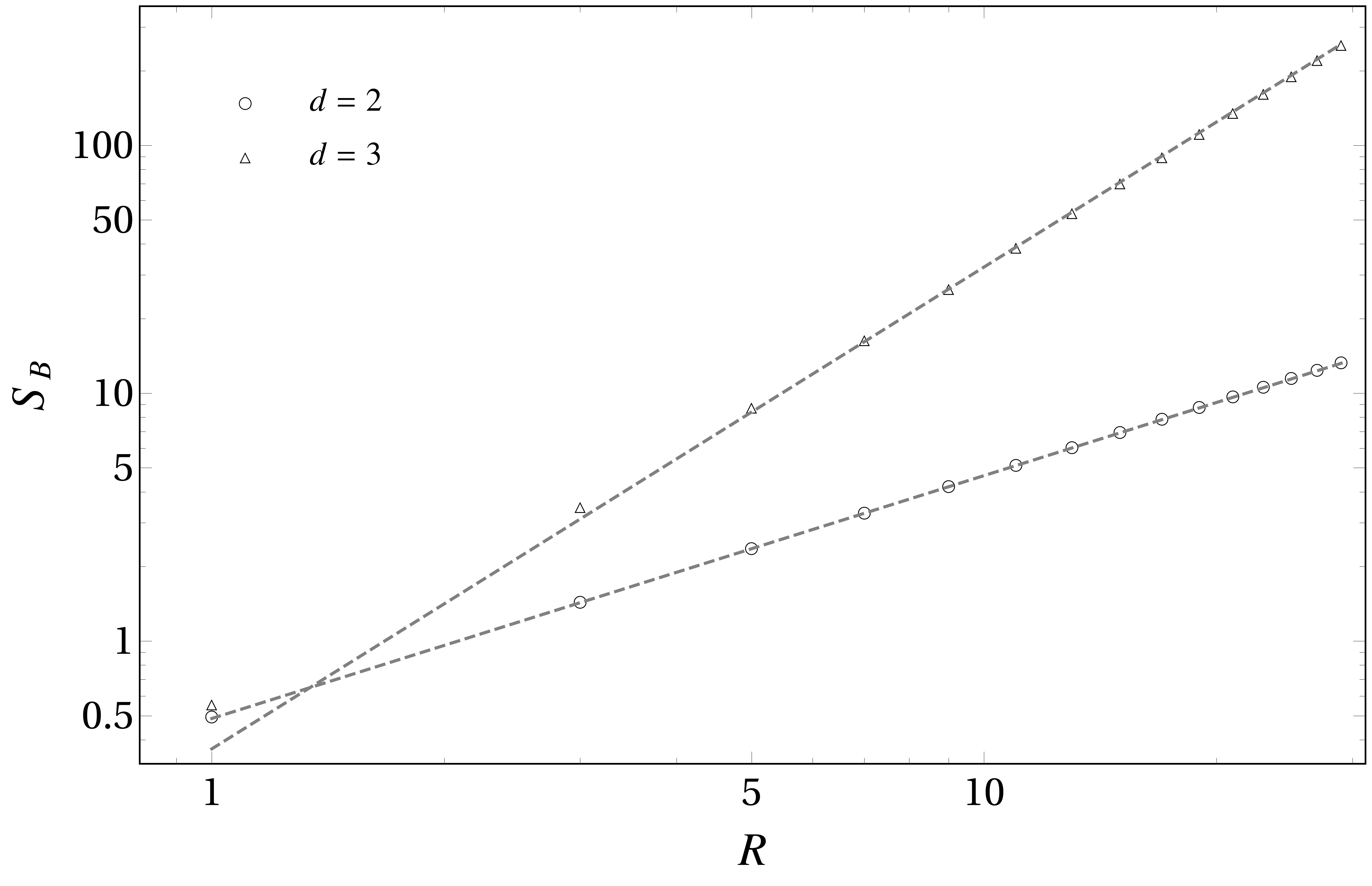}
\end{subfigure}
\hfill
\caption{The quantity $S_{B,l}$ (left panel) and
the entanglement entropy $S_B$ (right panel),
defined in (\ref{EE-mode-sum}), when $\omega =0$ and for some values of $d$.
}
\label{fig:EE-various}
\end{figure}

In the left panel of Fig.\,\ref{fig:EE-various},
we show some numerical data for $S_{B,l}$ defined in  (\ref{EE-mode-sum})
which display a clear dependence on the dimensionality parameter $d$ for this quantity. 
In this panel  the dashed curve corresponds to $\frac{1}{6} \log(R)+\mathrm{const}$,
which is the entanglement entropy of a segment at the beginning of the semi-infinite line \cite{Calabrese:2004eu}
and it  coincides with $S_{B,0}$ when $d=3$, 
as expected from the fact that $\mu_1(0) =\mu_3(0) =0$.
Moreover, the data corresponding to $(d,l)=(2,1)$ 
coincides with ones corresponding to $(d,l)=(4,0)$ 
because (\ref{mu-d-def}) gives $3/4$ in both these cases. 
In the right panel of Fig.\,\ref{fig:EE-various} 
we display some numerical results for the entanglement entropy $S_B$
(obtained through (\ref{EE-mode-sum}) by restricting the sum over $l$ to $l \leqslant 1000$)
showing the area law behaviour first observed in \cite{Srednicki:1993im} in a clear way.
The dashed lines in this panel correspond to $\alpha(d)\,R^{d-1}$
and the fitted values of the slopes are $\alpha(2) = 0.46$ and $\alpha(3) = 0.30$, 
in agreement with the numerical results reported e.g. in \cite{Srednicki:1993im, Lohmayer:2009sq,Liu:2012eea}.

Many interesting results have been obtained for the entanglement entropy of the sphere and other interesting subregions in the continuum limit for generic $d$
through various quantum field theory mehods
\cite{Calabrese:2004eu, Solodukhin:2010pk,Hertzberg:2010uv, Casini:2010kt, Dowker:2010nq, Smolkin:2014hba,Hung:2014npa,Herzog:2014fra,Herzog:2016bhv}.
It would be instructive to explore whether these methods can be employed to study also the 
corresponding entanglement Hamiltonians.

\section{Entanglement Hamiltonian at $\omega>0$}
\label{sec_massive}

In this section we consider the entanglement Hamiltonian 
of the $d$ dimensional sphere $B$ 
in the regularised model of \cite{Srednicki:1993im} introduced  in Sec.\,\ref{sec-ham-eh}
when $\omega > 0$ and the entire system is in its ground state.
We adapt to this case the analysis performed in \cite{Eisler:2020lyn}
for the entanglement Hamiltonian of a block made by consecutive sites 
in the infinite and non-critical harmonic chain.
The numerical setup is the one described in Sec.\,\ref{sec_EH_ball_massless}
and, since the outcomes for $d=2$ and $d=3$ are very similar, 
in the following we report only the results corresponding to $d=3$.

\begin{figure}[t!]
\centering
\begin{subfigure}{.32\textwidth}
\hspace{-2cm}
\includegraphics[scale=.35]{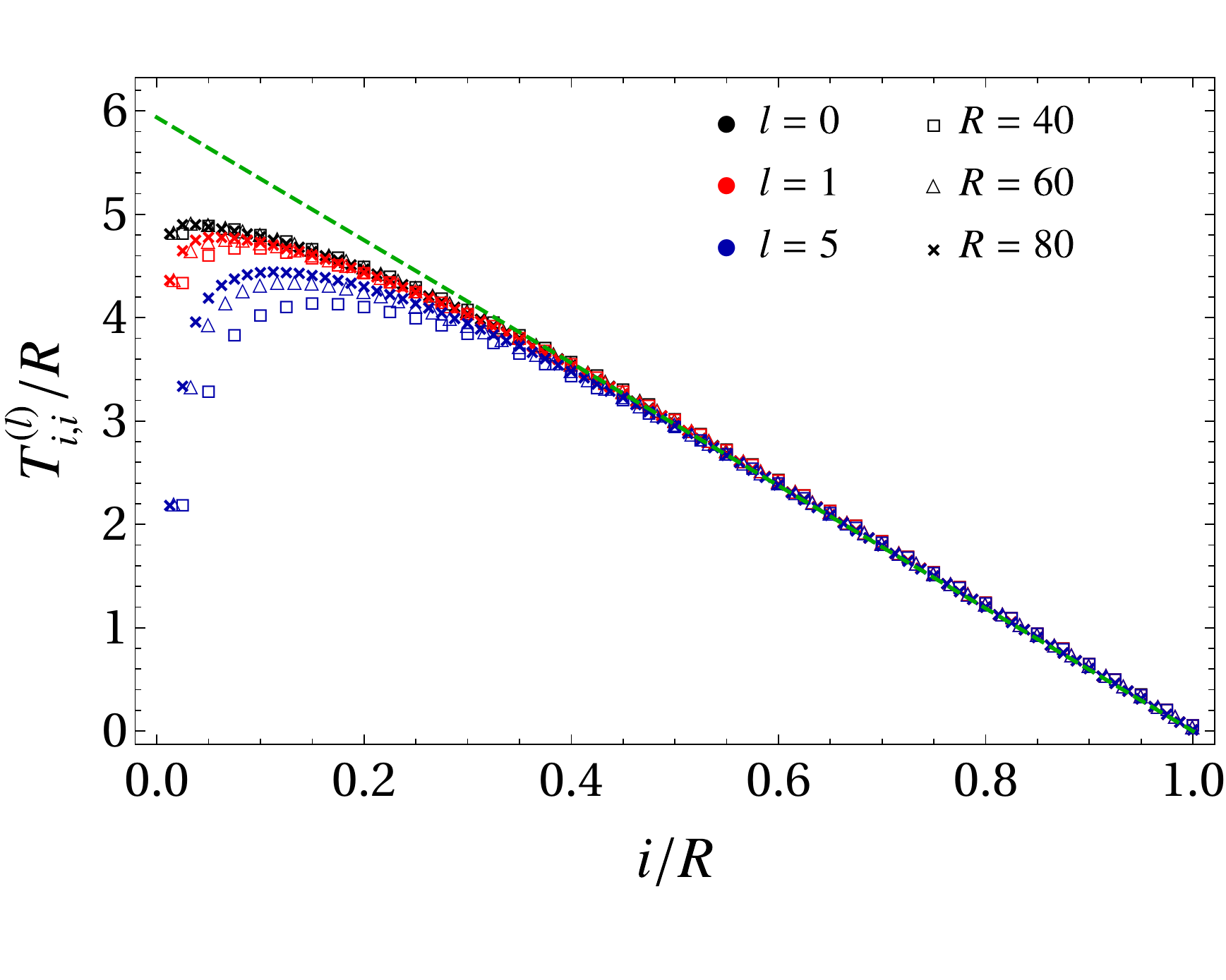}
\end{subfigure}
\hfill
\begin{subfigure}{.32\textwidth}
\hspace{-.9cm}
\includegraphics[scale=.35]{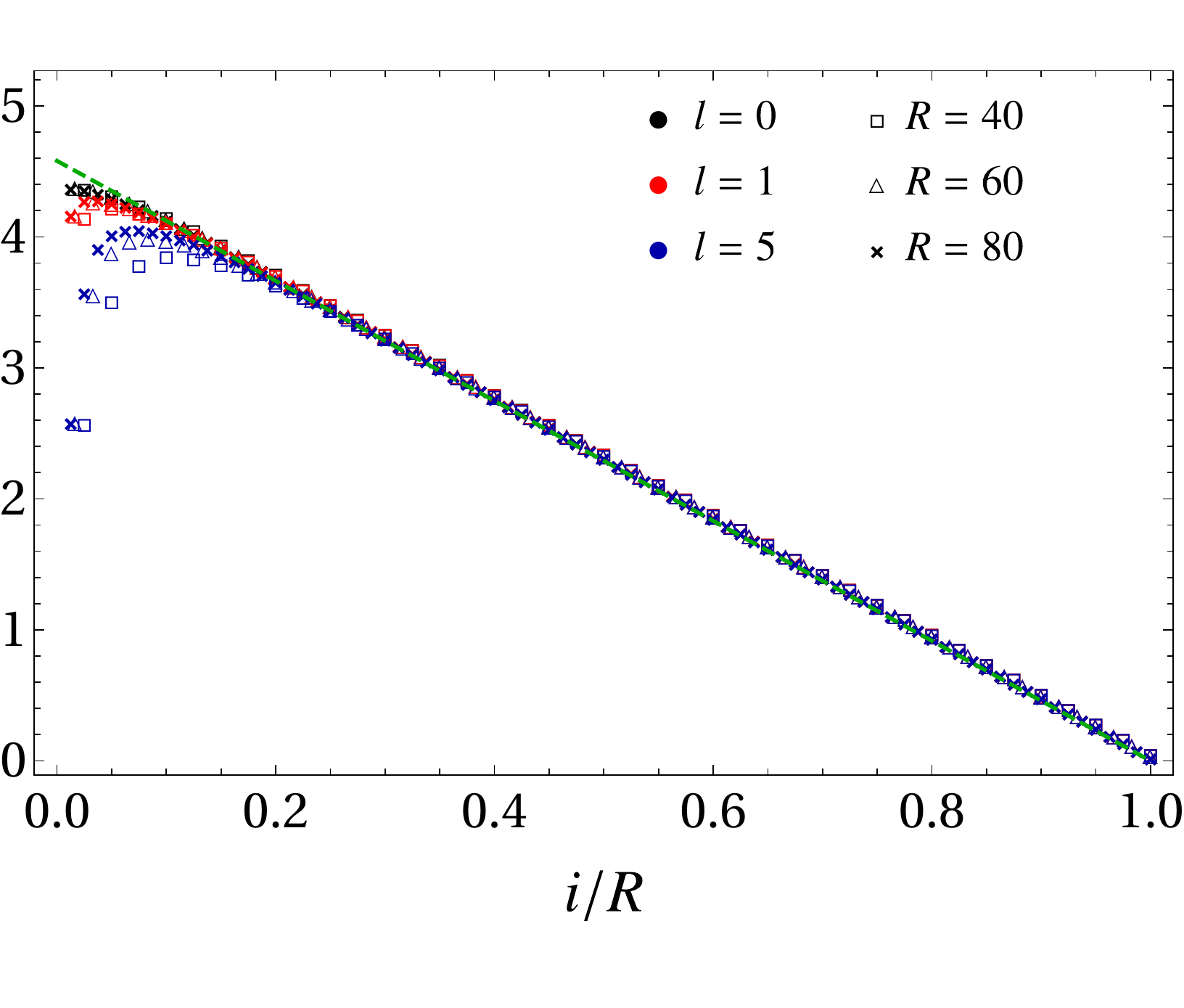}
\end{subfigure}
\hfill
\begin{subfigure}{.32\textwidth}
\hspace{1cm}
\includegraphics[scale=.35]{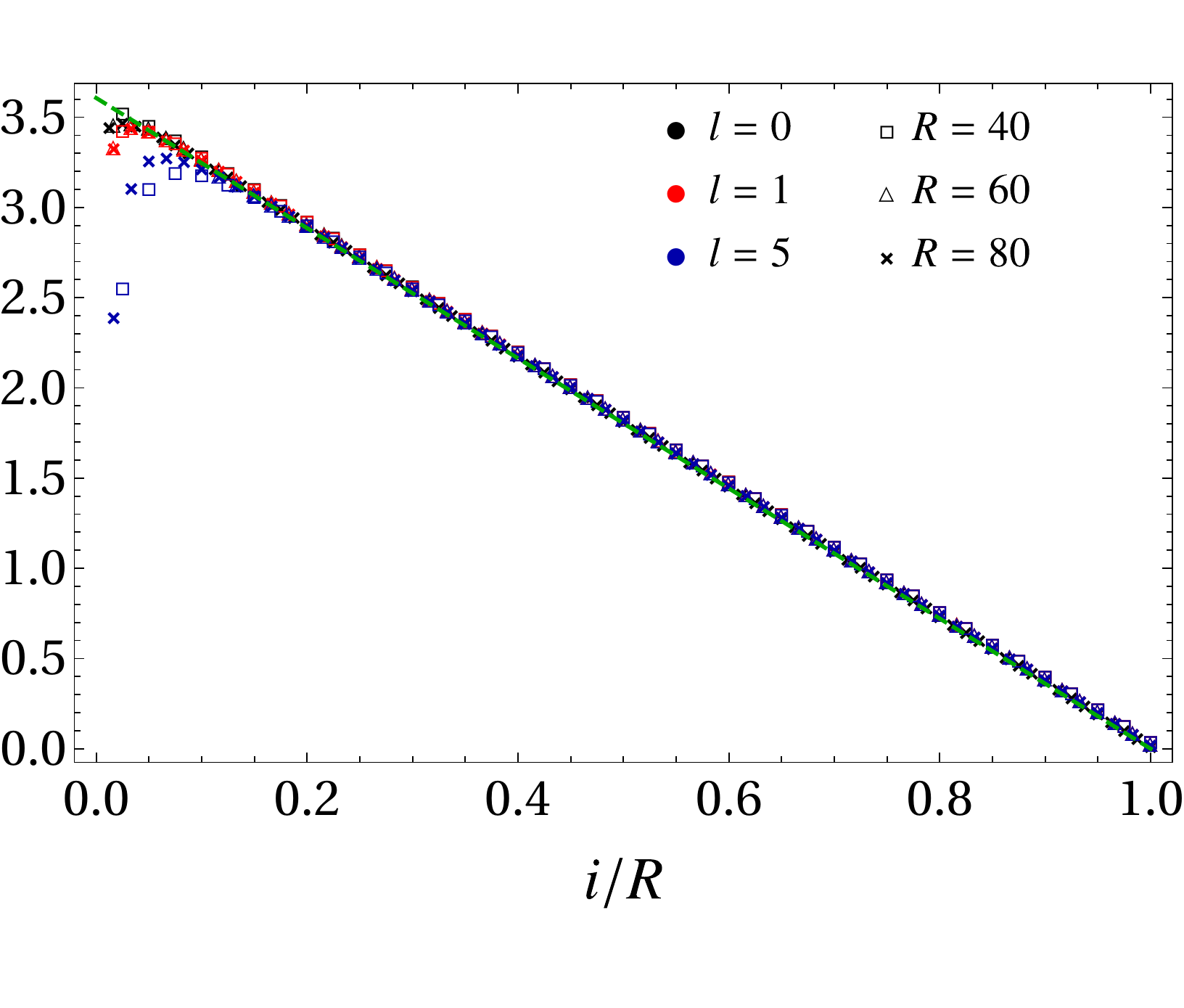}
\end{subfigure}
\hfill

\vspace{-.5cm}
\begin{subfigure}{.32\textwidth}
\hspace{-2cm}
\includegraphics[scale=.35]{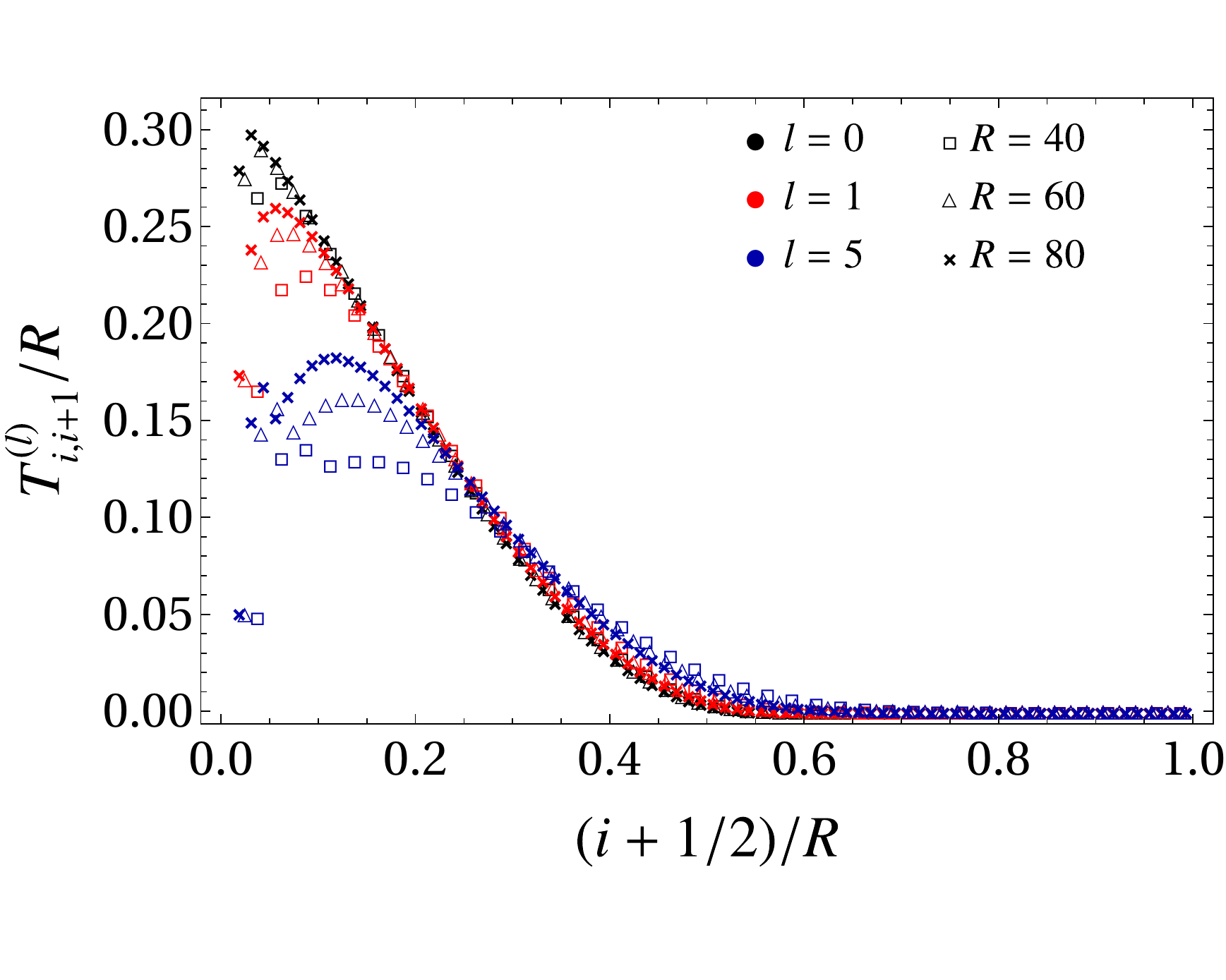}
\end{subfigure}
\hfill
\begin{subfigure}{.32\textwidth}
\hspace{-.9cm}
\includegraphics[scale=.35]{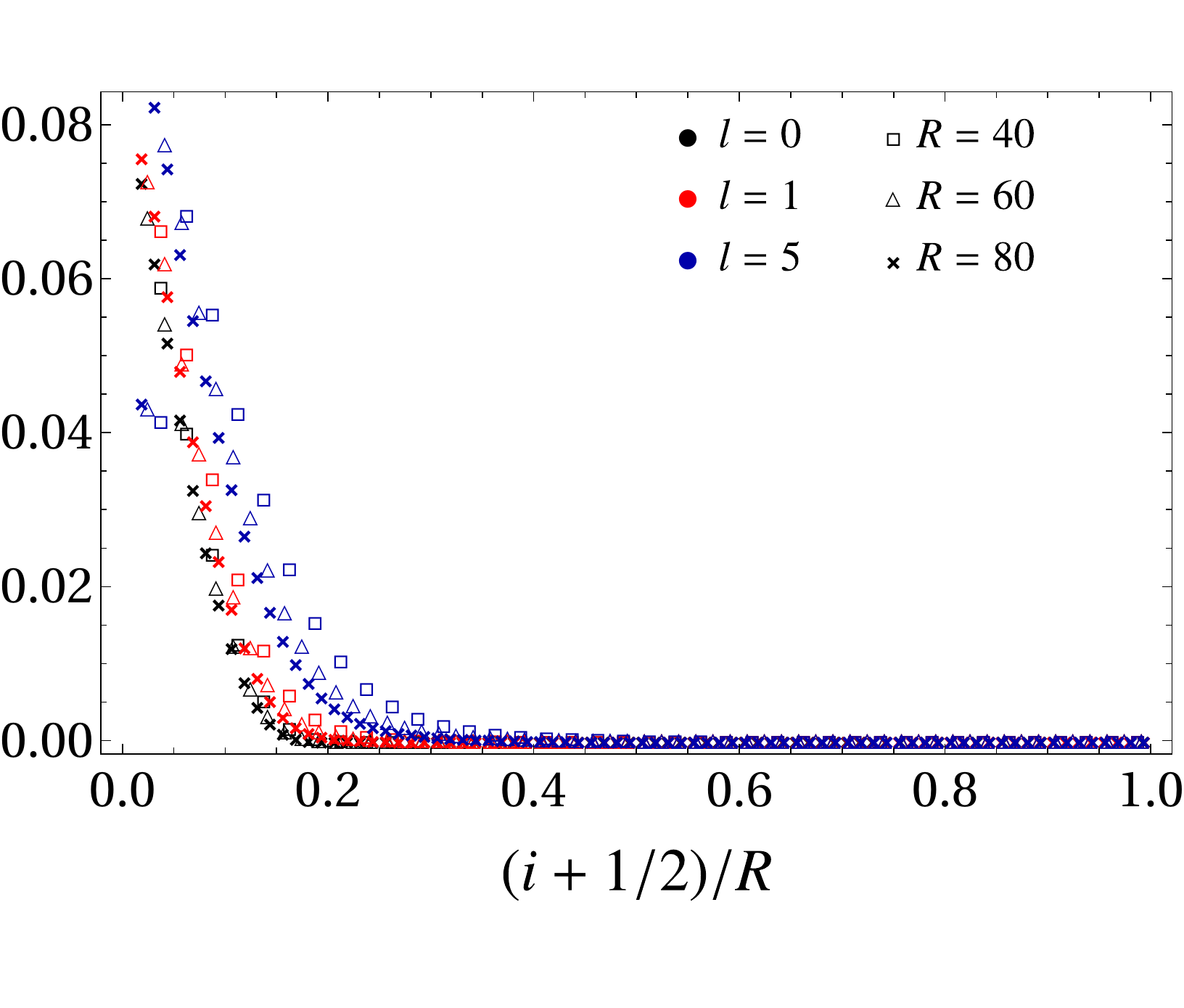}
\end{subfigure}
\hfill
\begin{subfigure}{.32\textwidth}
\hspace{1cm}
\includegraphics[scale=.35]{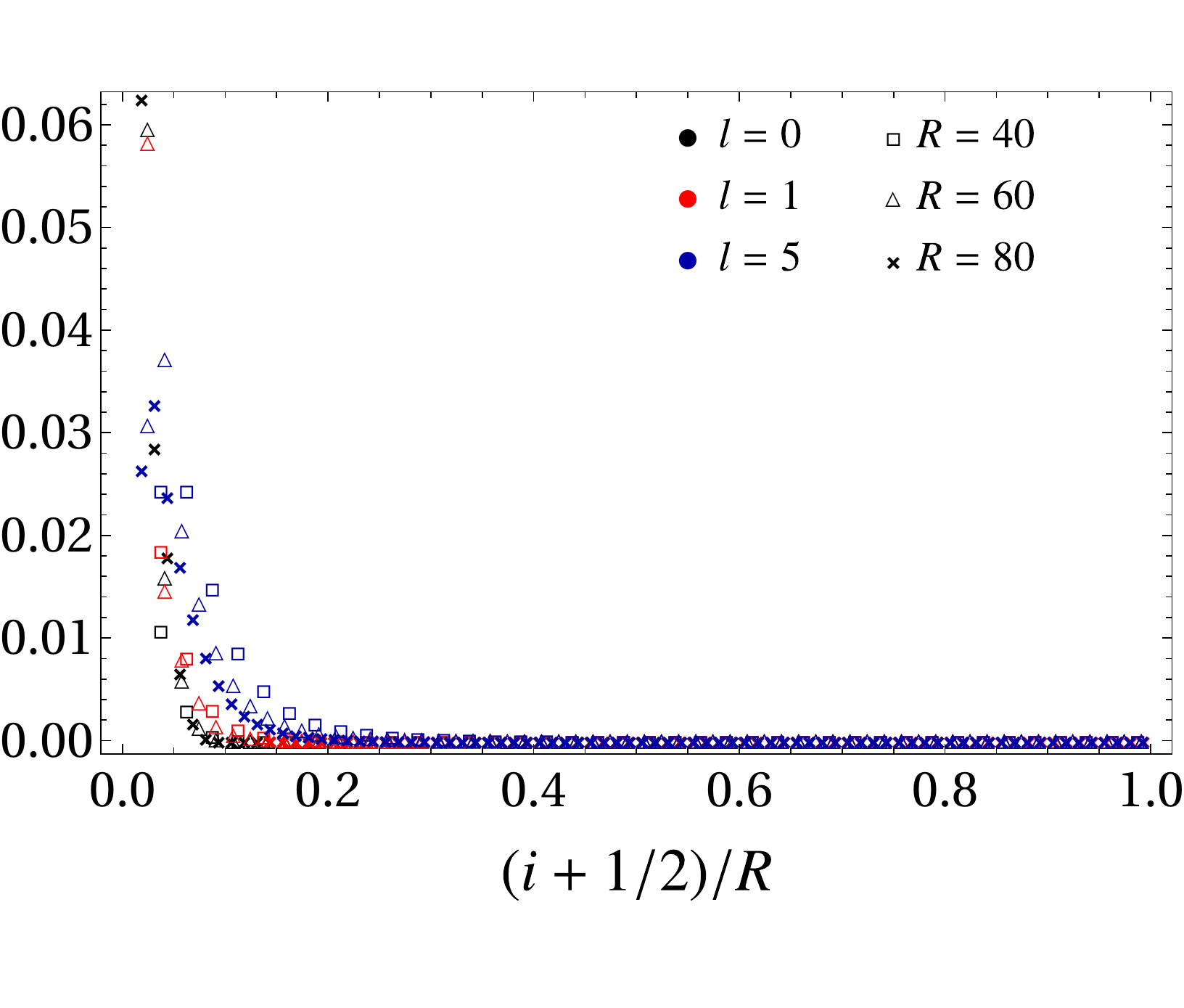}
\end{subfigure}
\hfill

\vspace{-.5cm}
\begin{subfigure}{.32\textwidth}
\hspace{-2cm}
\includegraphics[scale=.35]{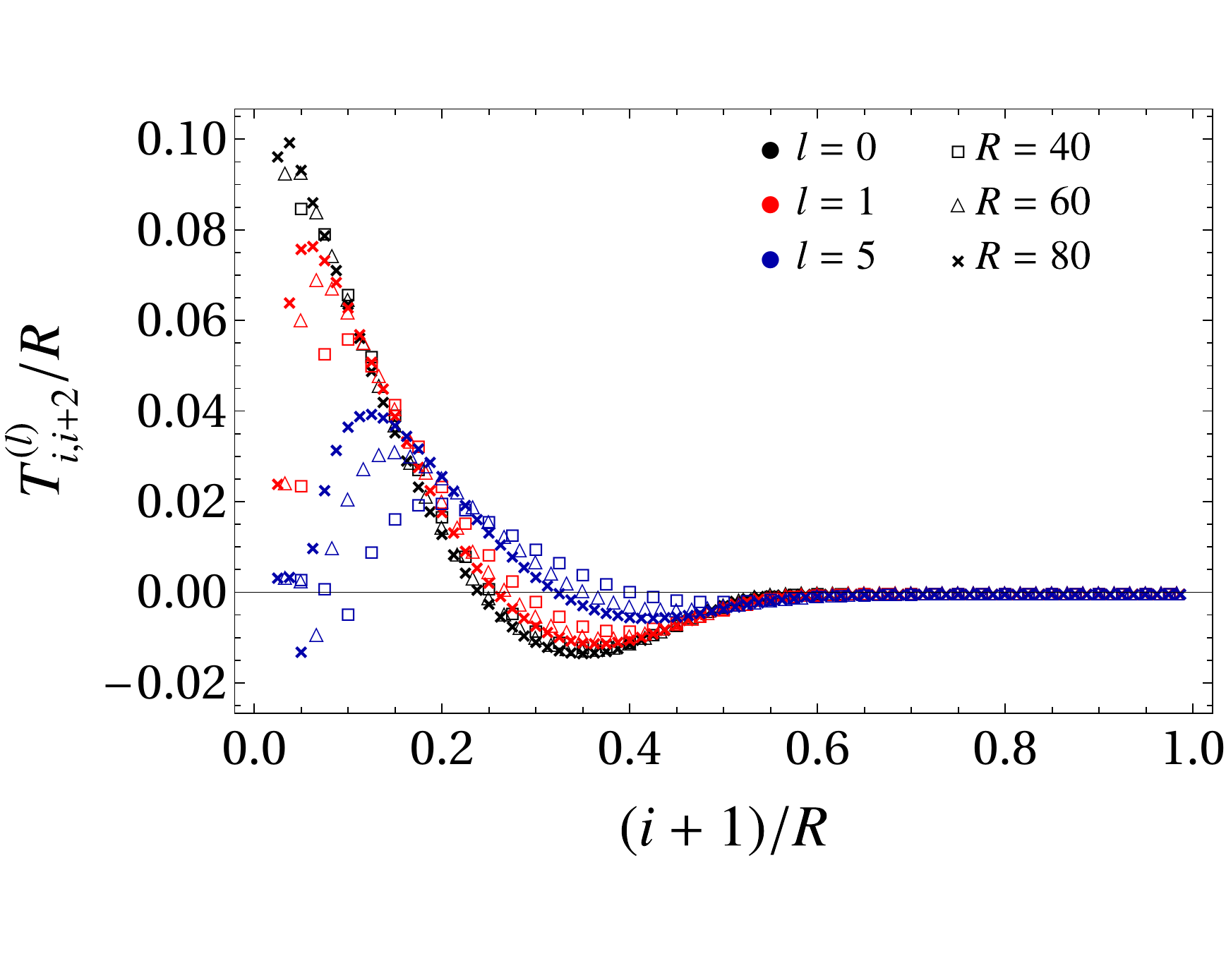}
\end{subfigure}
\hfill
\begin{subfigure}{.32\textwidth}
\hspace{-.9cm}
\includegraphics[scale=.35]{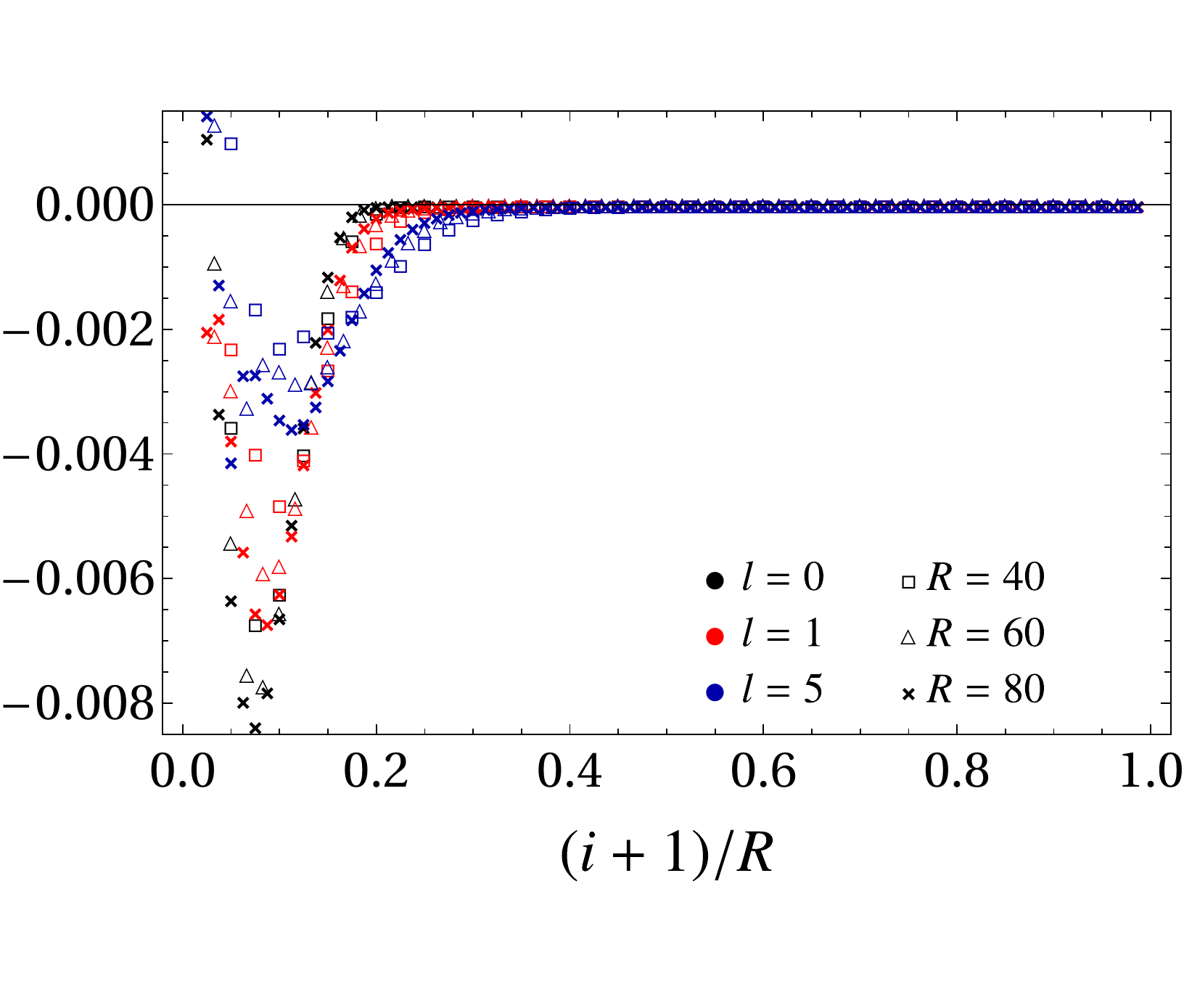}
\end{subfigure}
\hfill
\begin{subfigure}{.32\textwidth}
\hspace{1cm}
\includegraphics[scale=.35]{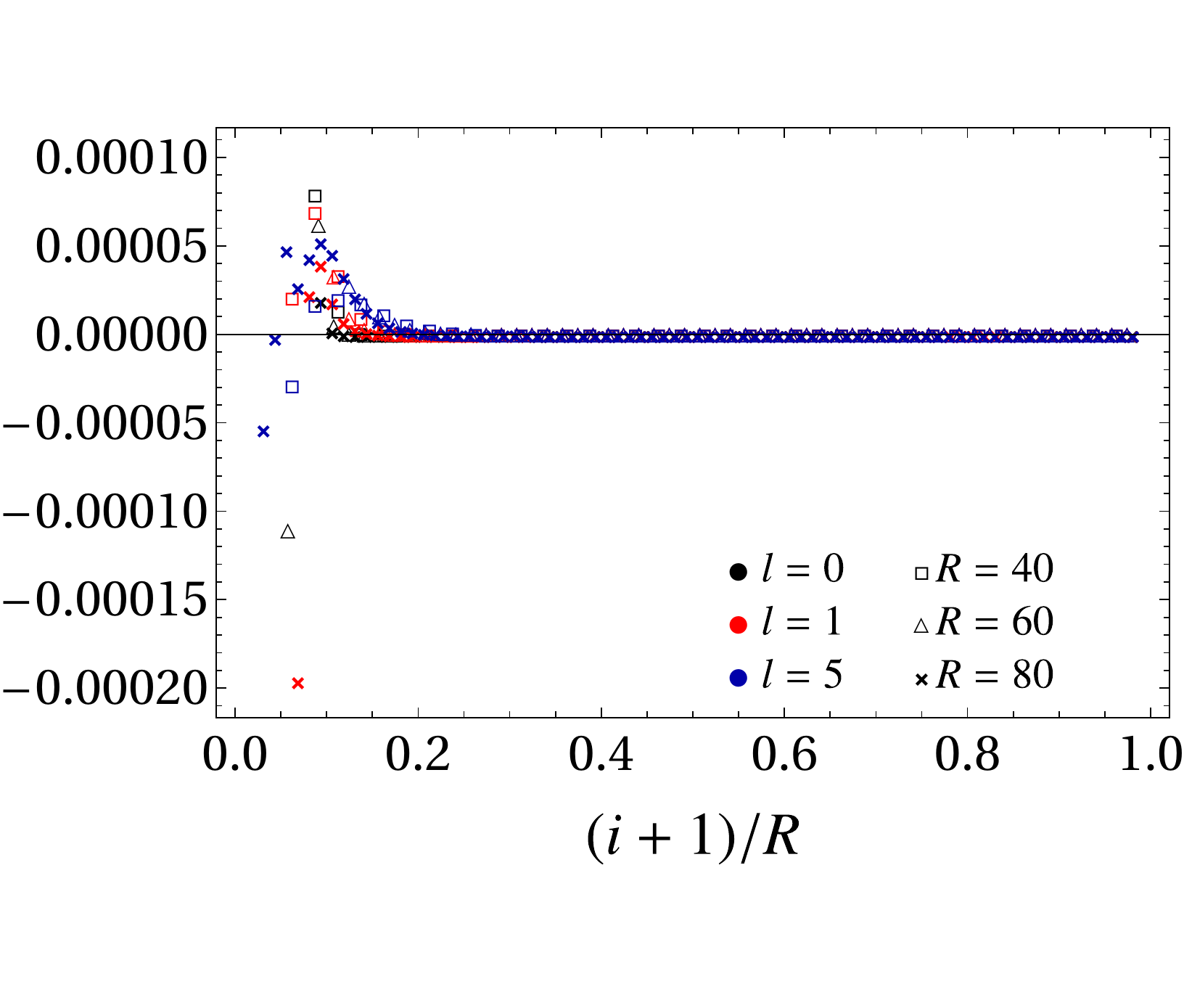}
\end{subfigure}
\vspace{-.5cm}
\caption{
\label{beta-T-triangle-diagonals}
Short-range couplings in the matrices 
$T^{(l)}/R$
(see (\ref{H_T and H_V operators}) and (\ref{munuk def})) 
when $\omega > 0$ and $d=3$,
for $\omega = 1$ (left panels), $\omega = 3$ (middle panels) and $\omega = 5$ (right panels), 
for some small $l$'s and different sizes $R$.
The dashed green lines in the top panels correspond to (\ref{tridiagonal-T}).
}
\end{figure}

\begin{figure}[t!]
\centering
\begin{subfigure}{.32\textwidth}
\hspace{-2cm}
\includegraphics[scale=.35]{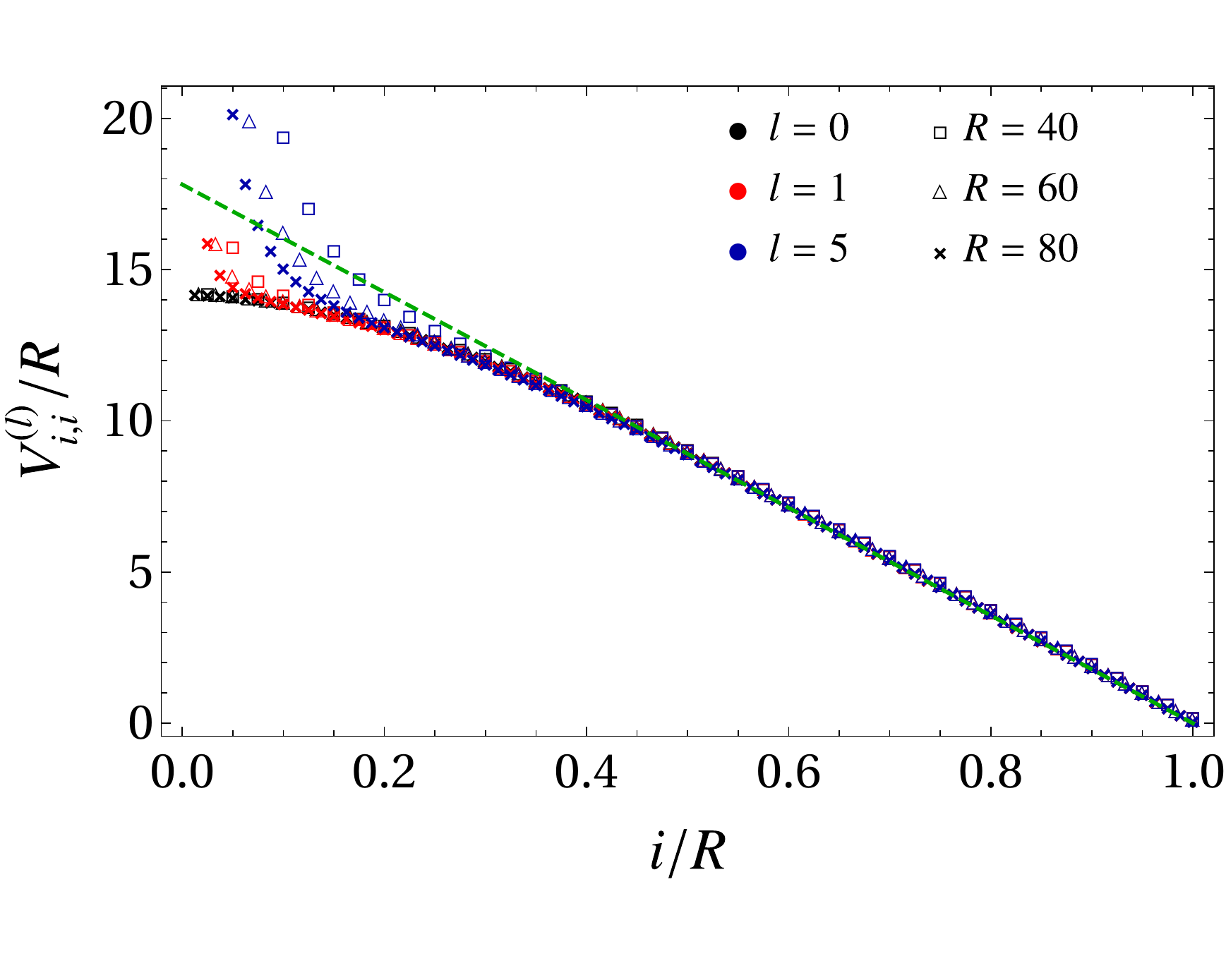}
\end{subfigure}
\hfill
\begin{subfigure}{.32\textwidth}
\hspace{-.9cm}
\includegraphics[scale=.35]{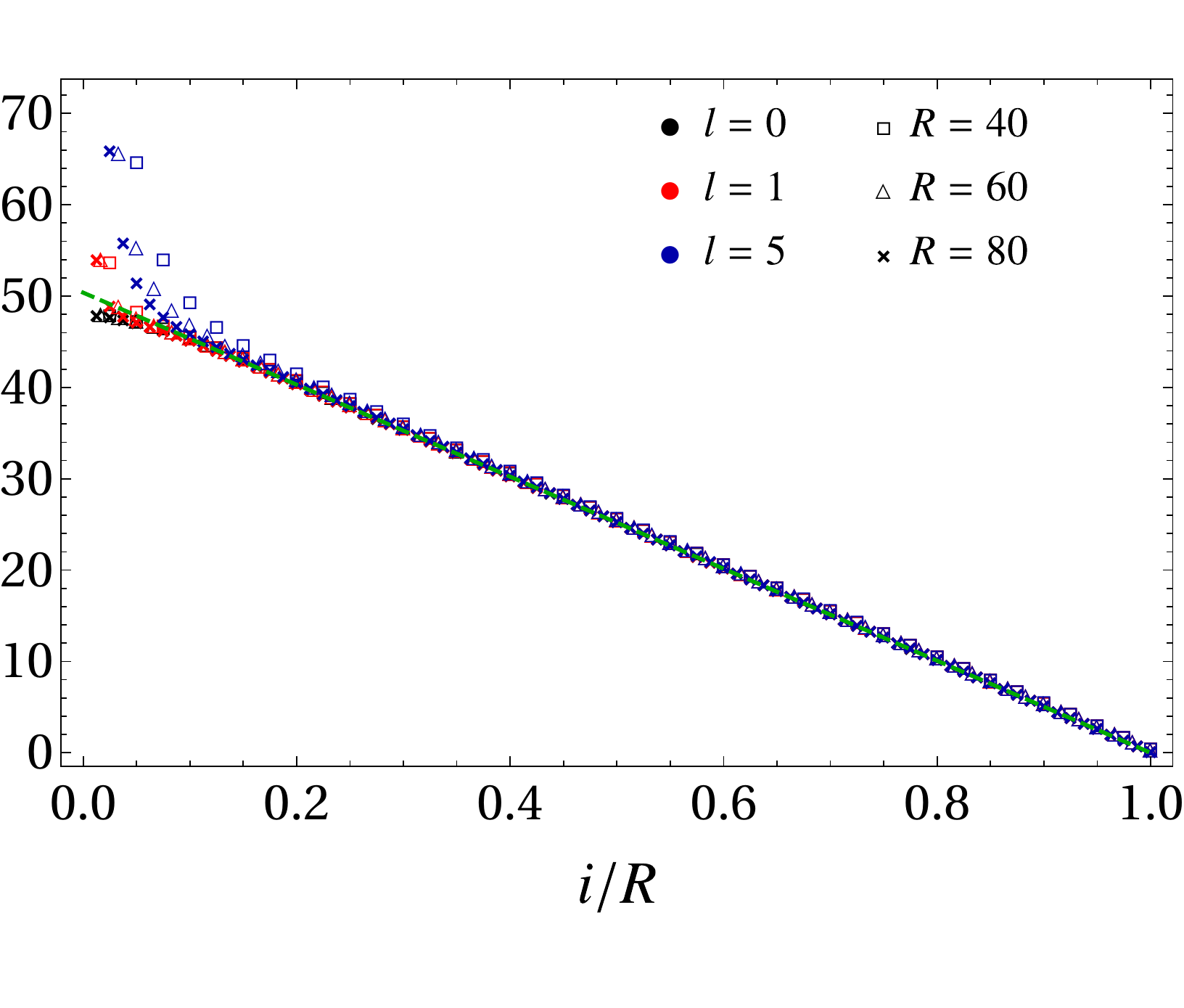}
\end{subfigure}
\hfill
\begin{subfigure}{.32\textwidth}
\hspace{1cm}
\includegraphics[scale=.35]{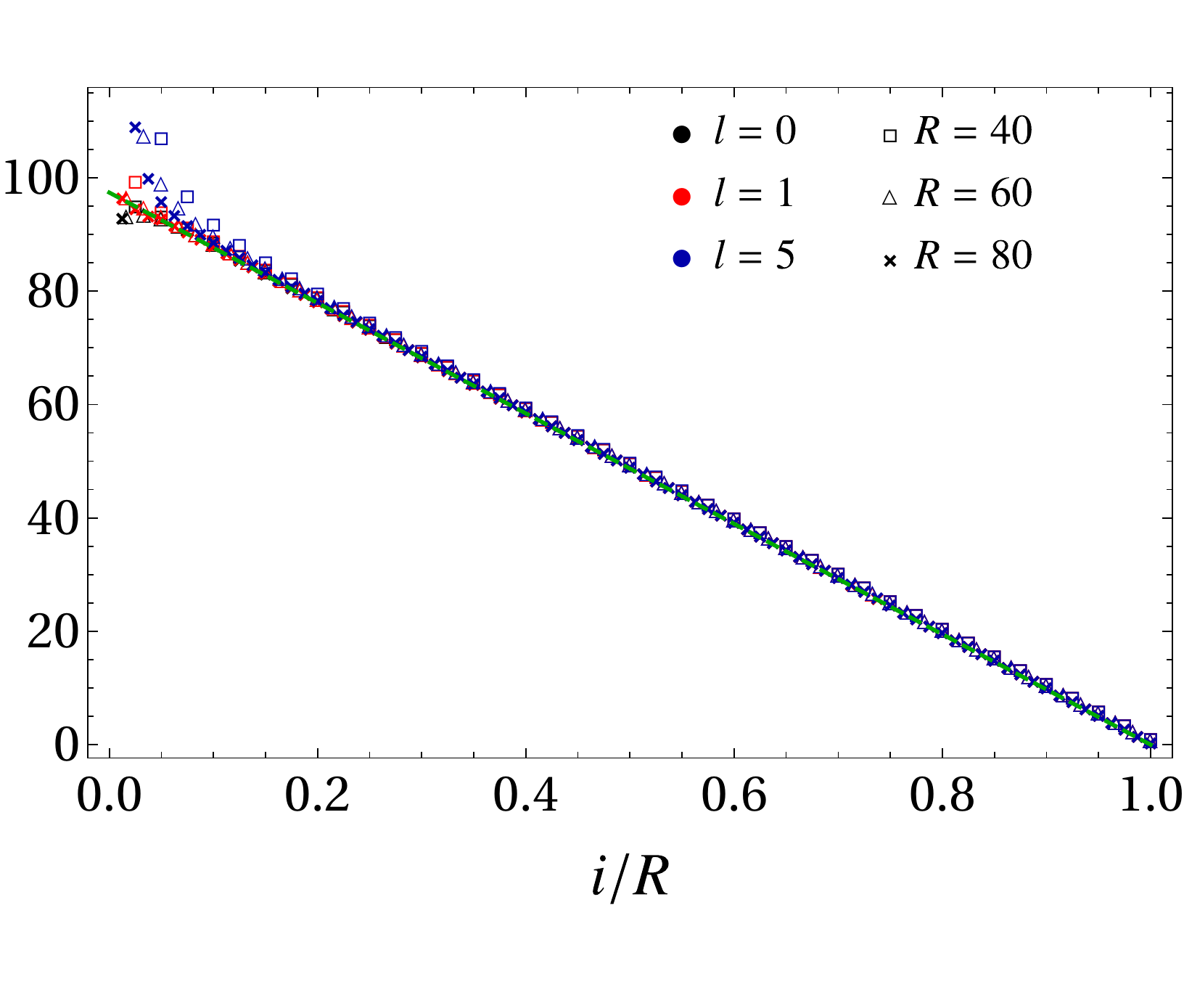}
\end{subfigure}
\hfill

\vspace{-.5cm}
\begin{subfigure}{.32\textwidth}
\hspace{-2cm}
\includegraphics[scale=.35]{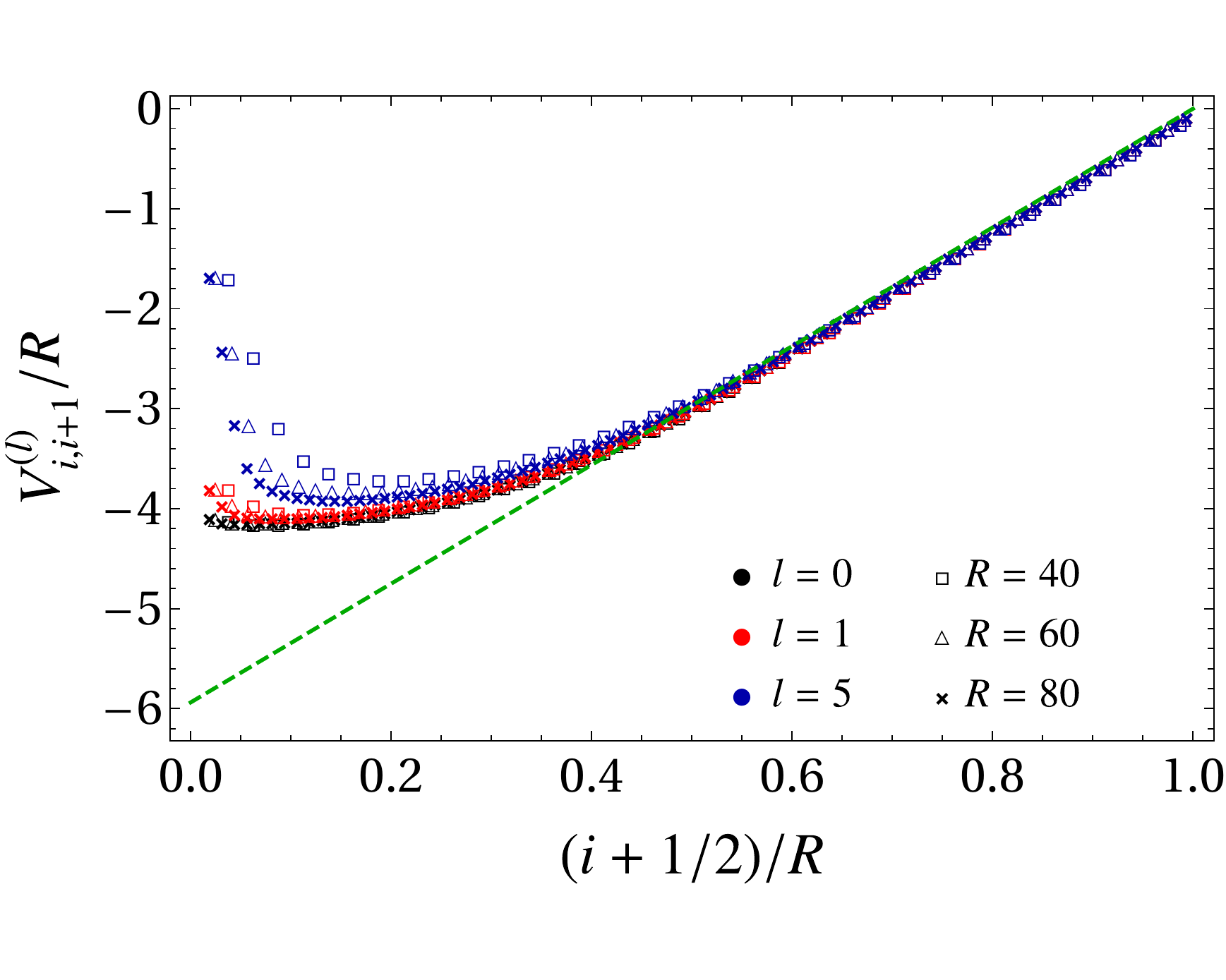}
\end{subfigure}
\hfill
\begin{subfigure}{.32\textwidth}
\hspace{-.9cm}
\includegraphics[scale=.35]{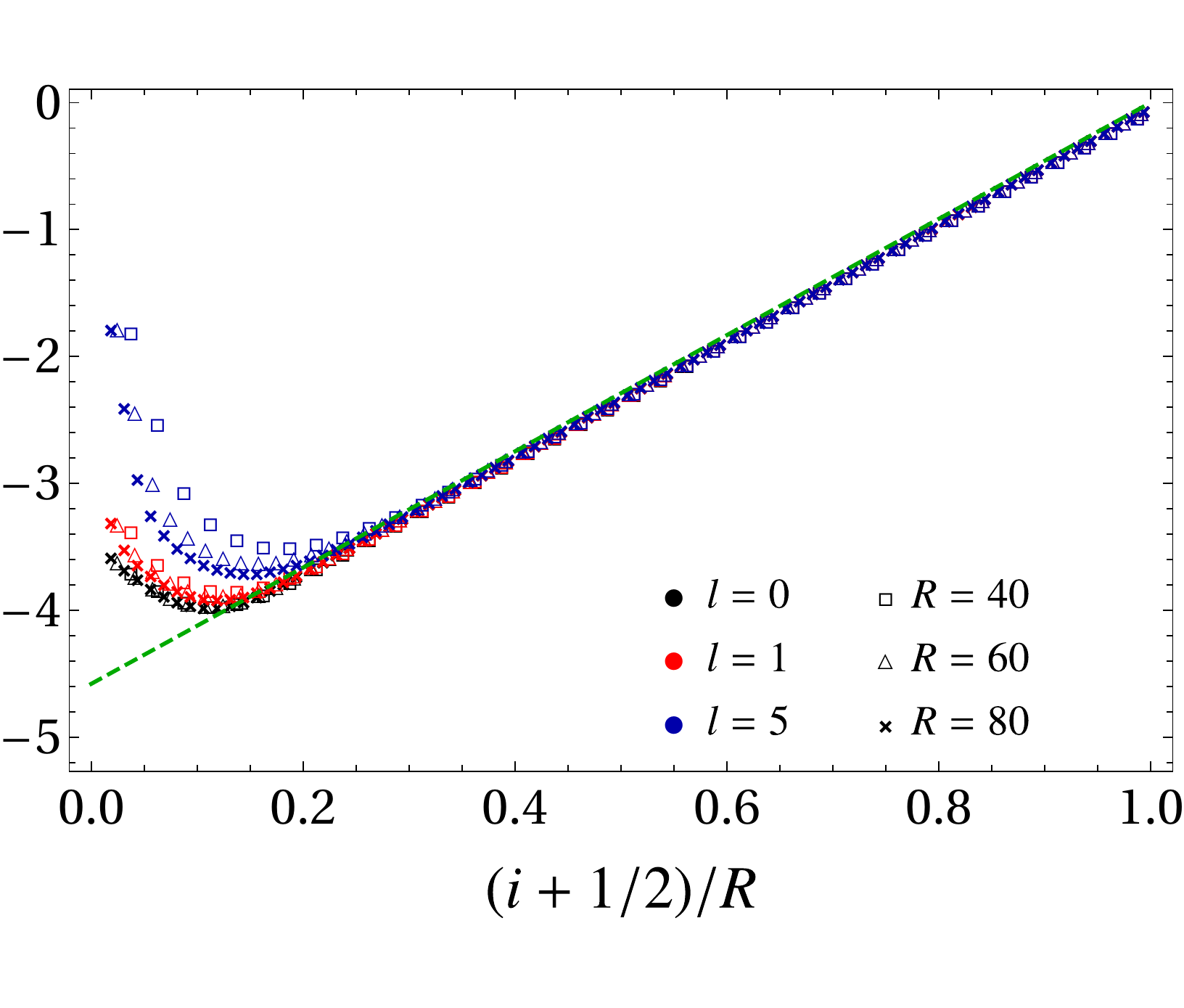}
\end{subfigure}
\hfill
\begin{subfigure}{.32\textwidth}
\hspace{1cm}
\includegraphics[scale=.35]{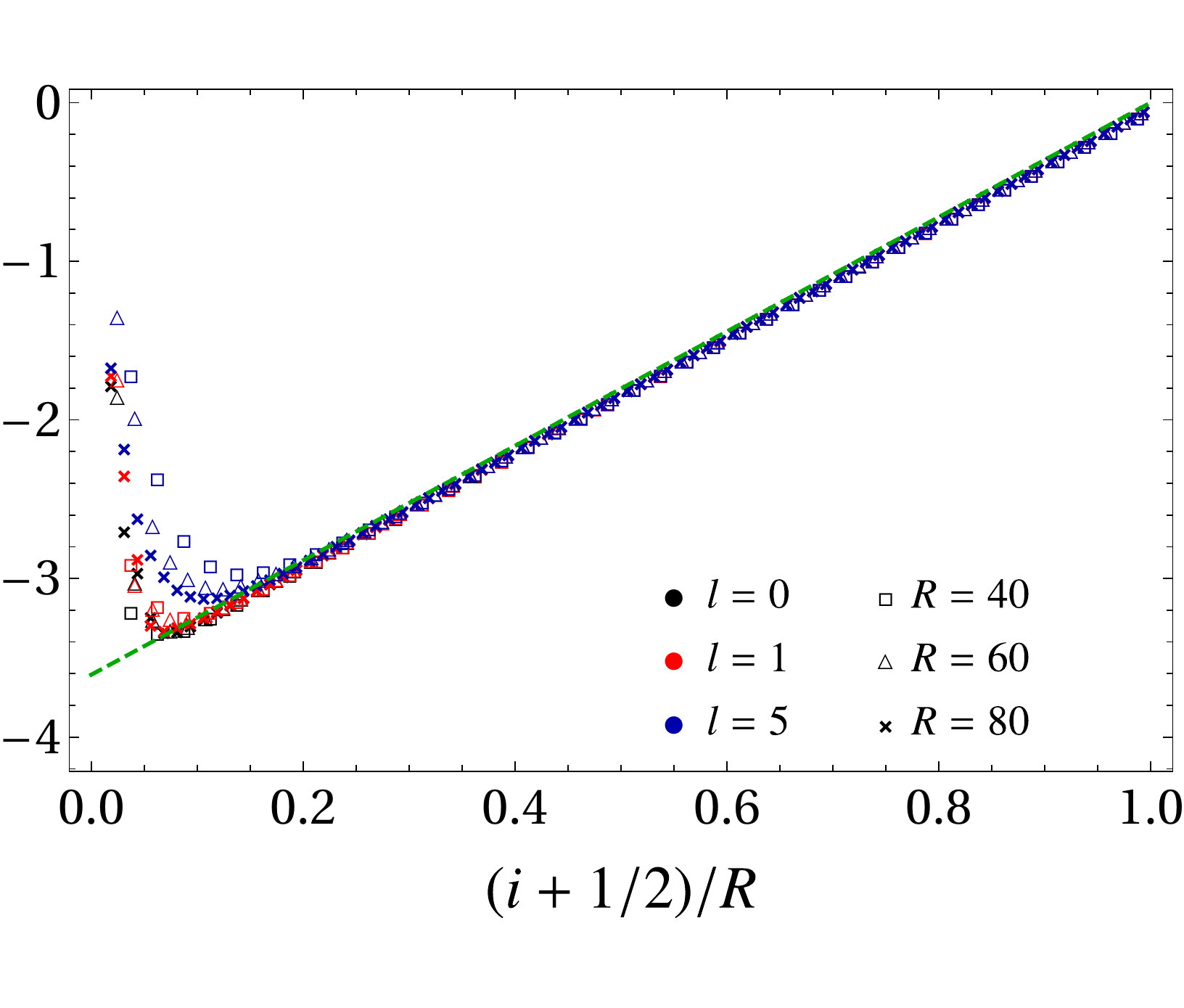}
\end{subfigure}
\hfill

\vspace{-.5cm}
\begin{subfigure}{.32\textwidth}
\hspace{-2cm}
\includegraphics[scale=.35]{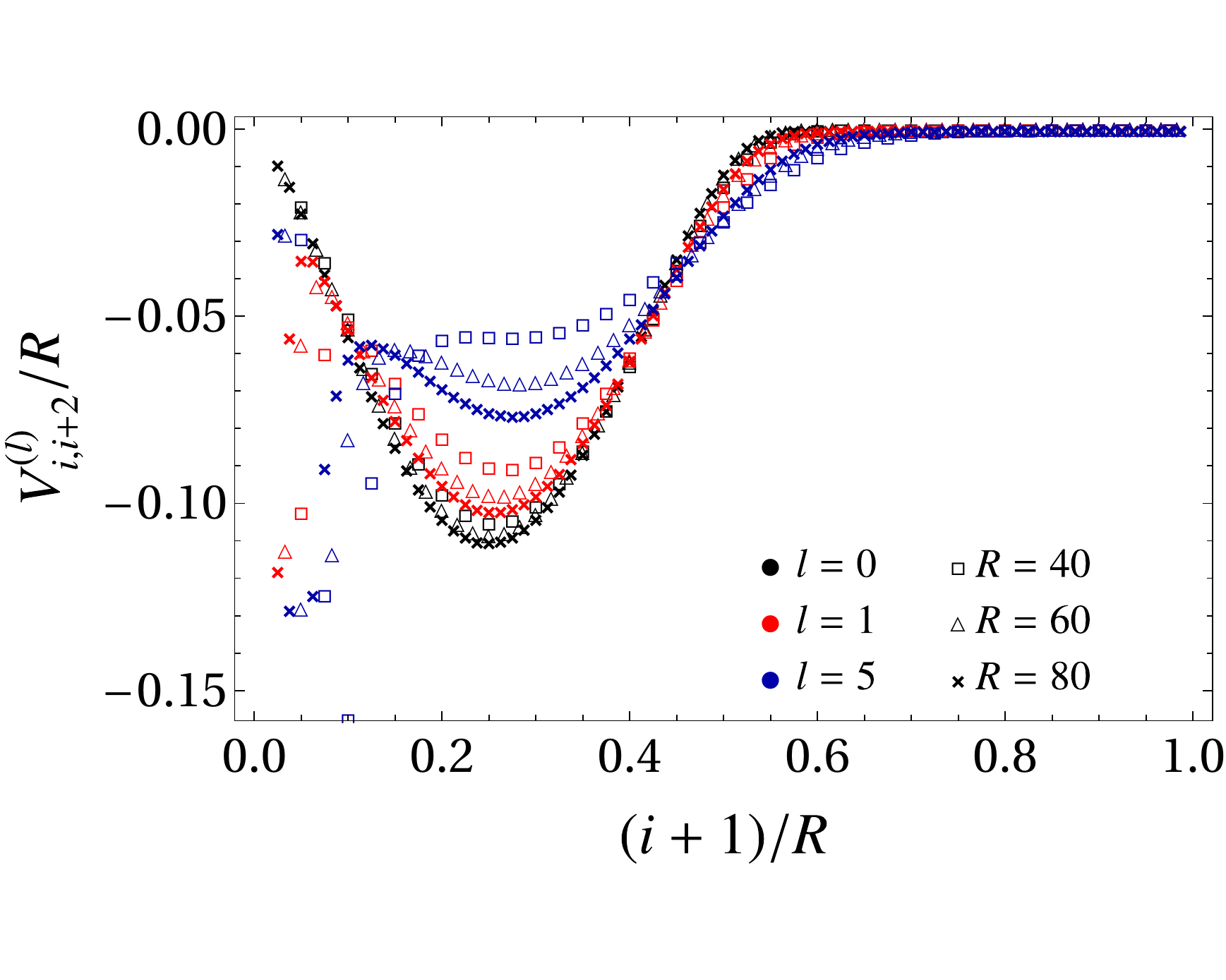}
\end{subfigure}
\hfill
\begin{subfigure}{.32\textwidth}
\hspace{-.9cm}
\includegraphics[scale=.35]{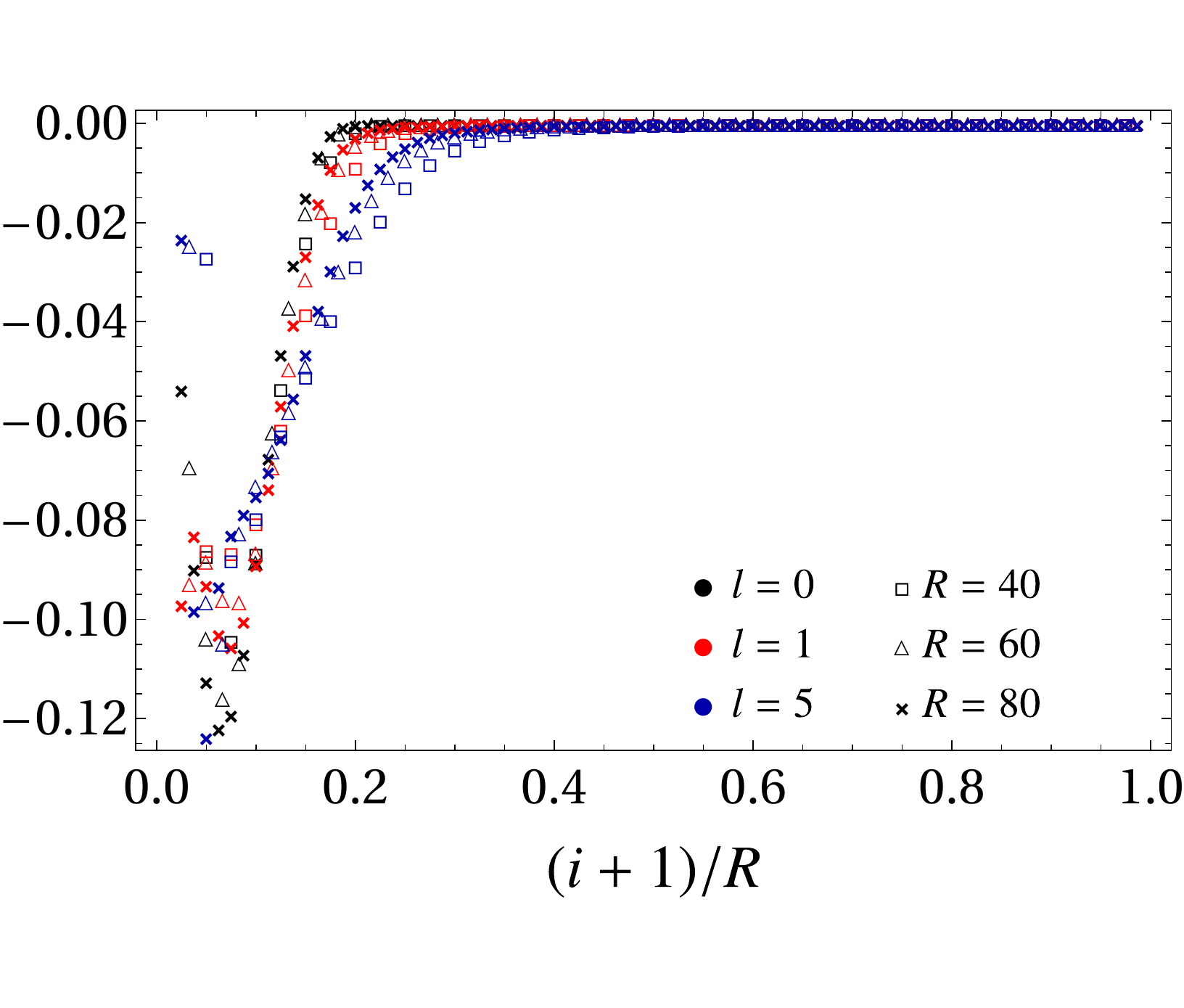}
\end{subfigure}
\hfill
\begin{subfigure}{.32\textwidth}
\hspace{1cm}
\includegraphics[scale=.35]{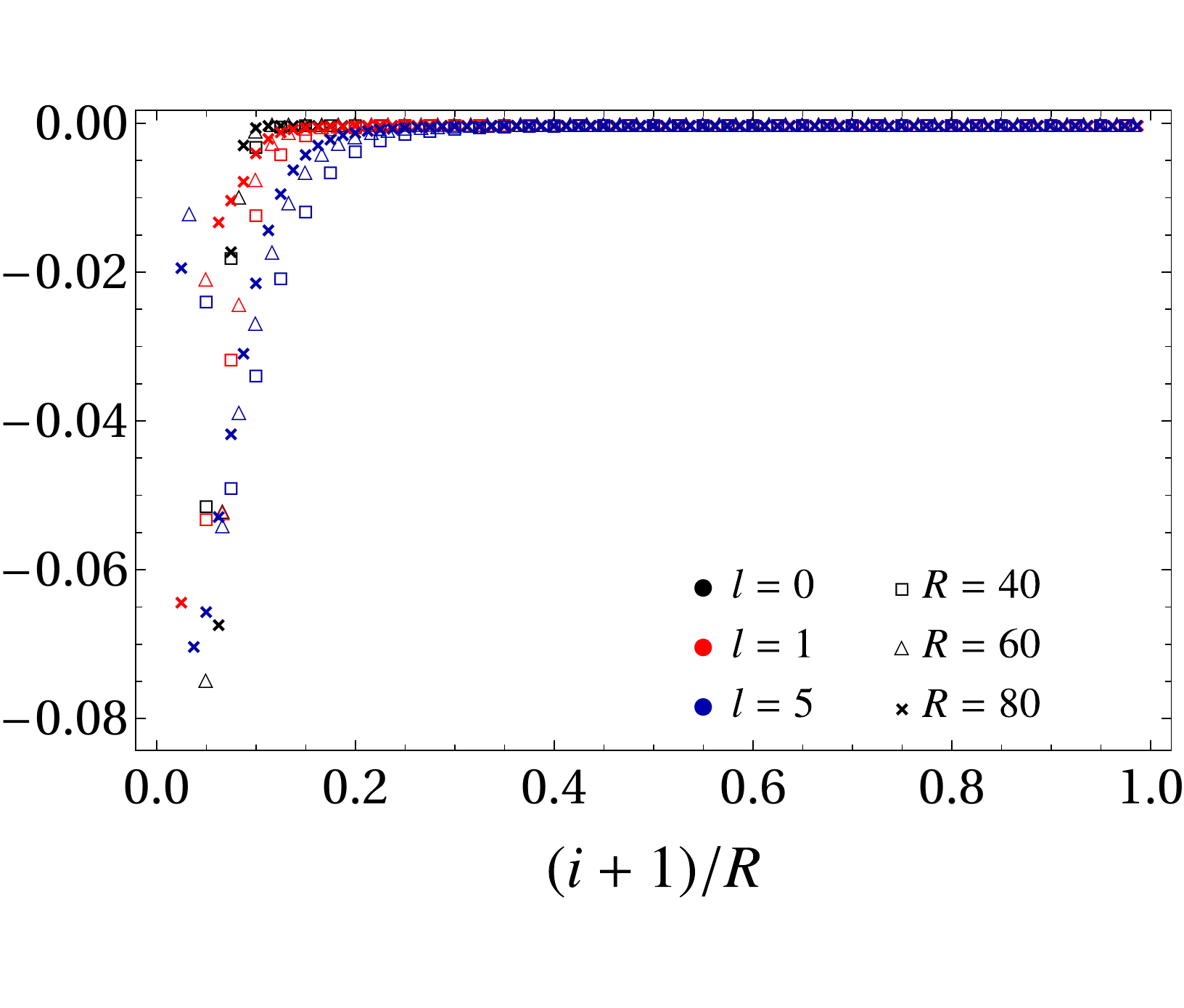}
\end{subfigure}
\vspace{-.5cm}
\caption{
\label{beta-V-triangle-diagonals}
Short-range couplings in the matrices
$V^{(l)}/R$
(see (\ref{H_T and H_V operators}) and (\ref{munuk def})) 
when $\omega > 0$ and $d=3$,
for $\omega = 1$ (left panels), $\omega = 3$ (middle panels) and $\omega = 5$ (right panels), 
for some $l$'s and different $R$.
The dashed green lines in the top and middle panels correspond to (\ref{tridiagonal-V}).\\
}
\end{figure}

In Fig.\,\ref{beta-T-triangle-diagonals} and Fig.\,\ref{beta-V-triangle-diagonals}
we show the numerical data points for the 
diagonals of the matrices $T^{(l)}$ and $V^{(l)}$ respectively,
evaluated from (\ref{eh-block-ch-version}) when $d=3$
and for $\omega=1$ (left panels), $\omega=3$ (central panels) or  $\omega=5$ (right panels).
It is instructive to compare these results against the corresponding ones obtained for $\omega = 0$
and displayed in Fig.\,\ref{fig-diagonals-T} and Fig.\,\ref{fig-diagonals-V}.
Also for $\omega > 0$, data collapses are observed for small values of $l$.
Presumably, 
larger systems are needed to observe these collapses also for higher values of $l$.

The considerations made in \cite{Eisler:2020lyn} 
for the entanglement Hamiltonian of a block of consecutive sites 
in the infinite and non-critical harmonic chain
can be adapted to this case in a straightforward way,
with the crucial difference that
only one point separates the subsystem from its complementary region
along the radial direction of the sphere $B$.
When $\omega > 0$ and $l$ are fixed,
for all the diagonals $T^{(l)}_{i,i+k}$ and $V^{(l)}_{i,i+k}$ with $k \geqslant 0$
except for $T^{(l)}_{i,i}$, $V^{(l)}_{i,i}$ and $V^{(l)}_{i,i+1}$
we can identify a region closed to the boundary of the sphere where the diagonal vanishes.
For a given $l$, the size of this region increases with $k$ at fixed $\omega$
and it also increases with $\omega$ at a given $k$.
Thus, in the large mass regime only the three diagonals 
$T^{(l)}_{i,i}$, $V^{(l)}_{i,i}$ and $V^{(l)}_{i,i+1}$ 
are non vanishing. 
%
Furthermore, 
when $\omega\gg l$ and for the small values of $l$ considered,
we find that the analytic results found in \cite{Eisler:2020lyn}
(which are based on \cite{Peschel-Truong91})
can be easily adapted to the case of the sphere.
In particular, the dominant matrix elements 
are well approximated by 
\be
\label{tridiagonal-T}
\frac{T^{(l)}_{i,i}}{R}
\,=\,
2 b(\kappa)\,\big(1- i /R \big)
\ee
and
\be
\label{tridiagonal-V}
\frac{V^{(l)}_{i,i}}{R}
\,=\,
\big(\omega^2+2\big)\,2b(\kappa)\,\big(1 - i / R \big)
\;\;\;\qquad\;\;\;
\frac{V^{(l)}_{i,i+1}}{R}
\,=\,
-\,2 b(\kappa)\,\big(1 - i / R \big)
\ee
where
\be
\label{b-kappa-def}
b(\kappa)
\,\equiv\,
2\,I(\tilde{\kappa})\,\sqrt{\kappa}
\ee
with $I(\kappa)$ being the complete elliptic integral of the first kind\footnote{The integral representation of the complete elliptic integral of the first kind is
\be
I(\kappa) = \int_0^{\frac{\pi}{2}}\,\frac{1}{\sqrt{1-\kappa^2 (\sin\theta)^2}}\; \textrm{d}\theta\,.
\ee
} 
and 
\be
\label{kappa-def}
\kappa
\,\equiv\,
\frac14\Big(\sqrt{\omega^2+4}-\omega\Big)^2
\;\;\;\qquad\;\;\;
\tilde{\kappa} \equiv \sqrt{1-\kappa^2}\,.
\ee

The analytic results in (\ref{tridiagonal-T}) and (\ref{tridiagonal-V}) 
correspond to the straight green dashed lines in 
Fig.\,\ref{beta-T-triangle-diagonals} and in Fig.\,\ref{beta-V-triangle-diagonals}
respectively. 
The massive scalar field in the continuum limit is described by taking 
$\omega \to 0$ and $a \to 0$ while $\omega / a$ is kept fixed.
When $\omega \to 0$, we have that $\kappa \to 1$ and $\tilde{\kappa} \to 0$;
hence $I(\tilde{\kappa}) \to \pi / 2$ and therefore $2b(\kappa) \to 2\pi$,
which is the value predicted by Bisognano and Wichmann \cite{Bisognano:1975ih, Bisognano:1976za},
as already remarked in \cite{Eisler:2020lyn}.
Notice that the analytic expressions in (\ref{tridiagonal-T}) and (\ref{tridiagonal-V}) 
are independent both of $d$ and of $l$.
Our numerical results for $d=2$ and $d=3$ confirm this observation
(in this section we show only the data corresponding to $d=3$).
In Fig.\,\ref{beta-T-triangle-diagonals} and in Fig.\,\ref{beta-V-triangle-diagonals}
one observes a dependence on the angular momentum parameter $l$
only in the central region of the sphere, 
where the numerical data do not follow
the straight lines given by  (\ref{tridiagonal-T}) and (\ref{tridiagonal-V}).
%

\begin{figure}[t!]
\vspace{-.5cm}
\begin{subfigure}{.4\textwidth}
\hspace{-1.4cm}
\vspace{.4cm}
\includegraphics[scale=.3]{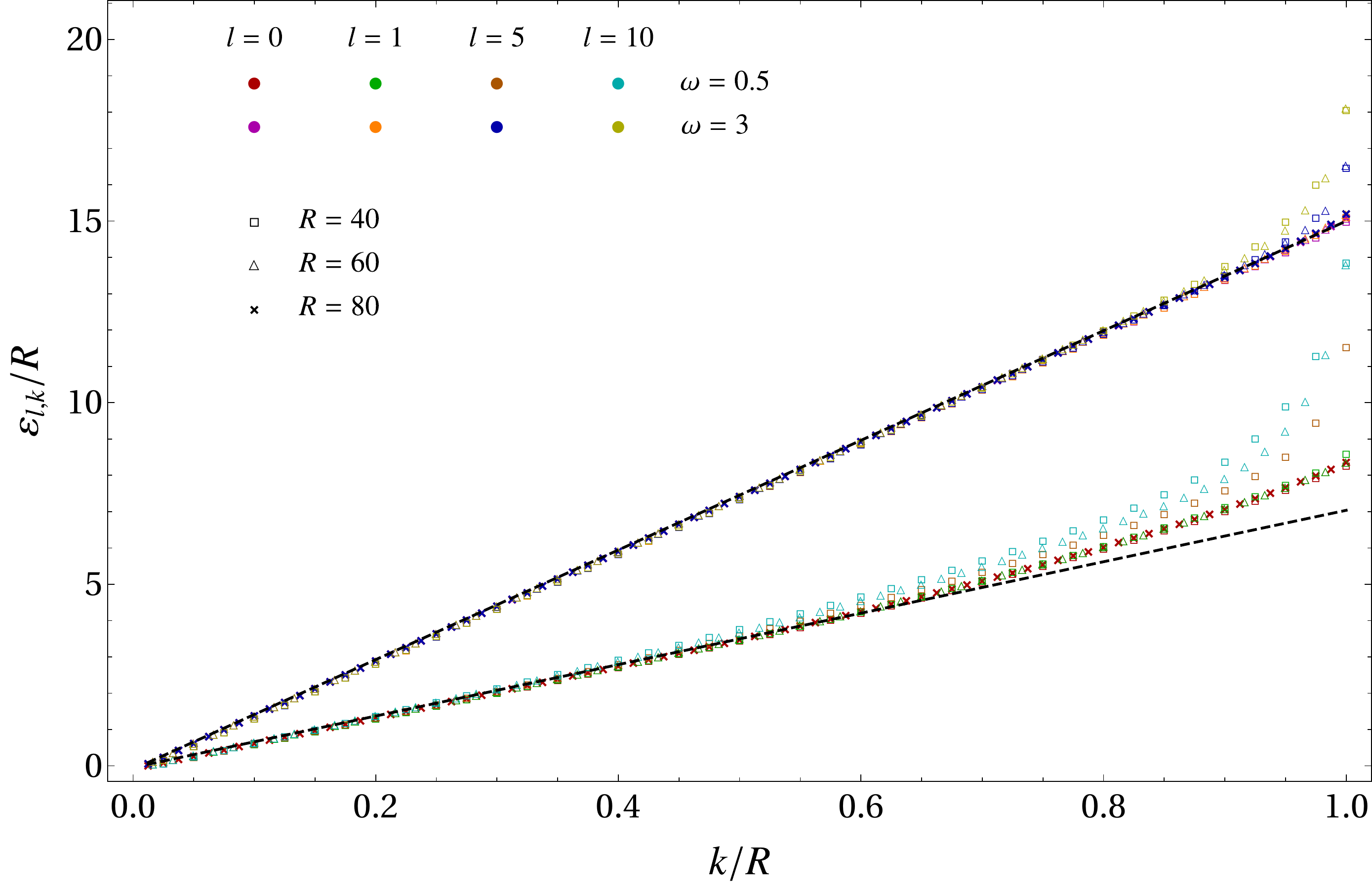}
\end{subfigure}\hfill
\begin{subfigure}{.4\textwidth}
\hspace{-1.4cm}
\vspace{.4cm}
\includegraphics[scale=.3]{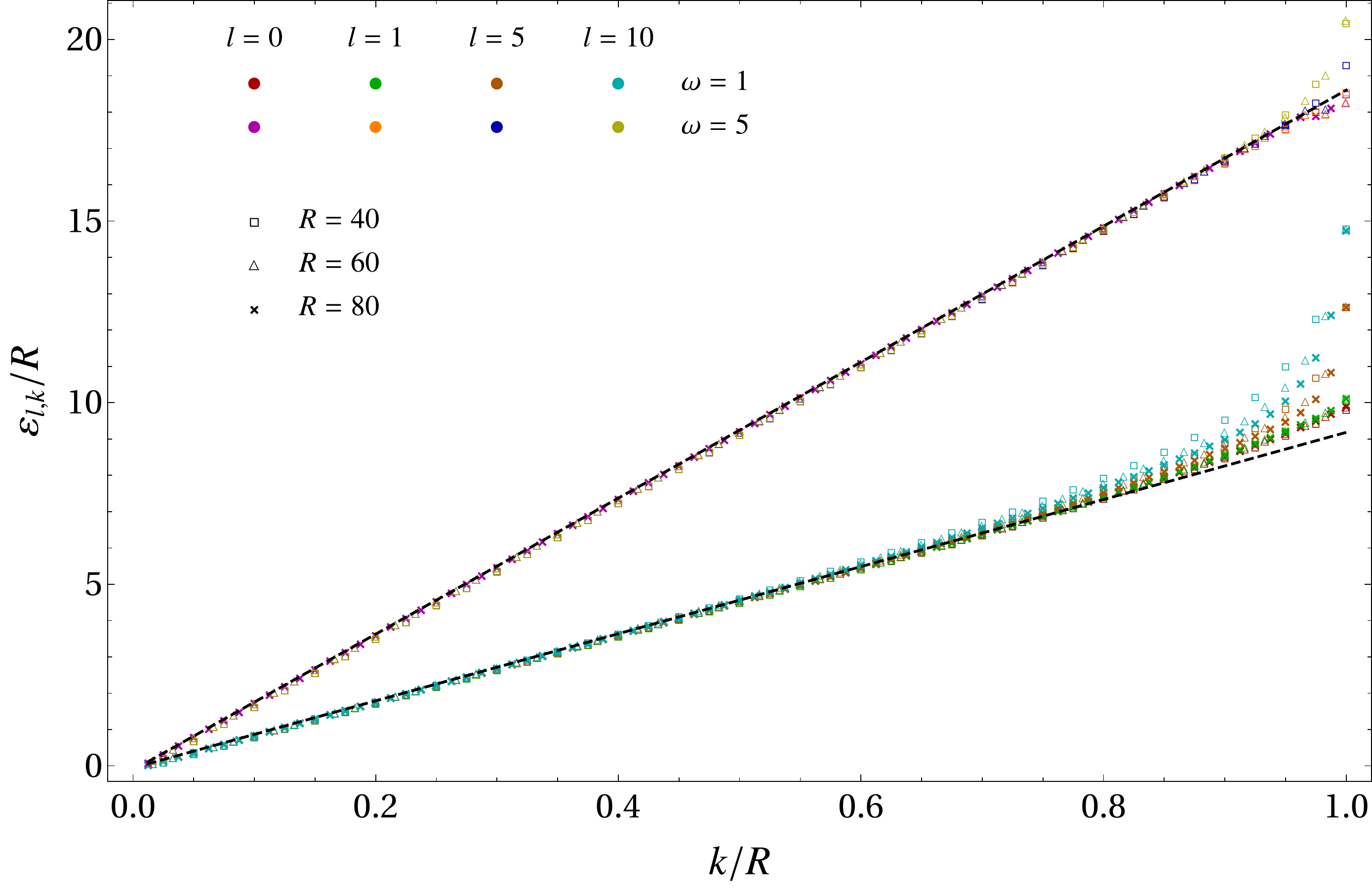}
\end{subfigure}
\vspace{-.3cm}
\caption{
Single particle entanglement energies at fixed $l$ for $d=3$
and $\omega \in \{0.5\,, 1\,, 3\,, 5\}$.
The dashed lines correspond to (\ref{epsk}).
}
\label{fig:epss}
\end{figure}


The symplectic spectrum of the reduced covariance matrix $Q_{l,B} \oplus P_{l,B}$
at a given $l$ provides the corresponding 
single particle entanglement energies $\varepsilon_{l,k}$ at fixed $l$
as discussed in Sec.\,\ref{subsec-eh}.
In Fig.\,\ref{fig:epss} we show numerical data for $\varepsilon_{l,k}$ (in the case of $d=3$)
when $l$ is small for some values of $\omega$.
By comparing these results with the corresponding ones 
for $\omega =0$ displayed in Fig.\,\ref{fig:eps0},
we observe that $\varepsilon_{l,k}$
as function of $k/R$ are well described by a straight line as $\omega$ increases. 
This feature has been already highlighted for $d=1$ in the case of the 
block of consecutive sites in the infinite harmonic chain on the line \cite{Eisler:2020lyn}.
Furthermore, the slope of this straight line is given by 
the same expression found in \cite{Eisler:2020lyn} for the $d=1$ case, 
namely \cite{Peschel_1999} 
\be
\label{epsk}
\varepsilon_{l,k} = (2k-1)\, \varepsilon
\;\;\;\qquad \;\;\;
\varepsilon \equiv \pi \, \frac{I(\tilde{\kappa})}{I(\kappa)}
\ee
where $\kappa$ and $\tilde{\kappa}$ have been defined in (\ref{kappa-def}).

\begin{figure}[t!]
\begin{subfigure}[b]{.4\textwidth}
\hspace{-1.4cm}
\includegraphics[scale=.35]{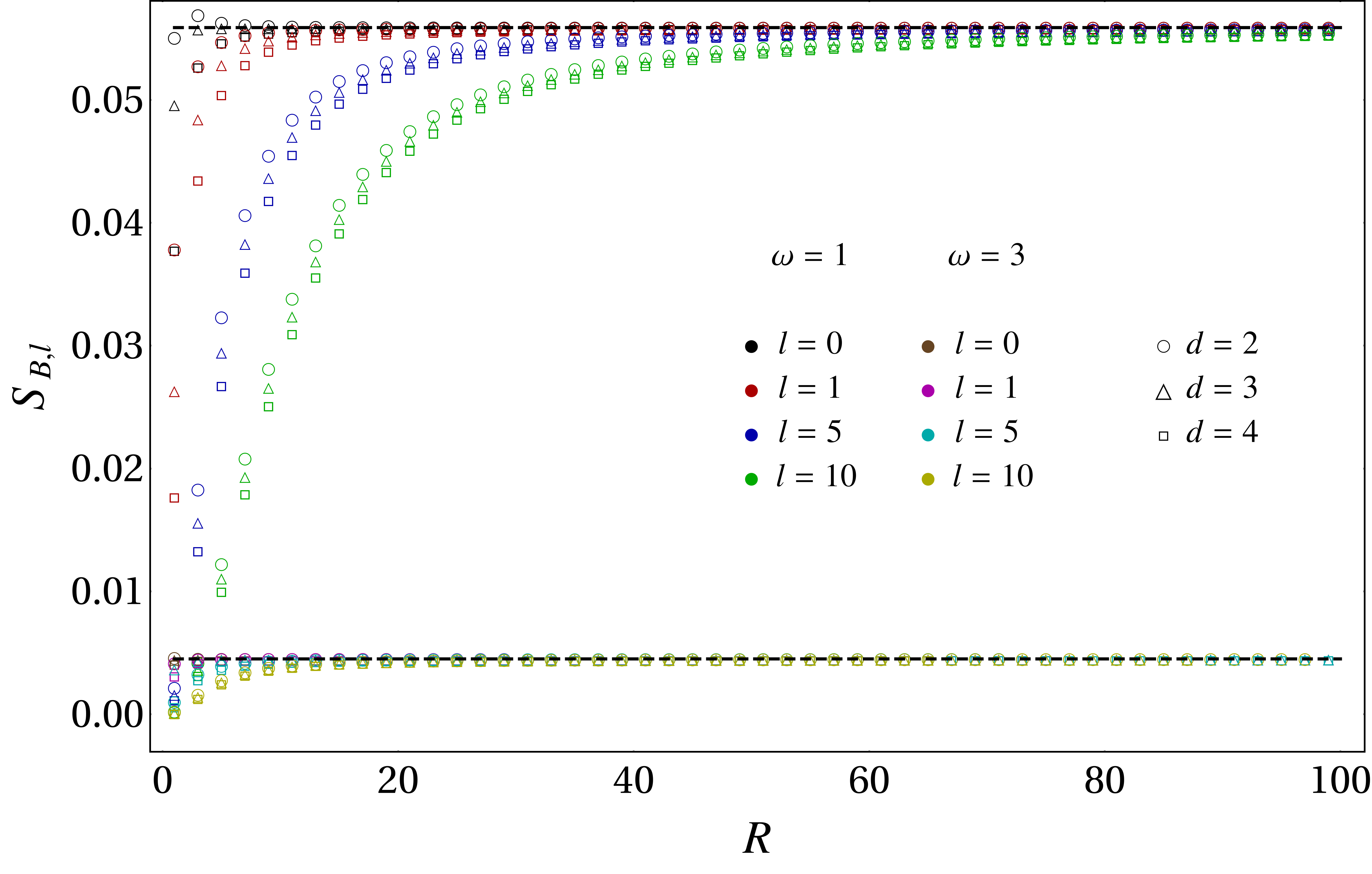}
\end{subfigure}
\hfill
\begin{subfigure}[b]{.4\textwidth}
\hspace{-1.6cm}
\includegraphics[scale=.35]{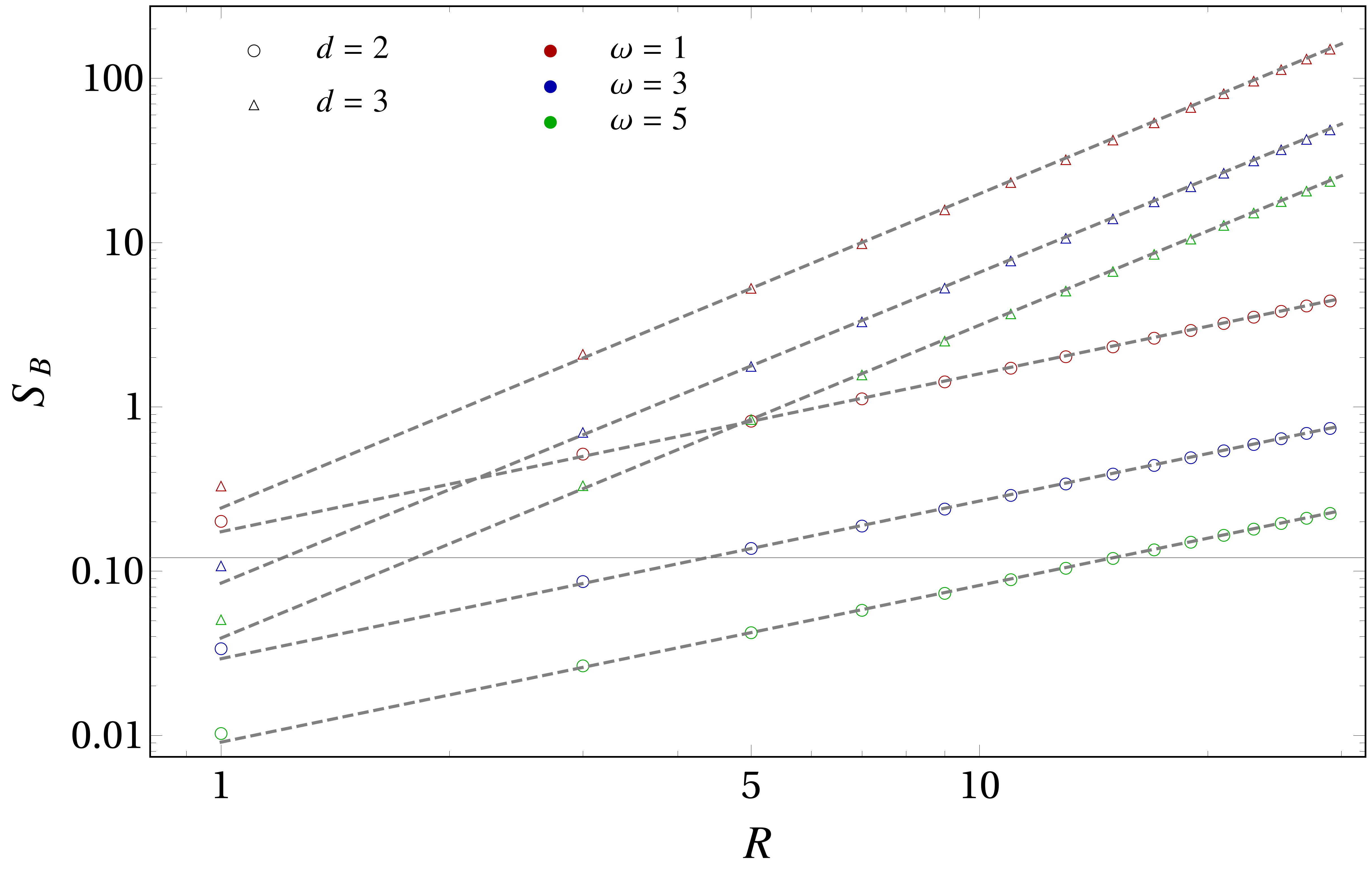}
\end{subfigure}
\hfill
\caption{
The entanglement entropy $S_B$ (right panel) 
and $S_{B,l}$ (left panel) defined in (\ref{EE-mode-sum}) when $\omega > 0$.
The horizontal dashed lines in the left panel correspond to (\ref{S0massive}).
}
\label{fig:EE-various-w1}
\end{figure}

Numerical results for the entanglement entropy $S_B$ of the sphere when $\omega > 0$
are shown in the right panel of Fig.\,\ref{fig:EE-various-w1}
(see the right panel of Fig.\,\ref{fig:EE-various} for $S_B$ when $\omega = 0$).
They have been obtained through (\ref{EE-mode-sum}),
with the sum over $l$ restricted 
to $l \leqslant 500$ for $\omega=1$ and $\omega=3$,
and to $l \leqslant 1000$ for $\omega=5$. 
The dashed lines correspond to $ \alpha(d,\omega)\,R^{d-1}$ 
for some fitted values of the constants $\alpha(d,\omega)$;
hence these data just highlight the expected area law behaviour of the entanglement entropy.

Similarly to the massless case,
also when $\omega > 0$ the dependence on $d$ and $l$ is more visible 
in the quantity $S_{B,l}$ defined in (\ref{EE-mode-sum}). 
In the left panel of Fig.\,\ref{fig:EE-various-w1} we show $S_{B,l}$ in terms of $R$
for $\omega=1$ and $\omega=3$,
finding a qualitatively different behaviour 
with respect to the massless case (see the left panel of Fig.\,\ref{fig:EE-various}).
In particular, the horizontal dashed line 
providing the asymptotic value of $S_{B,l}$
in the left panel of Fig.\,\ref{fig:EE-various-w1}
is given by \cite{EislerPeschel:2009review, Eisler:2020lyn}
\be
\label{S0massive}
S_0 \equiv -\frac{1}{24}
\left[\,
\log\!\left(\frac{16\,\tilde{\kappa}^4}{\kappa^2}\right)
-(1+\kappa^2)\,\frac{4\,I(\kappa)I(\tilde{\kappa})}{\pi} 
\,\right] .
\ee

We remark that the above numerical analysis 
does not correspond to the continuum limit of the entanglement Hamiltonian 
for the massive scalar field, 
where $\omega \to 0$ and $R \to \infty$ while $\omega R$ is kept fixed.
In this regime we have encountered the same difficulties discussed in \cite{Eisler:2020lyn}
for the entanglement Hamiltonians of a block of consecutive sites 
in non-critical free chains on the infinite line.

\section{Conclusions}
\label{sec_conclusions}

We have explored the continuum limit 
of the entanglement Hamiltonian of a sphere
in the $d+1$ dimensional Minkowski spacetime for a massless scalar field.
We have employed a numerical analysis based 
on the radial lattice discretisation introduced in \cite{Srednicki:1993im},
on the results found in \cite{Srednicki:1993im, Casini:2009sr} 
and on the procedure already introduced
\cite{Arias:2016nip, Eisler:2017cqi, Eisler:2019rnr, DiGiulio:2019cxv, Eisler:2018ugn} 
to study some entanglement Hamiltonians of a block of consecutive sites in 
one-dimensional free chains.
Our main results for the massless scalar field are the conjectures 
given by (\ref{conj-infty-beta}), (\ref{V-infty-beta-new-term}) and (\ref{conj-infty-zero}),
which are supported by the numerical data 
reported  in Fig.\,\ref{beta-TV-parabola}, 
Fig.\,\ref{beta-V0-horizontal-lines-mode} and Fig.\,\ref{beta-Vd1-zero}
in the cases of $d=2$ and $d=3$ and for small values of the total angular 
momentum parameter $l$.
By employing these conjectured results into (\ref{MplusNoperator expansion bdy}),
we have obtained the expression (\ref{EH-ball-massless-scalar}) in the continuum,
which can be derived also 
from  the entanglement Hamiltonian of the sphere for a generic CFT in its ground state 
\cite{Hislop:1981uh,Casini:2011kv} (see (\ref{EH-intro}))
specialised to the massless scalar field.
An explicit dependence on the dimensionality parameter $d$
and on the  total angular momentum parameter $l$ 
occurs in the term (\ref{V-infty-beta-new-term}),
which vanishes identically when $d=1$.
In the special case of $d=1$, 
we recover the results of  \cite{DiGiulio:2019cxv} 
for the entanglement Hamiltonian of an interval at the beginning of the semi-infinite line 
with Dirichlet boundary conditions at the origin.

As for the massive regime, 
we have discussed numerical results obtained for a given $\omega > 0$
(see Sec.\,\ref{sec_massive}).
At a generic value of $\omega \geqslant 0$, 
the matrix  (\ref{eh-block-ch-version}) characterising the quadratic entanglement Hamiltonian
contains long-range and inhomogeneous couplings 
(see Fig.\,\ref{beta-T-triangle-diagonals} and Fig.\,\ref{beta-V-triangle-diagonals}).
However, in the limit of $\omega \gg l$, 
only the nearest neighbour couplings are non vanishing 
and the corresponding weight function along the radial direction 
is well approximated by straight lines 
whose slopes are independent of $d$ and $l$
and can be determined analytically 
(see (\ref{tridiagonal-T}) and (\ref{tridiagonal-V})),
as already observed in \cite{Eisler:2020lyn}
also for the entanglement Hamiltonian of a block of consecutive sites
in the harmonic chain on the infinite line and in its ground state. 

Our analysis can be improved in various directions. 
All the numerical checks performed in this manuscript correspond to small values of 
the total angular momentum parameter $l$;
hence an improved numerical analysis is required 
to explore also the sectors corresponding to higher values of $l$.
For the massless scalar,  the existence of the functions 
$\tau_{l,\eta k}$ and $\nu_{l,\eta k}$ introduced in (\ref{munuk def})
is a crucial assumption throughout the derivation of (\ref{MplusNoperator expansion bdy}).
It would be interesting to obtain numerical data that support further these conjectures 
and also to find analytic expressions for these functions,
as done in \cite{Eisler:2017cqi} for the entanglement Hamiltonian 
of a block of consecutive sites in the chain of free fermions on the line. 
For the massive scalar, 
it is important to explore through these numerical methods the regime
characterised by a given value of $\omega R$ when $R \to \infty$,
where the entanglement Hamiltonian of the sphere in the continuum 
is fully non-local \cite{Arias:2016nip,Longo:2020amm}.

It would be interesting to employ the procedure discussed  in 
\cite{Arias:2016nip, Eisler:2017cqi, Eisler:2018ugn, Eisler:2019rnr, DiGiulio:2019cxv, Eisler:2020lyn}
and throughout this manuscript 
to study  the continuum limit of entanglement Hamiltonians where either multi-local 
\cite{Casini:2009vk, Longo:2009mn, Hollands:2019hje, Blanco:2019xwi, Fries:2019ozf, Mintchev:2020uom, Mintchev:2020jhc}
or fully non-local terms \cite{Arias:2018tmw, Longo:2020amm} occur,
in systems characterised by spatially inhomogeneity \cite{Tonni:2017jom}
and in systems driven out of equilibrium \cite{Cardy:2016fqc, Wen:2018svb, DiGiulio:2019lpb}.
It would be insightful also 
to identify possible boundary terms in the entanglement Hamiltonians
\cite{Herzog:2014fra,Lee:2014zaa,Herzog:2016bhv} through numerical analyses on the lattice.

It is interesting also to develop methods to write lattice operators 
including only nearest neighbour couplings
that approximate the entanglement Hamiltonians
\cite{Peschel_2004, Nienhuis_2009,Eisler:2017cqi,Eisler:2019rnr,Tonni:2017jom,Kim_2016,Dalmonte_2018,Zhang_2020}
and to understand their relation with the corresponding entanglement Hamiltonian,
as done e.g. in \cite{Peschel_2004,Eisler:2017cqi}.
A numerical approach to some entanglement Hamiltonains based on a quantum Monte Carlo method 
has been proposed in \cite{Parisen_Toldin_2018}.

In a generic number of spatial dimensions,
we find it worth exploring further 
also the relations between the entanglement Hamiltonians
and other insightful entanglement quantifiers
like the entanglement spectrum
\cite{Lauchli:2013jga, Torlai_2014, Cardy:2016fqc, Kim_2016, Tonni:2017jom, Alba:2017bgn, DiGiulio:2019lpb, Surace:2019mft, Roy:2020frd, Robertson:2021wpp},
the contour functions for the entanglement entropies
\cite{Chen_2014, Frerot2015, Coser:2017dtb}
and the logarithmic negativity 
\cite{Peres_1996, Vidal_2002, Plenio_2005, Calabrese:2012nk, 
Calabrese:2012ew, Calabrese:2013mi, Calabrese:2014yza, DeNobili:2015dla, Eisler_2016, DeNobili:2016nmj} 
or other related quantities like the circuit complexity of mixed states 
\cite{Caceres:2019pgf,DiGiulio:2020hlz}.

\vskip 20pt 
\centerline{\bf Acknowledgments} 
\vskip 5pt 

We are grateful to Filiberto Ares, Giuseppe Di Giulio, Viktor Eisler, 
Giuseppe Mussardo, Ingo Peschel, Diego Pontello, Benjamin Walter
and in particular to Marina Huerta and Mihail Mintchev 
for helpful discussions or correspondence. 
ET’s research has been conducted within the framework of the 
Trieste Institute for Theoretical Quantum Technologies (TQT).

\vskip 30pt


\appendix

\section{Correlators in the continuum}
\label{app_corr_check}

In this appendix we report the two-point functions in the continuum 
computed through quantum field theory methods \cite{Frolov:1999an,Saharian:2000mw}
and compare them with the correlators on the lattice (\ref{corr-matrices-mode})
obtained for different discretizations of the Hamiltonian of the scalar field.

In the spherical coordinates introduced in Sec.\,\ref{subsec-ham},
the two-point function of the free massive scalar field $\Phi(t,r,\Omega)$ 
in the $d+1$ dimensional Minkowski spacetime reads \cite{Saharian:2000mw}
\be
\label{wightman-d-dim}
\langle\, \Phi(t_1, r_1,\Omega_1)\, \Phi(t_2,r_2,\Omega_2)\,\rangle
\,=\,
\frac{1}{(r_1 r_2)^{\frac{d-1}{2}}}
\sum_{l = 0}^{\infty}
\frac{2l + d-2}{(d-2) \,\mathcal{S}_d}\; 
C_l^{\frac{d-2}{2}}\!(\cos \xi)
\; \mathcal{G}_l(t_1, r_1 ; t_2, r_2)
\ee
where $\mathcal{S}_d \equiv \tfrac{2\pi^{d/2}}{\Gamma(d/2)}$ is
the area of the $d-1$ dimensional unit sphere and
\be
\label{G_l-def}
\mathcal{G}_l(t_1, r_1 ; t_2, r_2)
\,\equiv\,
\frac{\sqrt{r_1 r_2}}{2}
\int_0^{\infty}\!\!
\frac{\gamma}{ 
\sqrt{\gamma^2 + m^2} }\;
J_{\tilde{\mu}}(\gamma r_1)\, J_{\tilde{\mu}}(\gamma r_2)
\; e^{\textrm{i} \sqrt{\gamma^2 + m^2} \,(t_2 - t_1)}
\,\textrm{d}\gamma
\ee
with
\be
\label{mu-tilde-def}
\tilde{\mu} \equiv \sqrt{\frac{1}{4} + \mu_d(l)} 
\,=\,
l +\frac{d - 2}{2}\,.
\ee
In (\ref{wightman-d-dim})
the angle between directions identified by $\Omega_1$ and $\Omega_2$ is denoted by $\xi$ 
and  $C_p^q(x)$ is the Gegenbauer polynomial  of degree $p$ and order $q$,
which can be expressed also through the hypergeometric function as follows 
(see e.g. section\;15.4 in \cite{Abramowitz:1964:HMF})
\be
C_p^q(x) = \frac{(2q)_p}{p!}\;
_2F_1\bigg(\!\! -\!p,2q+p\,;q+\frac{1}{2}\,;\frac{1-x}{2}\bigg)
\ee
where $(a)_b$ is the Pochhammer symbol.

It is instructive to consider the special case of $d=2$,
which is the lowest value of $d$ where the sum over $l$ 
in (\ref{wightman-d-dim}) is non trivial. 
Taking the limit $d\to 2$ in the infinite sum
obtained by isolating the $l=0$ term in (\ref{wightman-d-dim})
and using that $\tilde{\mu} = l$ when $d=2$ (from (\ref{mu-tilde-def})),  
one finds
\be
\label{d2-green}
\langle \Phi(t_1, r_1,\phi_1)\, \Phi(t_2,r_2,\phi_2) \rangle
=
\frac{1}{2\pi \sqrt{r_1 r_2}} \;
\bigg\{ \mathcal{G}_0(t_1, r_1 ; t_2, r_2)
+
2\sum_{l \geqslant 1}
\cos\! \big(l [ \phi_2-\phi_1] \big)\,\mathcal{G}_l(t_1, r_1 ; t_2, r_2)\bigg\}\,.
\ee

For a generic value of $d$,
when $t_1 = t_2$ we have that (\ref{G_l-def}) simplifies to 
\be
\label{phi-phi-massive}
\mathcal{Q}_l(r_1, r_2)
\,=\,
\frac{\sqrt{r_1 r_2} }{2}
\int_0^{\infty}\!
\frac{\gamma}{ 
\sqrt{\gamma^2 + m^2} }\;
J_{\tilde{\mu}}(\gamma r_1)\, J_{\tilde{\mu}}(\gamma r_2)
\,\textrm{d}\gamma\,.
\ee
When $m=0$ and for $r_1 < r_2$
(which leads to introduce $\rho_{12} \equiv r_1 / r_2 <1$),
this integral reads
\be
\label{phi-phi}
\mathcal{Q}_l(r_1, r_2)
\,=\,
\frac{\Gamma(\tfrac{1}{2} + \tilde{\mu})}{2\, \Gamma(\tfrac{1}{2}) \; \Gamma(1+ \tilde{\mu})}\;
\rho_{12}^{\tilde{\mu} + 1/2}
\, _2F_1\! \left( \frac{1}{2}\,, \frac{1}{2} + \tilde{\mu}\,; 1+\tilde{\mu} \,;  \rho_{12}^2 \right)
\ee
which holds for $d\geqslant 2$;
indeed, for $d=1$ and $l=0$ we have $\tilde{\mu} = -1/2$,
where this expression is not well defined.
From (\ref{G_l-def})
we can compute also $\partial_{t_1} \partial_{t_2} \mathcal{G}_l(t_1, r_1 ; t_2, r_2)$ for $t_1 = t_2$
and the result is
\be
\label{pi-pi-massive}
\mathcal{P}_l(r_1, r_2)
\,=\,
\frac{\sqrt{r_1 r_2} }{2}
\int_0^{\infty}\!
\gamma\, \sqrt{\gamma^2 + m^2}
\; J_{\tilde{\mu}}(\gamma r_1)\, J_{\tilde{\mu}}(\gamma r_2)
\,\textrm{d}\gamma\,.
\ee
In the massless regime, this becomes
\bea
\label{pi-pi}
& &
\hspace{-1cm}
\mathcal{P}_l(r_1, r_2)
\,=\,
\frac{\Gamma(\frac32+\tilde{\mu})\,\rho_{12}^{\tilde{\mu}+\frac12}}{
\Gamma(\frac12)\,\Gamma(1+\tilde{\mu})\,(r_2^2-r_1^2)} \;
\Bigg\{ 2\big[\,\tilde{\mu}-\rho_{12}^2(1+\tilde{\mu})\,\big] \; 
_2F_1\!\left(\frac{1}{2},\frac{3}{2}+\tilde{\mu}\, ;1+\tilde{\mu}\, ;\rho_{12}^2\right)
\hspace{.5cm}
\\
& & \hspace{6.7cm}
-\,(1+2\tilde{\mu})\;_2F_1\!\left(-\frac{1}{2},\frac{3}{2}+\tilde{\mu}\, ;1+\tilde{\mu}\, ; \rho_{12}^2\right) \Bigg\}\,.
\nonumber
\eea

\begin{figure}[t!]
\begin{subfigure}{.45\textwidth}
\hspace{-1.3cm}
\includegraphics[scale=.35]{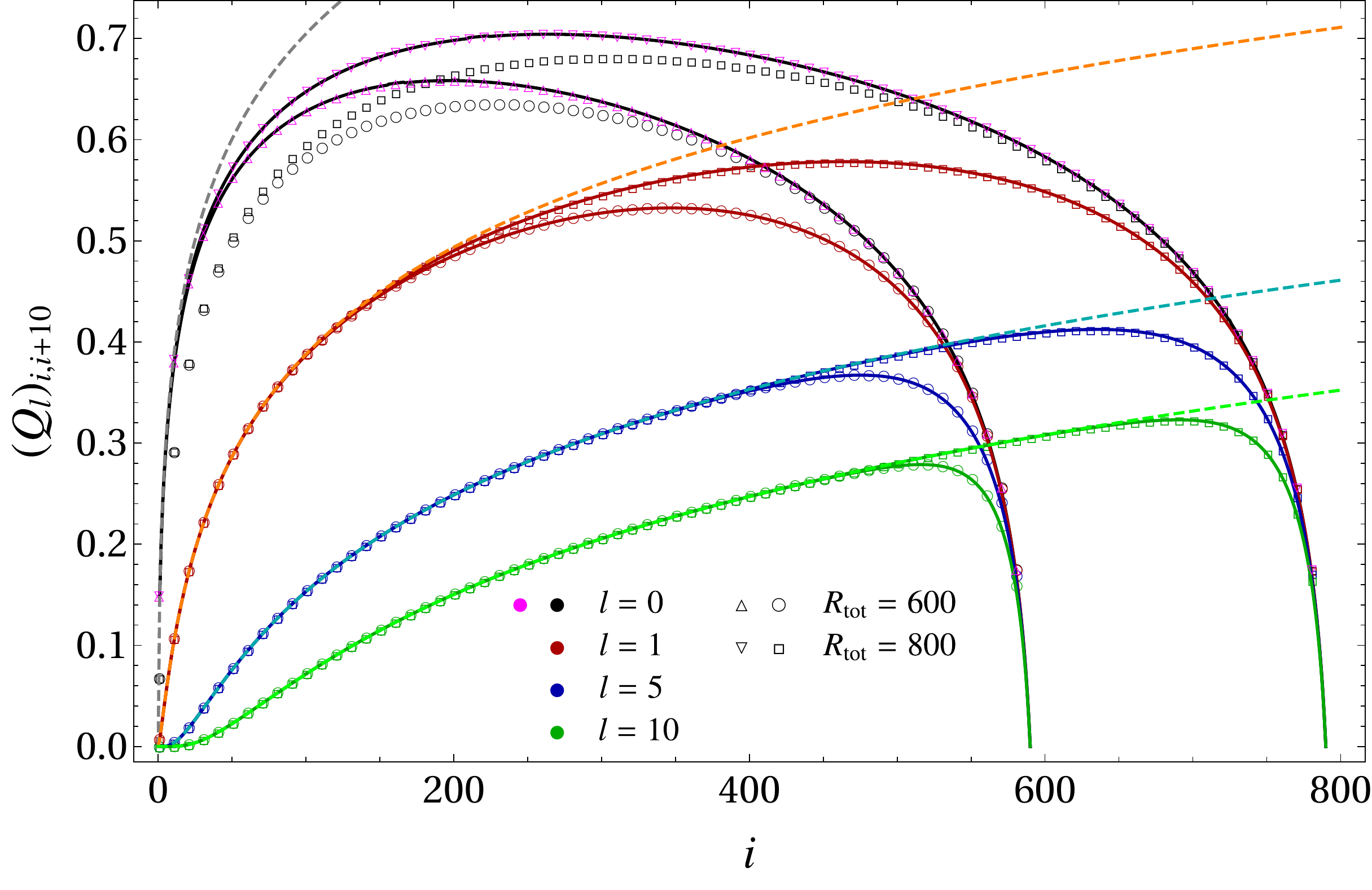}
\end{subfigure}
\hfill
\begin{subfigure}{.45\textwidth}
\hspace{-.7cm}
\includegraphics[scale=.35]{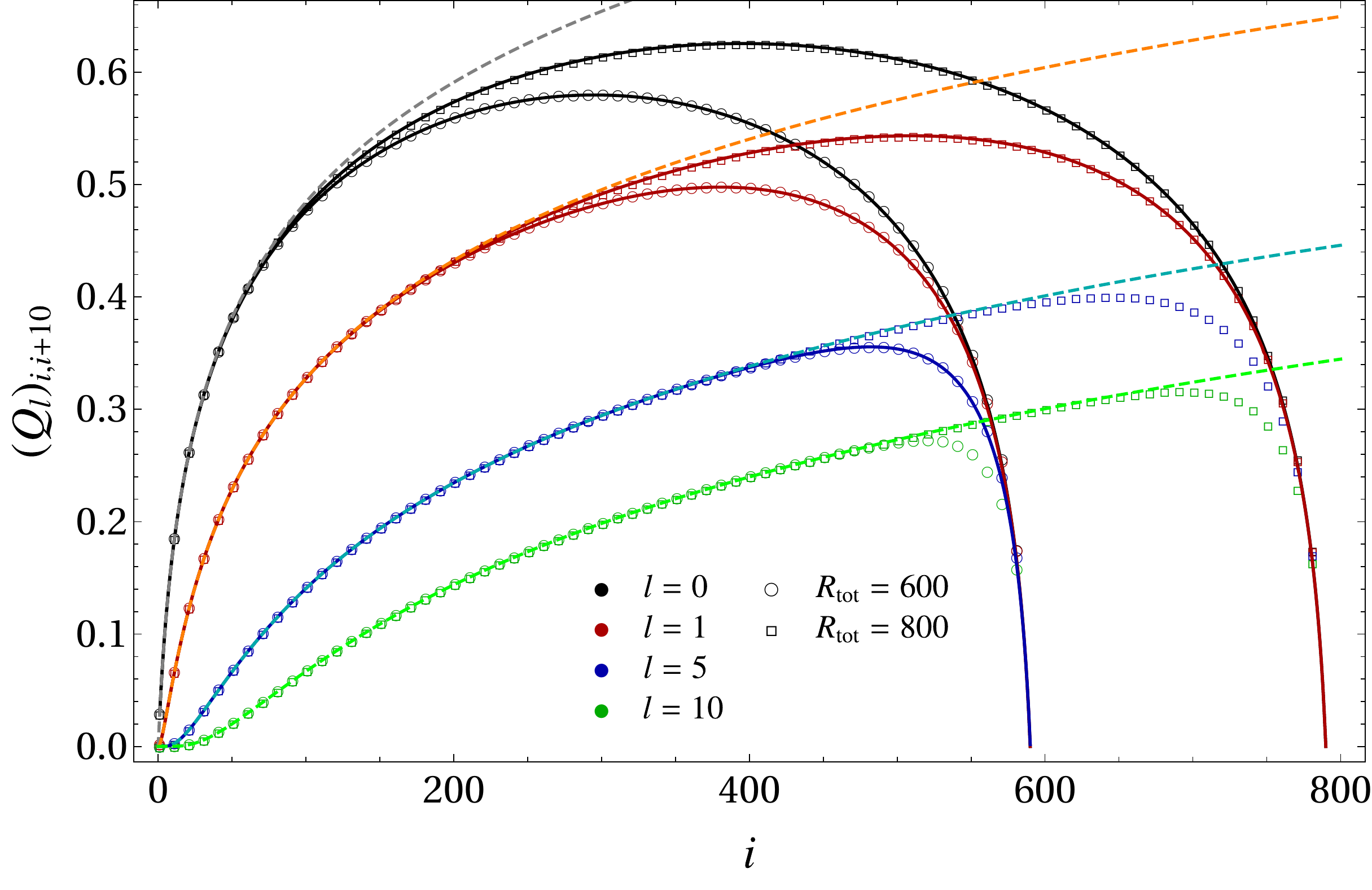}
\end{subfigure}
\begin{subfigure}{.45\textwidth}
\hspace{-1.4cm}
\includegraphics[scale=.354]{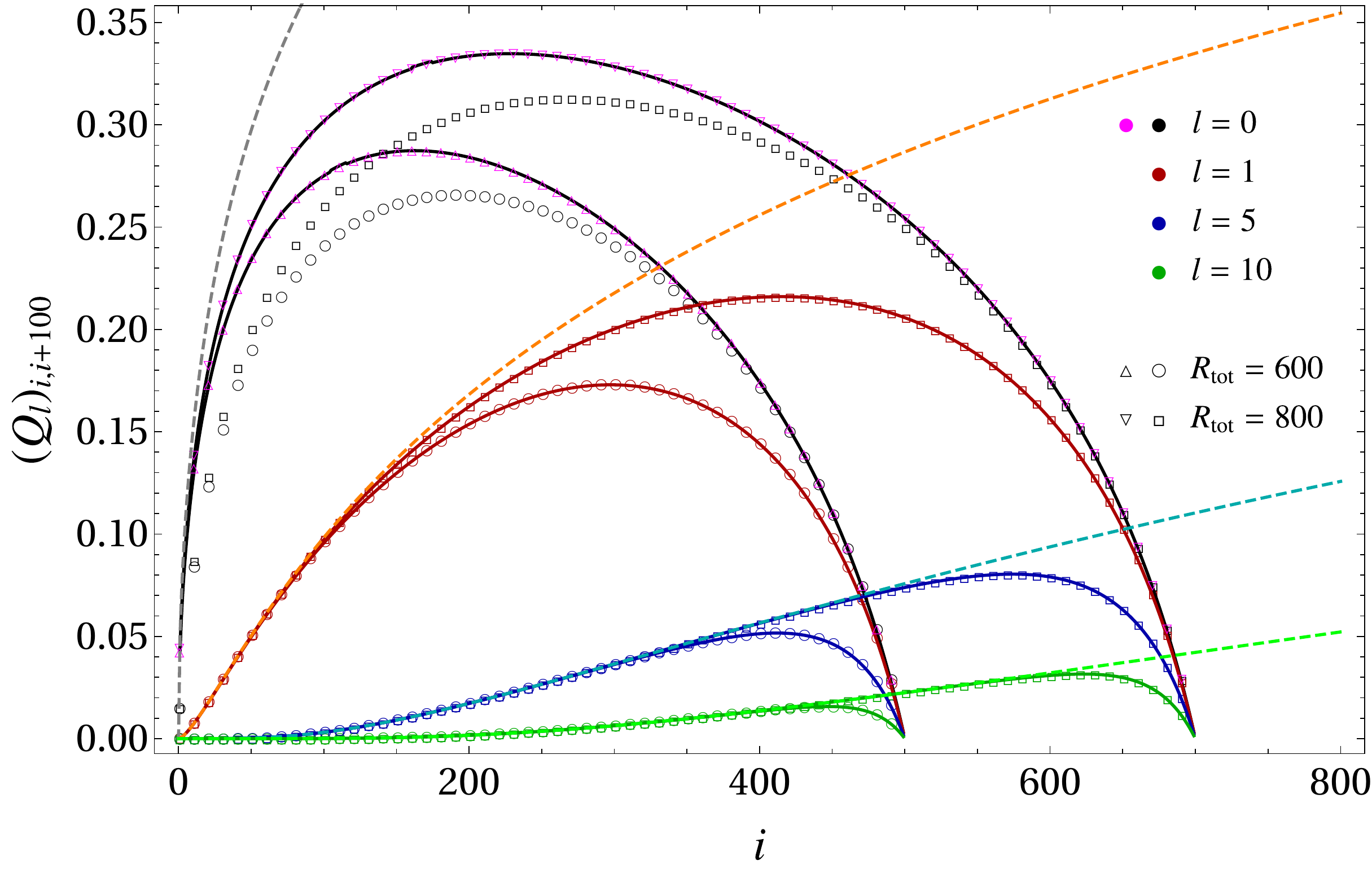}
\end{subfigure}
\hfill
\begin{subfigure}{.45\textwidth}
\hspace{-.7cm}
\includegraphics[scale=.355]{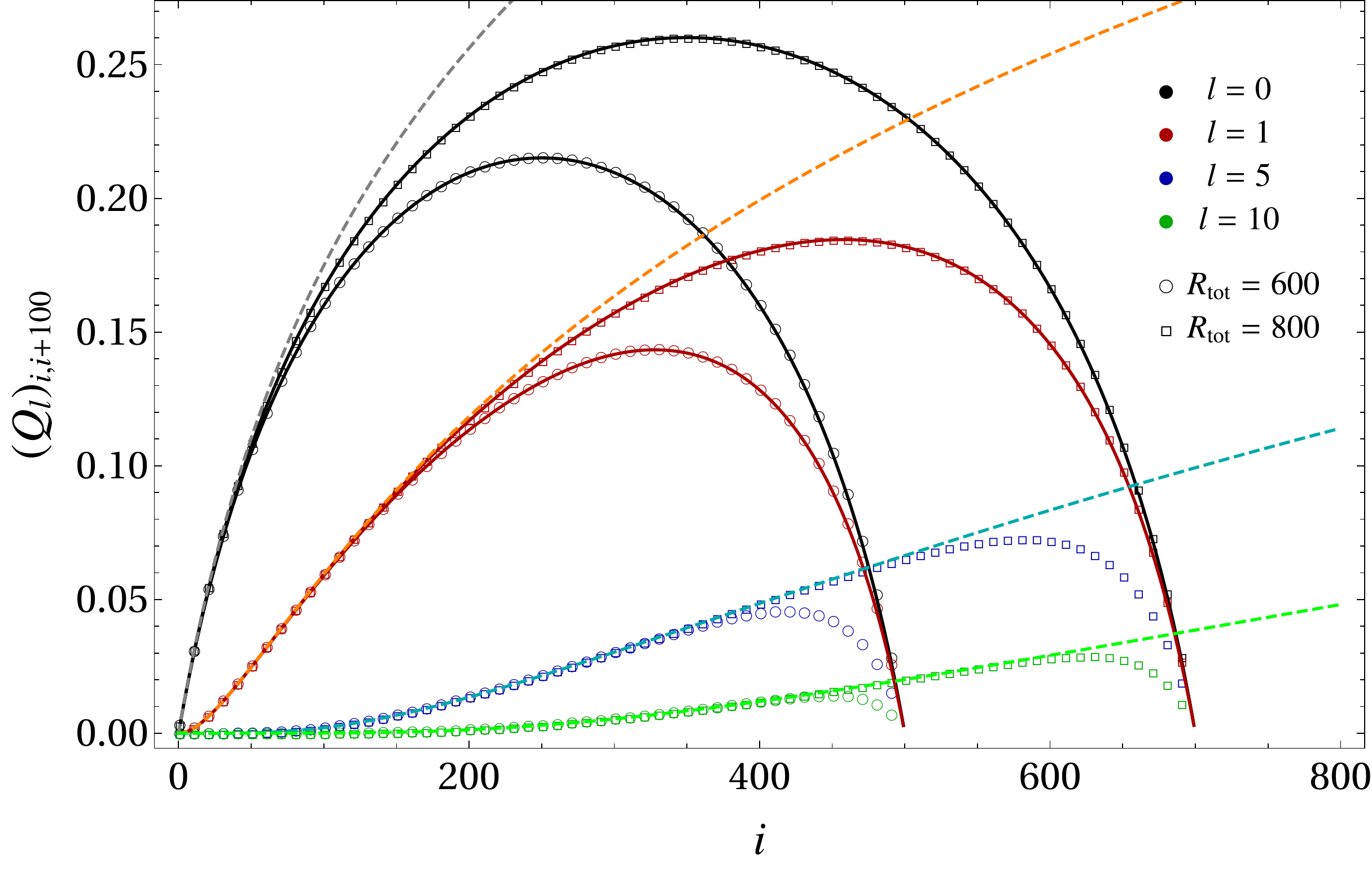}
\end{subfigure}
\hfill
\caption{
Correlators (\ref{corr-matrices-mode}) in the massless regime: 
$Q_{i,i+10}$ (top panels) and $Q_{i,i+100}$ (bottom panels) 
for $d=2$ (left panels) and $d=3$ (right panels)
evaluated from (\ref{matrix-M}) with $\omega = 0$ 
and for systems whose total size is given by $R_{\textrm{\tiny tot}}\in \{600, 800\}$.
The dashed lines correspond to (\ref{phi-phi}) and the solid ones to (\ref{Q_l-def-volume}) with $m=0$.
The magenta data points in the left panels are obtained from (\ref{M-def-app})  with $\omega = 0$.
}
\label{fig:Q}
\end{figure}

\begin{figure}[t!]
\begin{subfigure}{.45\textwidth}
\hspace{-1.5cm}
\includegraphics[scale=.36]{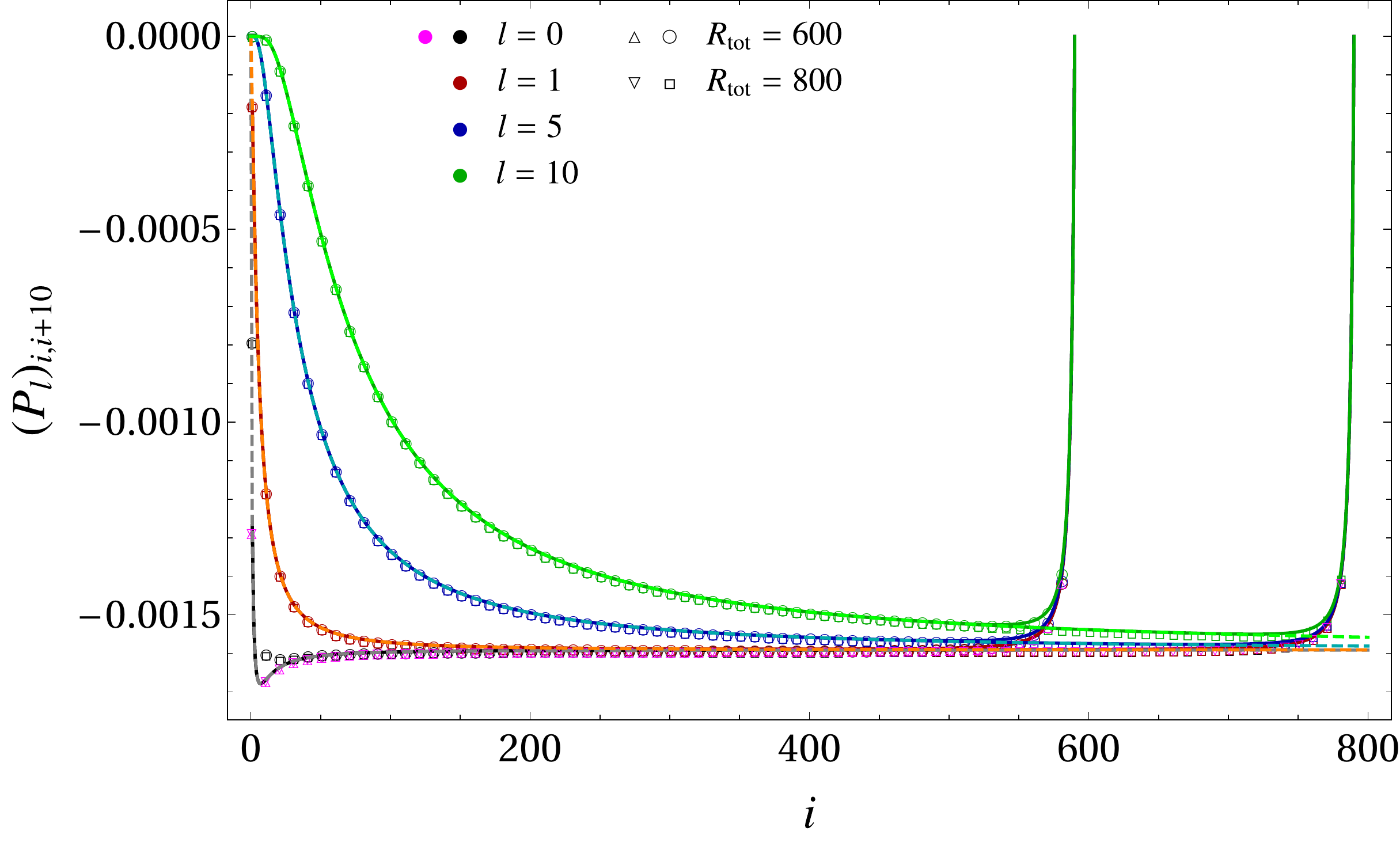}
\end{subfigure}
\hfill
\begin{subfigure}{.45\textwidth}
\hspace{-.7cm}
\includegraphics[scale=.36]{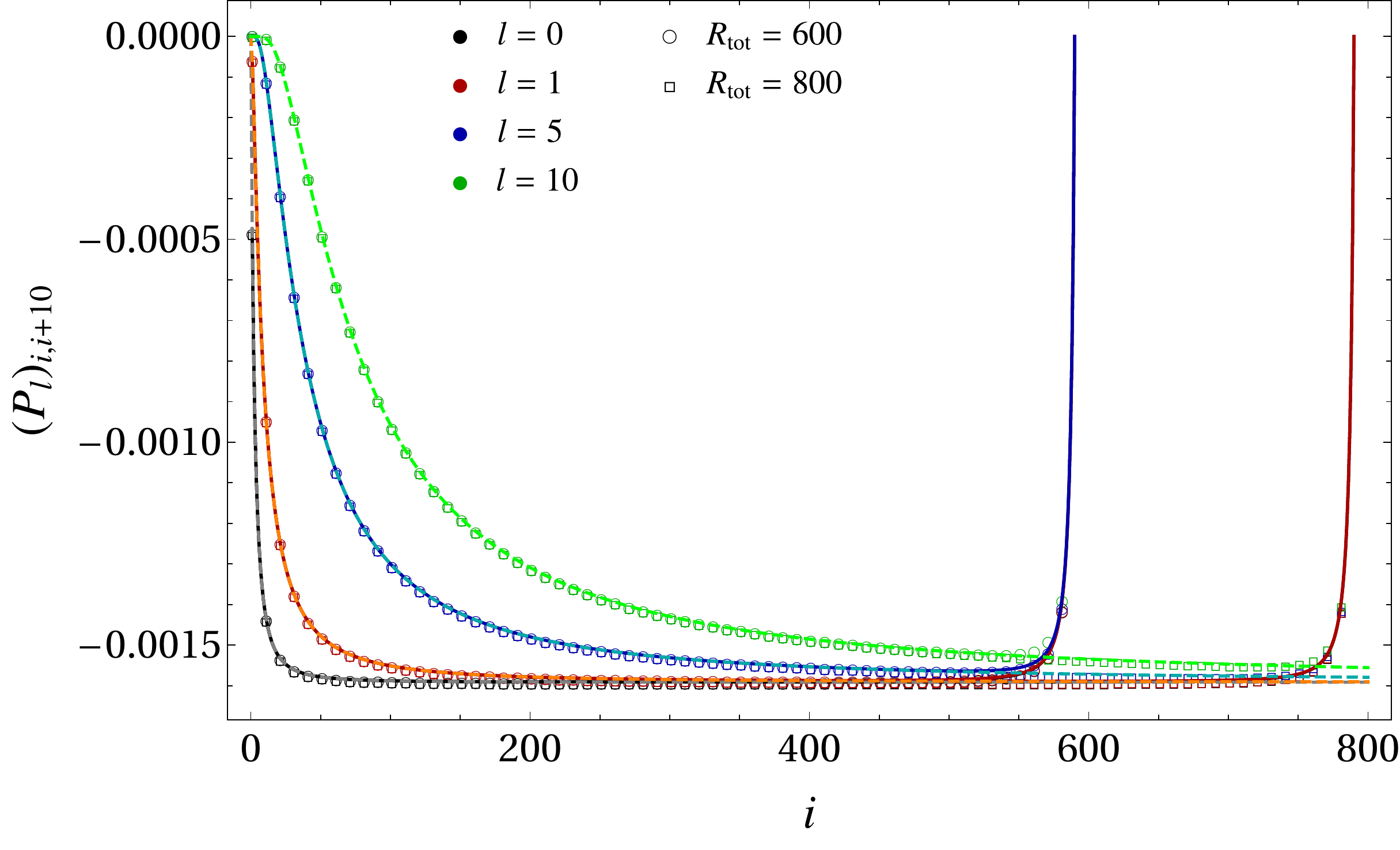}
\end{subfigure}
\hfill
\begin{subfigure}{.45\textwidth}
\hspace{-1.7cm}
\includegraphics[scale=.37]{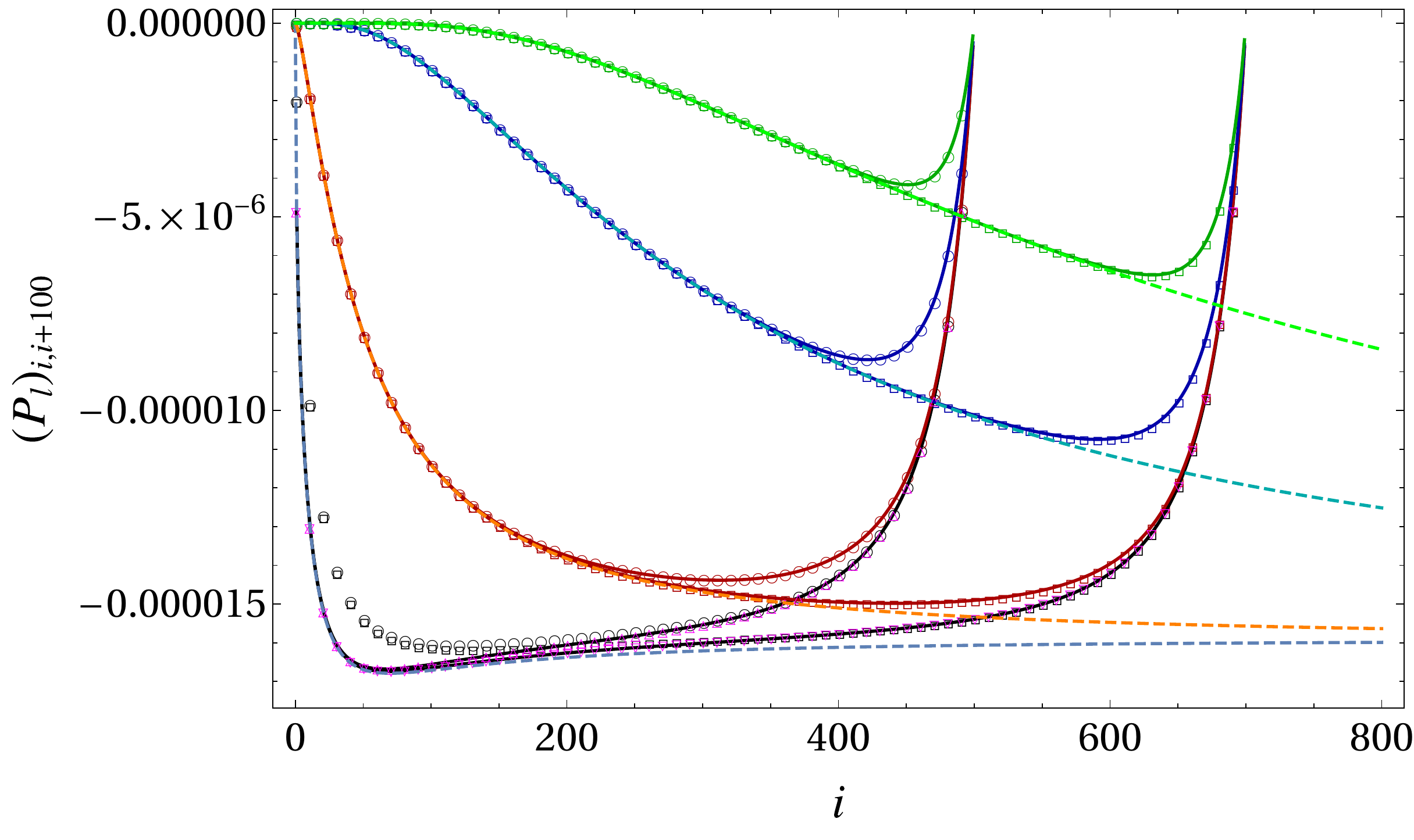}
\end{subfigure}
\hfill
\begin{subfigure}{.45\textwidth}
\hspace{-.8cm}
\includegraphics[scale=.37]{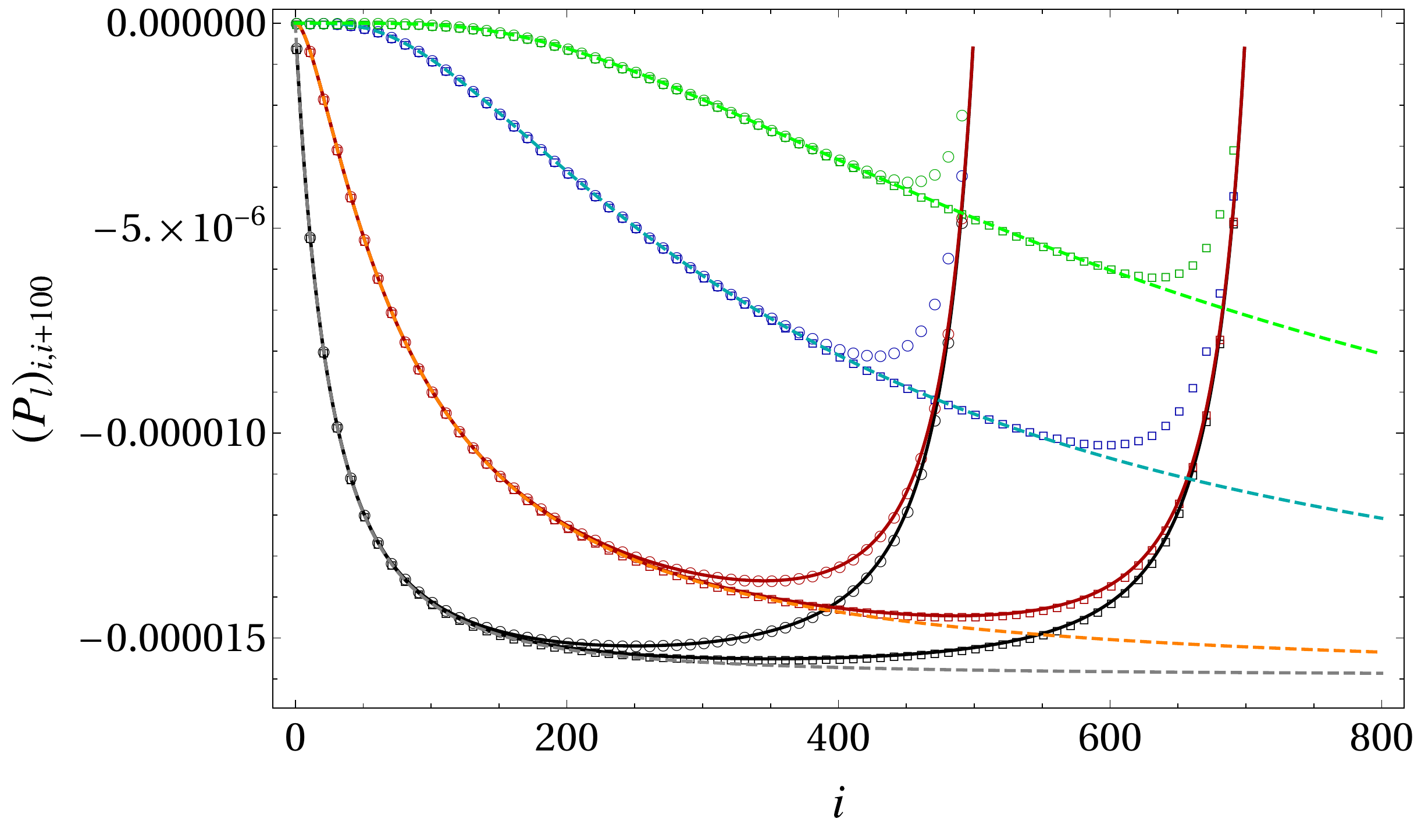}
\end{subfigure}
\hfill
\caption{
Correlators (\ref{corr-matrices-mode}) in the massless regime: 
$P_{i,i+10}$ (top panels) and $P_{i,i+100}$ (bottom panels) 
for $d=2$ (left panels) and $d=3$ (right panels)
evaluated from (\ref{matrix-M}) with $\omega = 0$ 
and for systems whose total size is given by $R_{\textrm{\tiny tot}}\in \{600, 800\}$.
The dashed lines correspond to (\ref{pi-pi}) and the solid ones to (\ref{P_l-def-volume}) with $m=0$.
The magenta data points are obtained from (\ref{M-def-app})  with $\omega = 0$.
}
\label{fig:P}
\end{figure}

When the free massive scalar field is defined 
inside a $d$ dimensional sphere whose radius is $\mathcal{R}_{\textrm{\tiny tot}}$
and Dirichlet boundary conditions are imposed,
the two-point function is (\ref{wightman-d-dim}) with \cite{Saharian:2000mw}
\bea
\label{G_l-def-volume}
\mathcal{G}_l(t_1, r_1 ; t_2, r_2)
&\equiv&
\\
\rule{0pt}{.8cm}
& &\hspace{-2.8cm}
\equiv \frac{\sqrt{r_1 r_2}}{2\, \mathcal{R}_{\textrm{\tiny tot}}^2}\,
\Bigg[\,
\int_0^{\infty}\!\!
\frac{\gamma}{ 
\sqrt{(\gamma/\mathcal{R}_{\textrm{\tiny tot}})^2 + m^2} }\;
J_{\tilde{\mu}}(\gamma r_1/\mathcal{R}_{\textrm{\tiny tot}})\, 
J_{\tilde{\mu}}(\gamma r_2/\mathcal{R}_{\textrm{\tiny tot}})
\; e^{\textrm{i} \sqrt{(\gamma/\mathcal{R}_{\textrm{\tiny tot}})^2 + m^2} \,(t_2 - t_1)}
\,\textrm{d}\gamma
\nonumber
\\
\rule{0pt}{.8cm}
& &\hspace{-1.6cm}
- \, \frac{2}{\pi}
\int_{m R_{\textrm{\tiny tot}}}^{\infty}
\frac{K_{\tilde{\mu}}(\gamma)}{I_{\tilde{\mu}}(\gamma)}\;
\frac{I_{\tilde{\mu}}(\gamma r_1/\mathcal{R}_{\textrm{\tiny tot}})\, 
I_{\tilde{\mu}}(\gamma r_2/\mathcal{R}_{\textrm{\tiny tot}})
}{
\sqrt{(\gamma/\mathcal{R}_{\textrm{\tiny tot}})^2 - m^2} }
\;\gamma\,
\cosh\!\big[ \sqrt{(\gamma/\mathcal{R}_{\textrm{\tiny tot}})^2 - m^2 }\,(t_2 - t_1)\big]
\,\textrm{d}\gamma\,
\Bigg]
\nonumber
\eea
which simplifies to (\ref{G_l-def}) when $\mathcal{R}_{\textrm{\tiny tot}} \to +\infty$, as expected. 
Setting $t_1 = t_2$ in (\ref{G_l-def-volume}), one obtains
\bea
\label{Q_l-def-volume}
\mathcal{Q}_l(r_1 , r_2)
&\equiv &
\frac{\sqrt{r_1 r_2}}{2\, \mathcal{R}_{\textrm{\tiny tot}}^2}\,
\Bigg[\,
\int_0^{\infty}\!\!
\frac{\gamma}{ 
\sqrt{(\gamma/\mathcal{R}_{\textrm{\tiny tot}})^2 + m^2} }\;
J_{\tilde{\mu}}(\gamma r_1/\mathcal{R}_{\textrm{\tiny tot}})\, 
J_{\tilde{\mu}}(\gamma r_2/\mathcal{R}_{\textrm{\tiny tot}})
\,\textrm{d}\gamma
\\
\rule{0pt}{.8cm}
& &\hspace{2.cm}
- \, \frac{2}{\pi}
\int_{m \mathcal{R}_{\textrm{\tiny tot}}}^{\infty}
\frac{K_{\tilde{\mu}}(\gamma)}{I_{\tilde{\mu}}(\gamma)}\;
\frac{I_{\tilde{\mu}}(\gamma r_1/\mathcal{R}_{\textrm{\tiny tot}})\, I_{\tilde{\mu}}(\gamma r_2/\mathcal{R}_{\textrm{\tiny tot}})}{\sqrt{(\gamma/\mathcal{R}_{\textrm{\tiny tot}})^2 - m^2} }
\;\gamma\,
\textrm{d}\gamma\,
\Bigg]\,.
\nonumber
\eea
In this finite volume case, 
we can evaluate $\partial_{t_1} \partial_{t_2} \mathcal{G}_l(t_1, r_1 ; t_2, r_2)$  for $t_1 = t_2$ 
from (\ref{G_l-def-volume}), finding 
\bea
\label{P_l-def-volume}
& & \hspace{-.5cm}
\mathcal{P}_l(r_1 , r_2)
\,\equiv\,
\frac{\sqrt{r_1 r_2}}{2\,\mathcal{R}^2_{\textrm{\tiny tot}}}\,
\Bigg[\,
\int_0^{\infty}\!\!
\gamma\,\sqrt{(\gamma / \mathcal{R}_{\textrm{\tiny tot}})^2+ m^2} \;
J_{\tilde{\mu}}(\gamma r_1/ \mathcal{R}_{\textrm{\tiny tot}})\, 
J_{\tilde{\mu}}(\gamma r_2/ \mathcal{R}_{\textrm{\tiny tot}})
\,\textrm{d}\gamma
\\
\rule{0pt}{.8cm}
& &\hspace{3.3cm}
+ \, \frac{2}{\pi}
\int_{m \mathcal{R}_{\textrm{\tiny tot}}}^{\infty}\!\!\!
\sqrt{(\gamma / \mathcal{R}_{\textrm{\tiny tot}})^2 - m^2 } \;
\frac{K_{\tilde{\mu}}(\gamma)}{I_{\tilde{\mu}}(\gamma)}\;
I_{\tilde{\mu}}(\gamma r_1/\mathcal{R}_{\textrm{\tiny tot}})\, I_{\tilde{\mu}}(\gamma r_2/\mathcal{R}_{\textrm{\tiny tot}})
\;\gamma\,
\textrm{d}\gamma\,
\Bigg]\,.
\nonumber
\eea

In Fig.\,\ref{fig:Q} and Fig.\,\ref{fig:P} 
we compare the numerical data points obtained 
for the correlation matrices 
given by (\ref{corr-matrices-mode}) and (\ref{matrix-M})
with  the corresponding expressions in the continuum,
which are given by (\ref{phi-phi}) and (\ref{pi-pi}) for 
the infinite volume regime (dashed lines)
and by (\ref{Q_l-def-volume}) and (\ref{P_l-def-volume}) 
for the finite volume space enclosed in a sphere of radius $\mathcal{R}_{\textrm{\tiny tot}}$
where Dirichlet boundary conditions are imposed on the boundary of the sphere (solid lines).
The solid lines that are missing in these figures correspond to the cases where 
our numerical integration of (\ref{Q_l-def-volume}) and (\ref{P_l-def-volume}) failed. 
A remarkable agreement between the numerical data points for the lattice correlators
and the curves in the continuum is observed,
except when $d=2$ and $l=0$ 
(see the black data points in the left panels of Fig.\,\ref{fig:Q} and Fig.\,\ref{fig:P}).

In order to explain this discrepancy for $(d,l)=(2,0)$,
consider $ \Delta  = \tfrac{1}{r^{d-1}}\,\partial_r(r^{d-1}\partial_r) + \tfrac{1}{r^2}\, \tilde{\Delta}$
in (\ref{hamiltonian}).
By employing the decomposition (\ref{rescaled-fields-inv}) first 
and then performing an integration by parts in the remaining radial integration, 
one obtains (\ref{phys-H-decomposed}) with 
\cite{Srednicki:1993im, Riera_2006, Huerta:2011qi, Cotler:2016acd,Nishioka:2018khk}
\bea
\label{T00-scalar-l-mode-app}
\widetilde{T}_{tt}^{(\boldsymbol{l})} 
&\equiv&
\frac{1}{2} \,\bigg\{\,
\Pi_{\boldsymbol{l}}^2
+
r^{d-1} \left[ \,\partial_r \! \left(\,\frac{\Phi_{\boldsymbol{l}} }{r^{\frac{d-1}{2}}}\right)\right]^2
+ 
\left(\! -\frac{\lambda_d(l)}{r^2} +  m^2  \right) \Phi_{\boldsymbol{l}}^2
\,\bigg\}
\\
\label{T00-scalar-l-mode-app-v2}
\rule{0pt}{.8cm}
&=&
\frac{1}{2} \,\bigg\{\,
\Pi_{\boldsymbol{l}}^2
+
\big(\partial_r \Phi_{\boldsymbol{l}} \big)^2
-
\frac{d-1}{r}\; \Phi_{\boldsymbol{l}}\, \partial_r \Phi_{\boldsymbol{l}}
+
\frac{(d-1)^2}{4\, r^2}\; \Phi_{\boldsymbol{l}}^2
+ 
\left(\! -\frac{\lambda_d(l)}{r^2} +  m^2  \right) \Phi_{\boldsymbol{l}}^2
\,\bigg\}
\hspace{1cm}
\eea
where $\lambda_d(l)$ has been defined in (\ref{spherical-harmonics-def}).
Notice that, 
by integrating (\ref{T00-scalar-l-mode-app-v2}) in the radial variable first
and then performing an integration by parts in the terms corresponding to 
$(\partial_r \Phi_{\boldsymbol{l}})^2$ 
and $\Phi_{\boldsymbol{l}}\, \partial_r \Phi_{\boldsymbol{l}}$,
the integral of (\ref{phys-H-decomposed}) is recovered.

Plugging (\ref{T00-scalar-l-mode-app}) into (\ref{phys-H-decomposed}),
we obtain $H_{\boldsymbol{l}}$.
Then, following the regularisation procedure discussed in Sec.\,\ref{subsec-ham},
we find that the corresponding operator in the lattice model along the radial direction reads
\be
\label{Ham-fixed-mode-app}
\widehat{H}_{\boldsymbol{l}}
\,=\, 
\frac{1}{2a}\, 
\sum_{j=1}^{R_{\textrm{\tiny tot}}}
\Bigg\{\,
\hat{p}_{\boldsymbol{l},j}^2
+\left(j+\frac{1}{2}\right)^{d-1}
\Bigg[\, 
\frac{\hat{q}_{\boldsymbol{l},j+1}}{(j+1)^{\frac{d-1}{2}}} 
- \frac{\hat{q}_{\boldsymbol{l},j}}{j^{\frac{d-1}{2}}}
\,\Bigg]^2
+
\left(\! -\frac{\lambda_d(l)}{r^2} +  m^2  \right)
\hat{q}_{\boldsymbol{l},j}^2
\,\Bigg\}
\ee
which can be written in the form (\ref{Ham-fixed-mode}) with 
\be
\label{M-def-app}
(M_l)_{j,j} \,\equiv\,
-\frac{\lambda_d(l)}{j^2} +  m^2 + \mathcal{M}_j
\;\;\qquad\;\;
(M_l)_{j,j+1} \,\equiv\, 
- \left( \frac{j+1/2}{\sqrt{j(j+1)}} \right)^{d-1}
\ee
where $\mathcal{M}_j$ is defined as
\be
\mathcal{M}_1 \equiv \left( \frac{3}{2} \right)^{d-1}
\;\;\qquad\;\;
\mathcal{M}_j \equiv \left(1+\frac{1}{2j}\right)^{d-1} \!\! + \left(1-\frac{1}{2j}\right)^{d-1}
\qquad
2 \leqslant j \leqslant R_{\textrm{\tiny tot}} \,.
\ee
The correlation matrices corresponding to (\ref{Ham-fixed-mode-app})
are (\ref{corr-matrices-mode}) with $M_l$ given by (\ref{M-def-app}).
For $\omega = 0$, we checked that the data points for these correlators perfectly agree with 
the ones coming from (\ref{matrix-M}) and (\ref{corr-matrices-mode})
reported in Fig.\,\ref{fig:Q} and Fig.\,\ref{fig:P}, 
except for $(d,l) = (2,0)$.
In this case they correspond to
the magenta data points in the left panels of Fig.\,\ref{fig:Q} and Fig.\,\ref{fig:P},
which are nicely reproduced by the curves obtained from 
(\ref{Q_l-def-volume}) and (\ref{P_l-def-volume}) respectively
with $m=0$ and $\tilde{\mu} = 0$.
In massive regime, we have considered e.g. the case of $(d,l) = (2,1)$
and checked numerically that the matrix correlator $Q_{i,i+10}$
evaluated from (\ref{matrix-M}) and (\ref{corr-matrices-mode})
agrees with (\ref{Q_l-def-volume}) 
for $\omega = m=0.1$.


\section{Comments on the role of $k_{\textrm{\tiny max}}$}
\label{app_kmax}

\begin{figure}[t!]
\centering
\begin{subfigure}{.28\textwidth}
\hspace{-2cm}
\includegraphics[scale=.35]{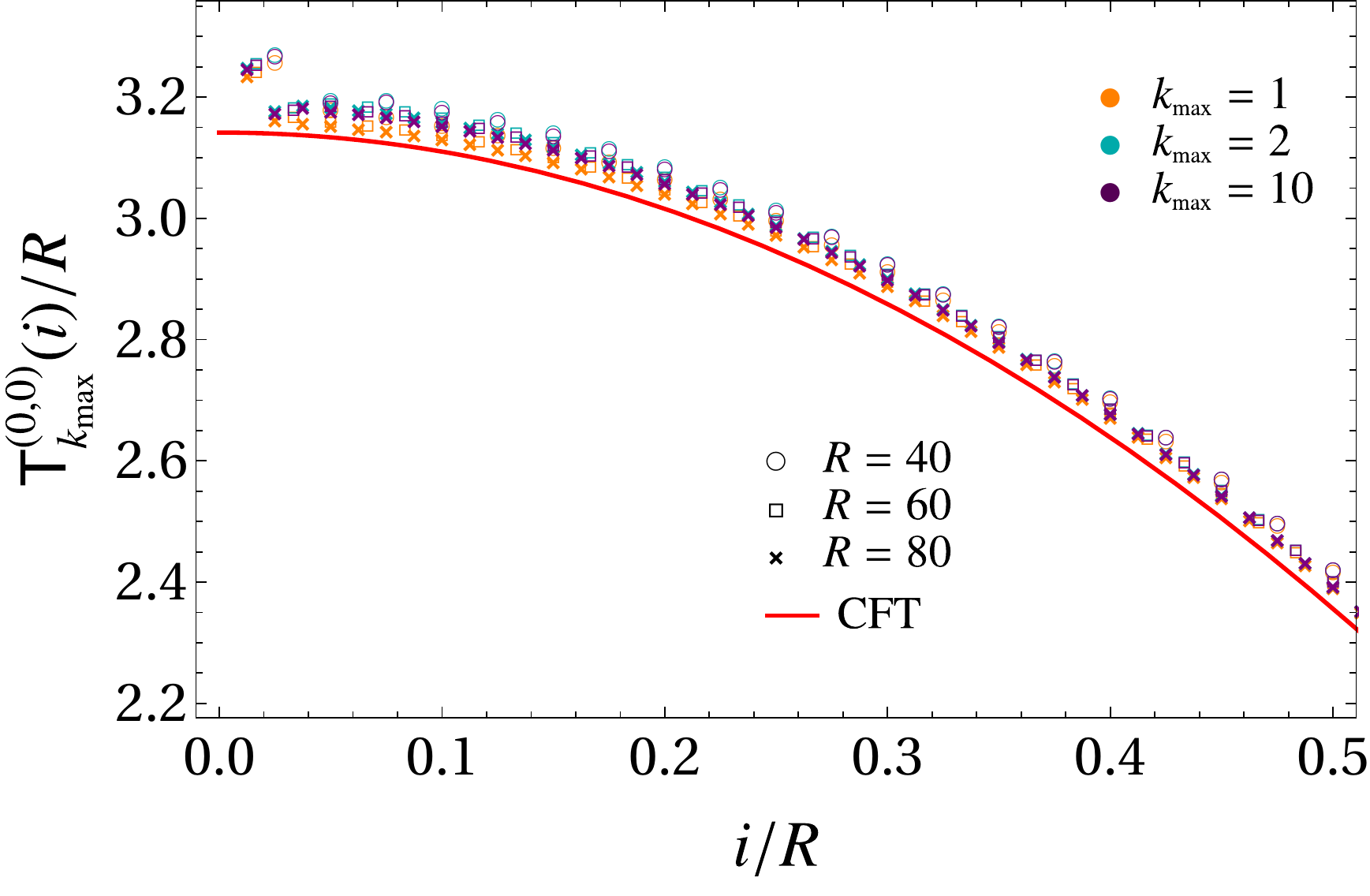}
\end{subfigure}
\hfill
\begin{subfigure}{.36\textwidth}
\hspace{-.3cm}
\includegraphics[scale=.35]{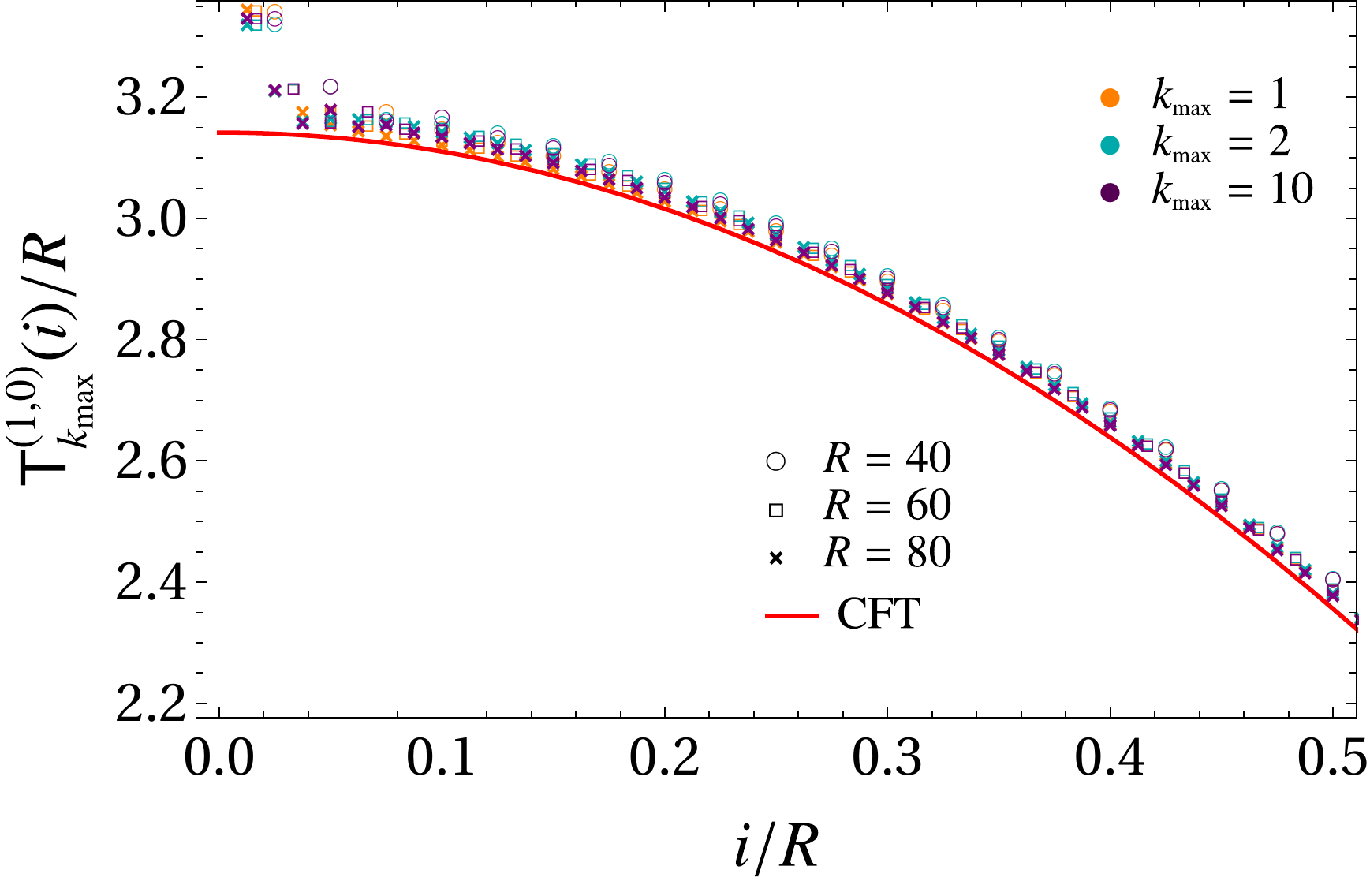}
\end{subfigure}
\hfill
\begin{subfigure}{.32\textwidth}
\includegraphics[scale=.35]{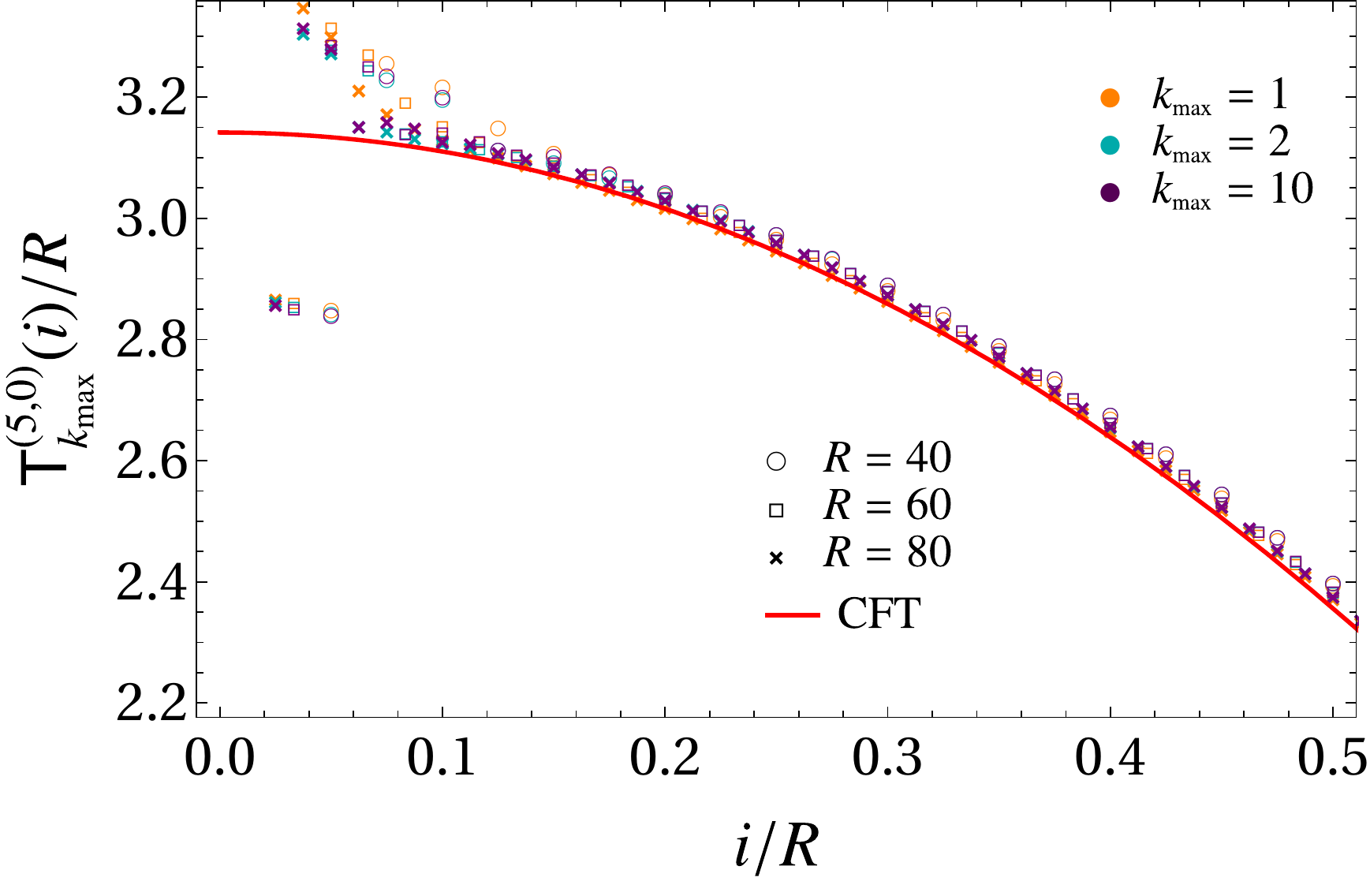}
\end{subfigure}
\vspace{-.5cm}
\begin{subfigure}{.28\textwidth}
\hspace{-2cm}
\includegraphics[scale=.35]{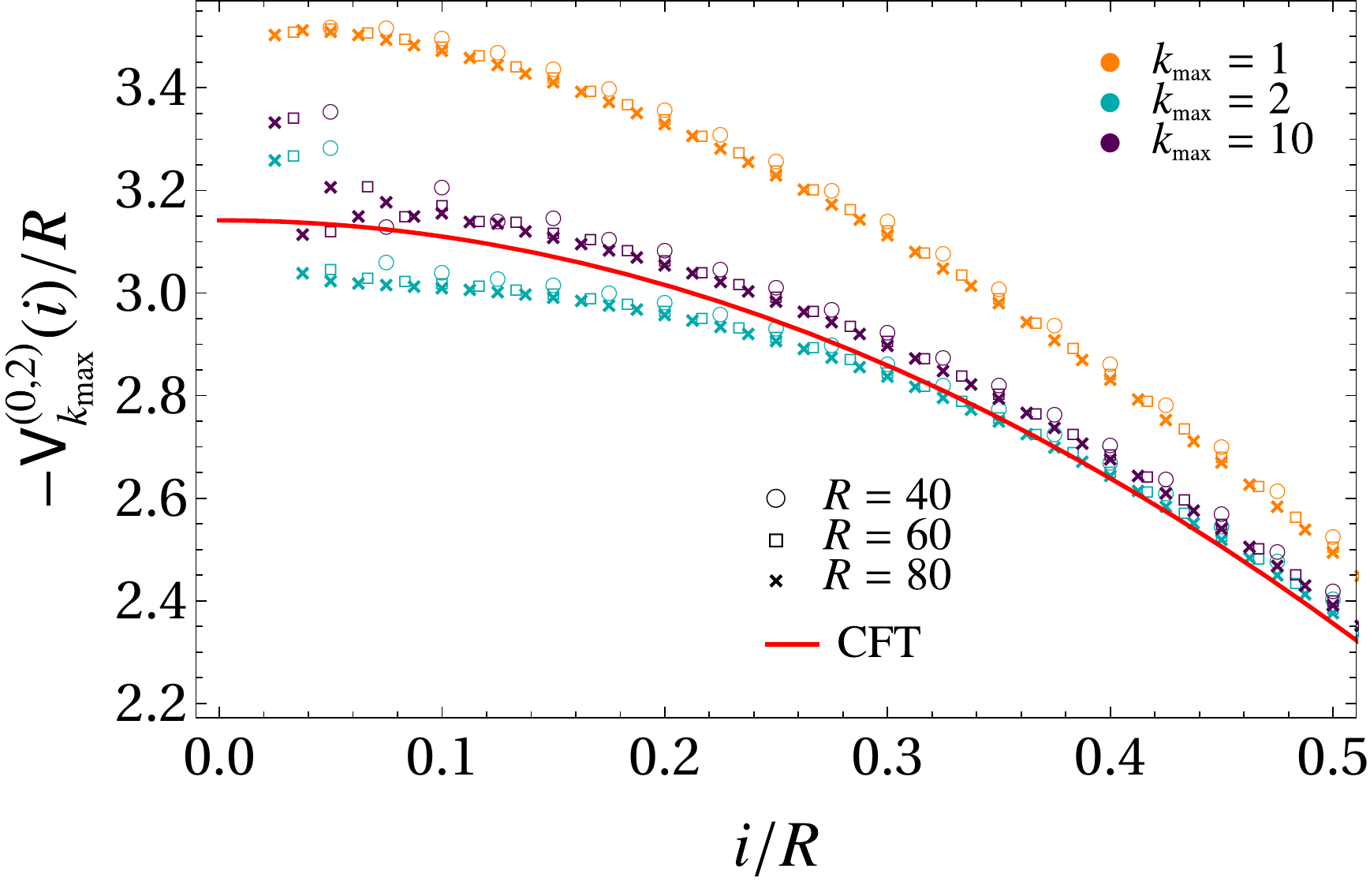}
\end{subfigure}
\hfill
\begin{subfigure}{.36\textwidth}
\hspace{-.15cm}
\includegraphics[scale=.35]{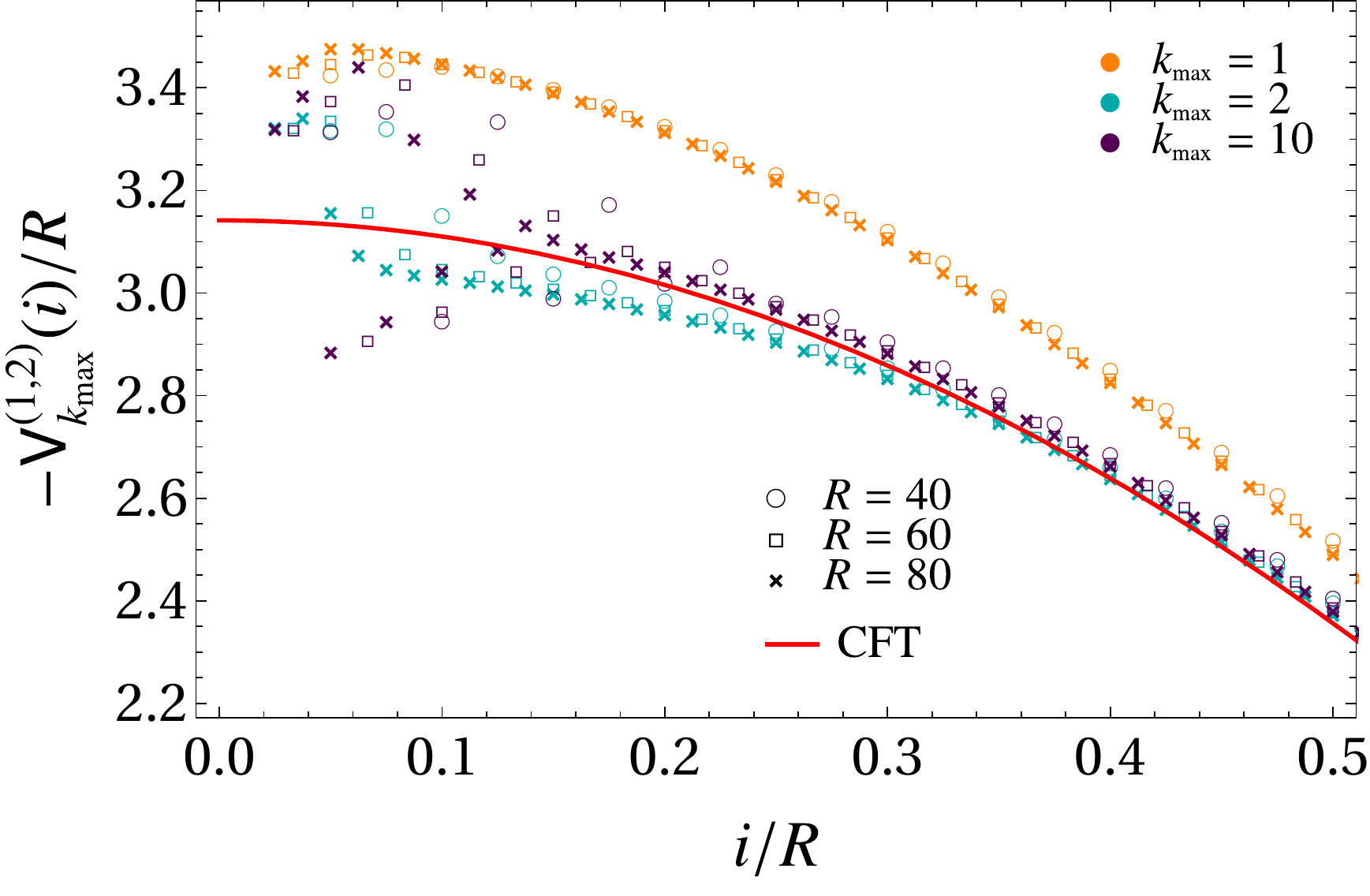}
\end{subfigure}
\hfill
\begin{subfigure}{.33\textwidth}
\hspace{.25cm}
\includegraphics[scale=.35]{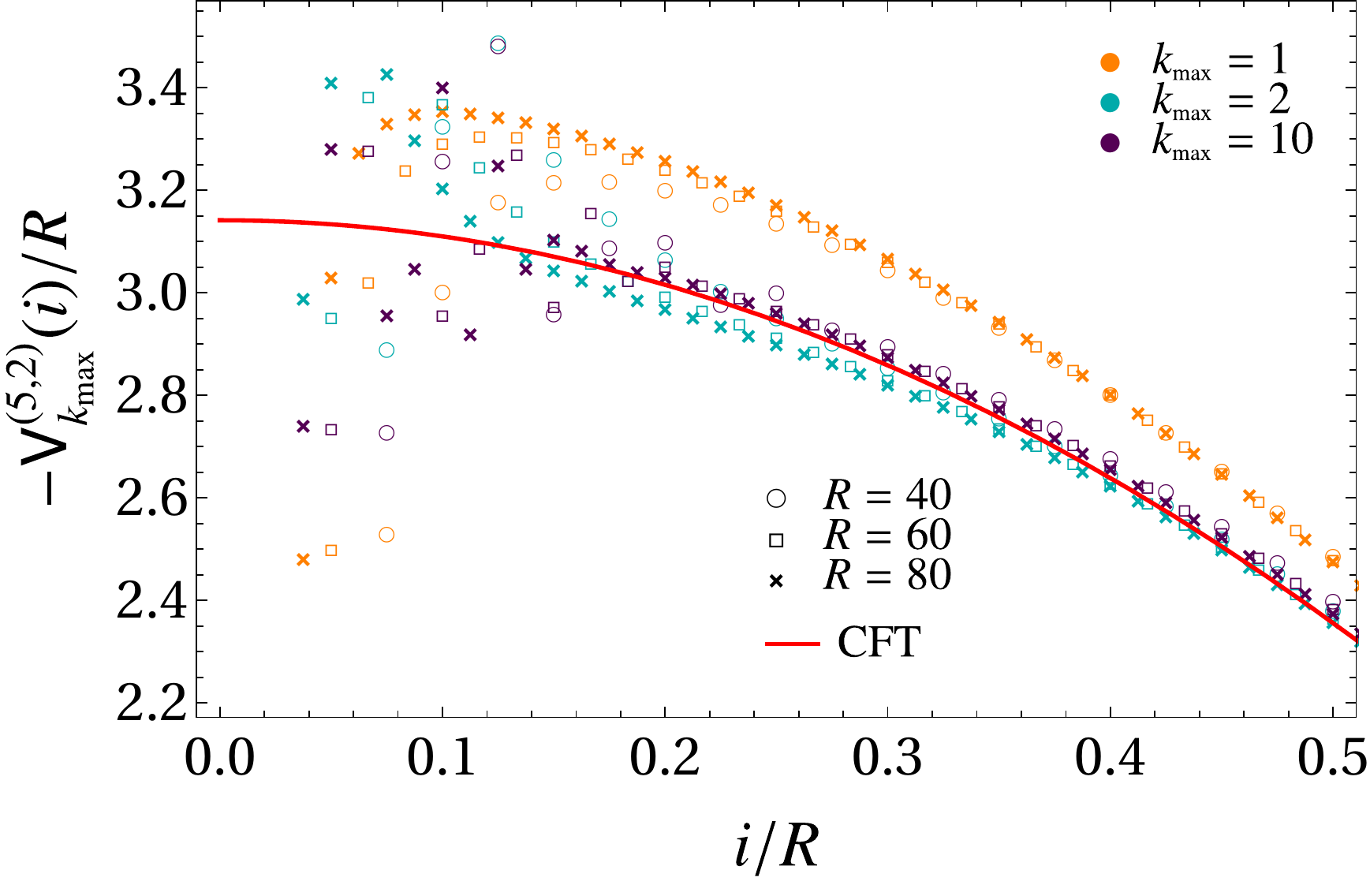}
\end{subfigure}
\vspace{.5cm}
\caption{
\label{fig:T0andV2app}
Combinations (\ref{T0-def}) and (\ref{V2-def}) when $\omega = 0$ in the case of $d=3$,
for various $k_{\textrm{\tiny{max}}}$.
These panels should be compared with the right panels of Fig.\,\ref{beta-TV-parabola}.
}
\end{figure}

In this appendix we briefly discuss the role of $k_{\textrm{\tiny max}}$ in the summations 
$\mathsf{T}_ {k_{\textrm{\tiny{max}}}}^{(l,0)}$,
$\mathsf{V}_ {k_{\textrm{\tiny{max}}}}^{(l,0)}$ 
and $\mathsf{V}_{k_{\textrm{\tiny{max}}}}^{(l,2)}$,
defined in (\ref{T0-def}), (\ref{V0-def}) and (\ref{V2-def}) respectively,
which are the crucial quantities explored in this manuscript,
together with (\ref{V12-def})  and (\ref{V22-def}).

\begin{figure}[t!]
\centering
\vspace{-.0cm}
\begin{subfigure}{.28\textwidth}
\hspace{-2cm}
\includegraphics[scale=.35]{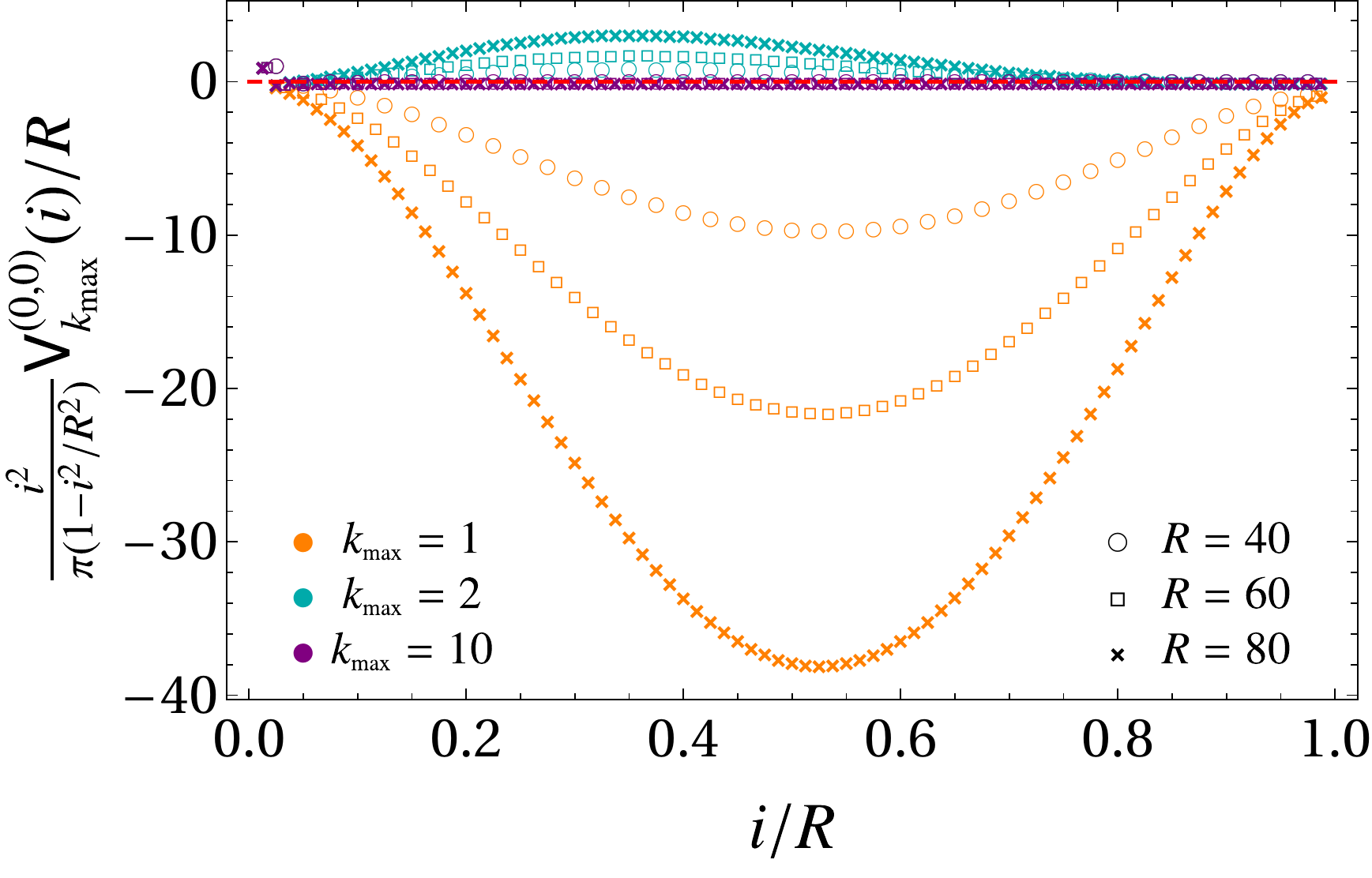}
\end{subfigure}
\hfill
\begin{subfigure}{.36\textwidth}
\hspace{-.4cm}
\includegraphics[scale=.35]{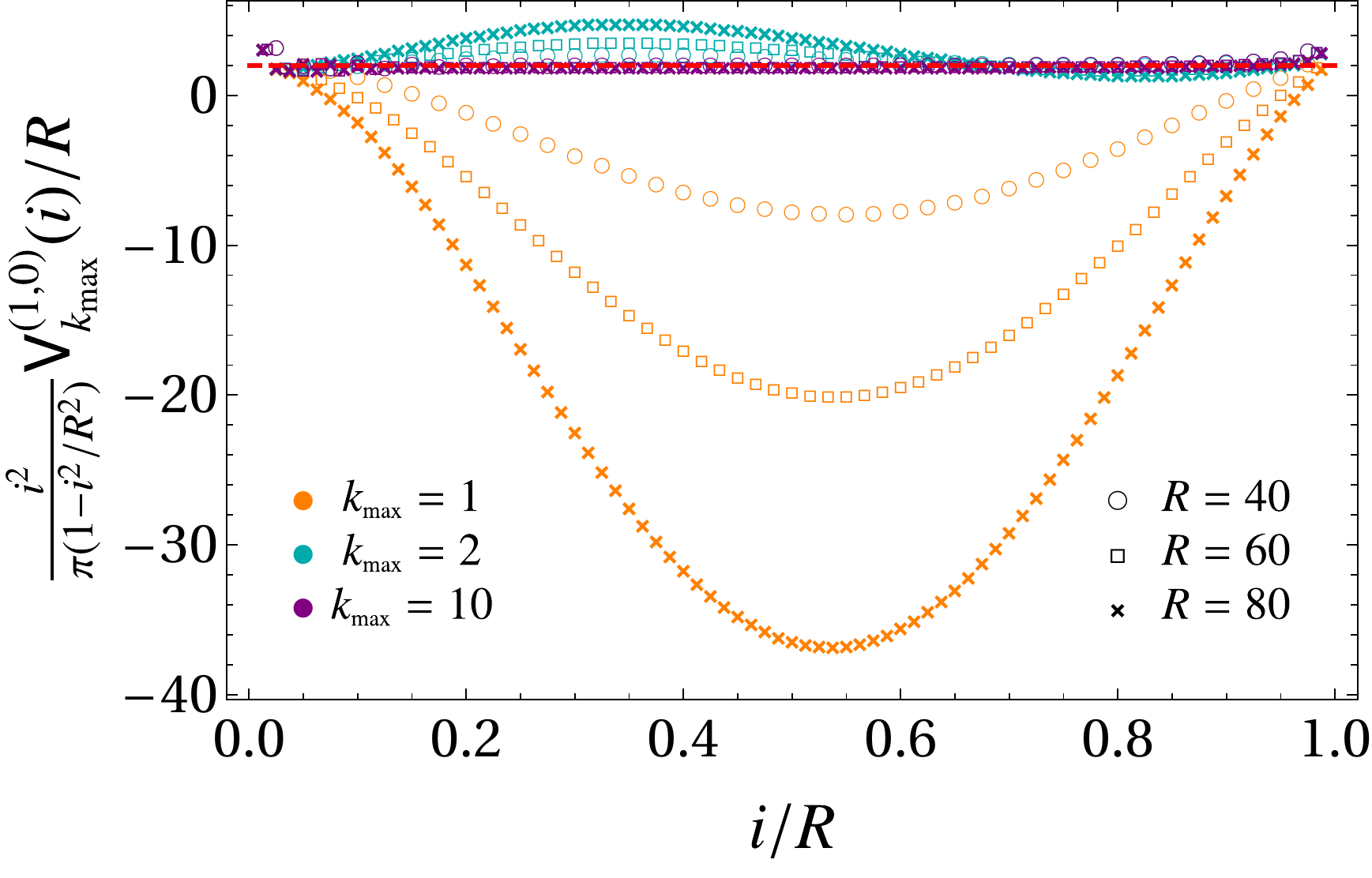}
\end{subfigure}
\hfill
\begin{subfigure}{.32\textwidth}
\includegraphics[scale=.35]{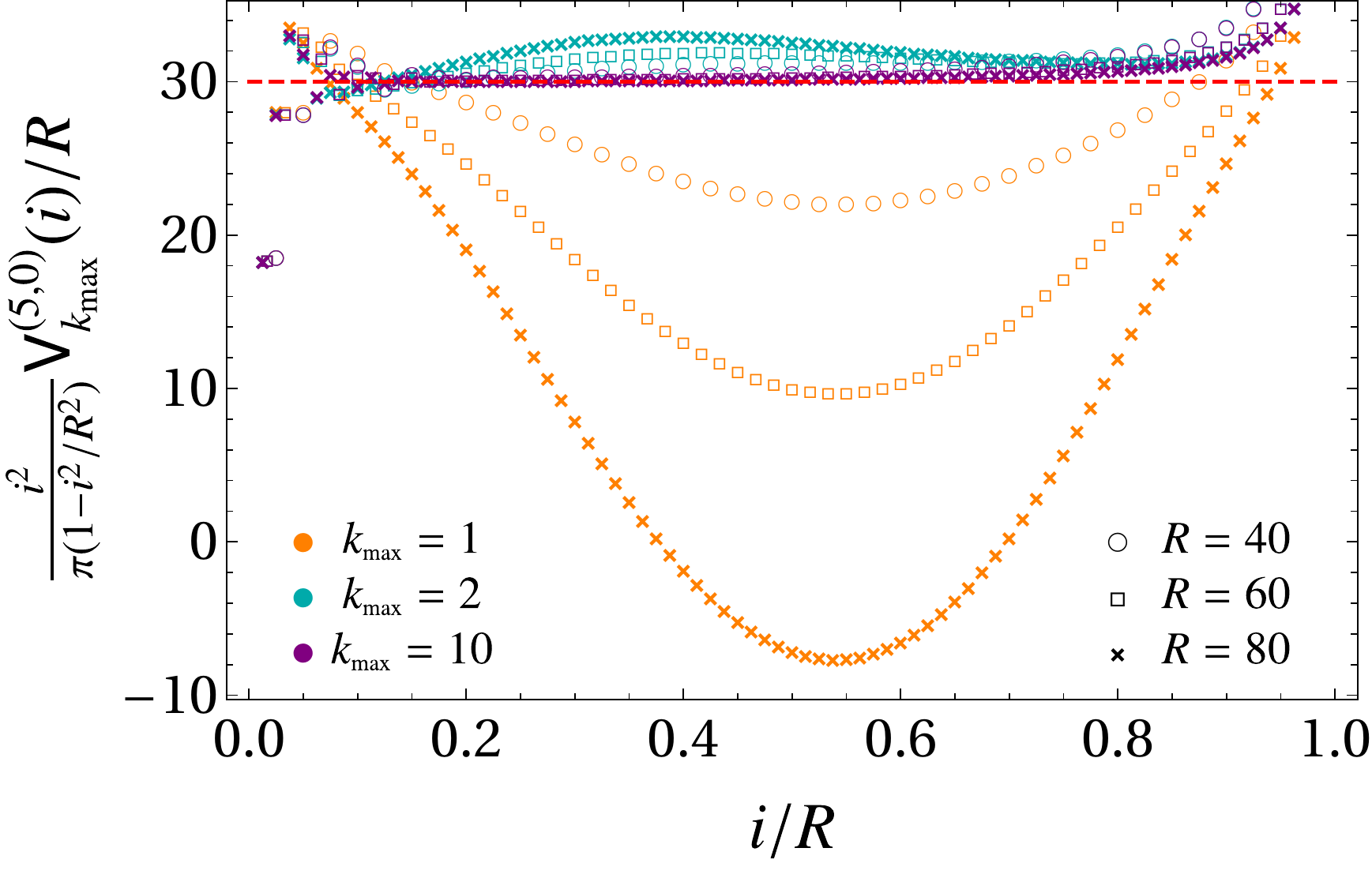}
\end{subfigure}
\hfill
\begin{subfigure}{.28\textwidth}
\hspace{-1.9cm}
\includegraphics[scale=.345]{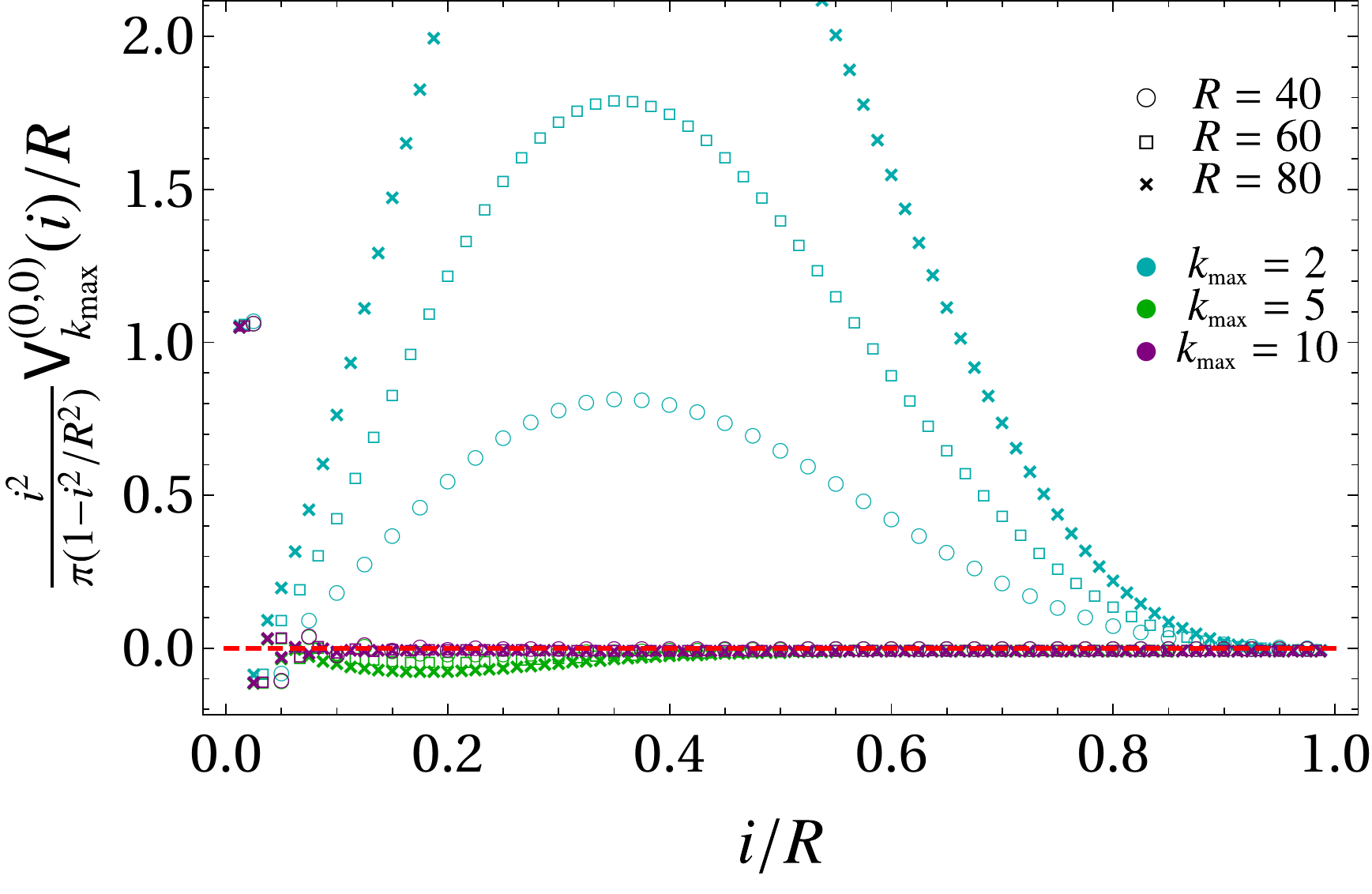}
\end{subfigure}
\hfill
\begin{subfigure}{.36\textwidth}
\hspace{.1cm}
\includegraphics[scale=.33]{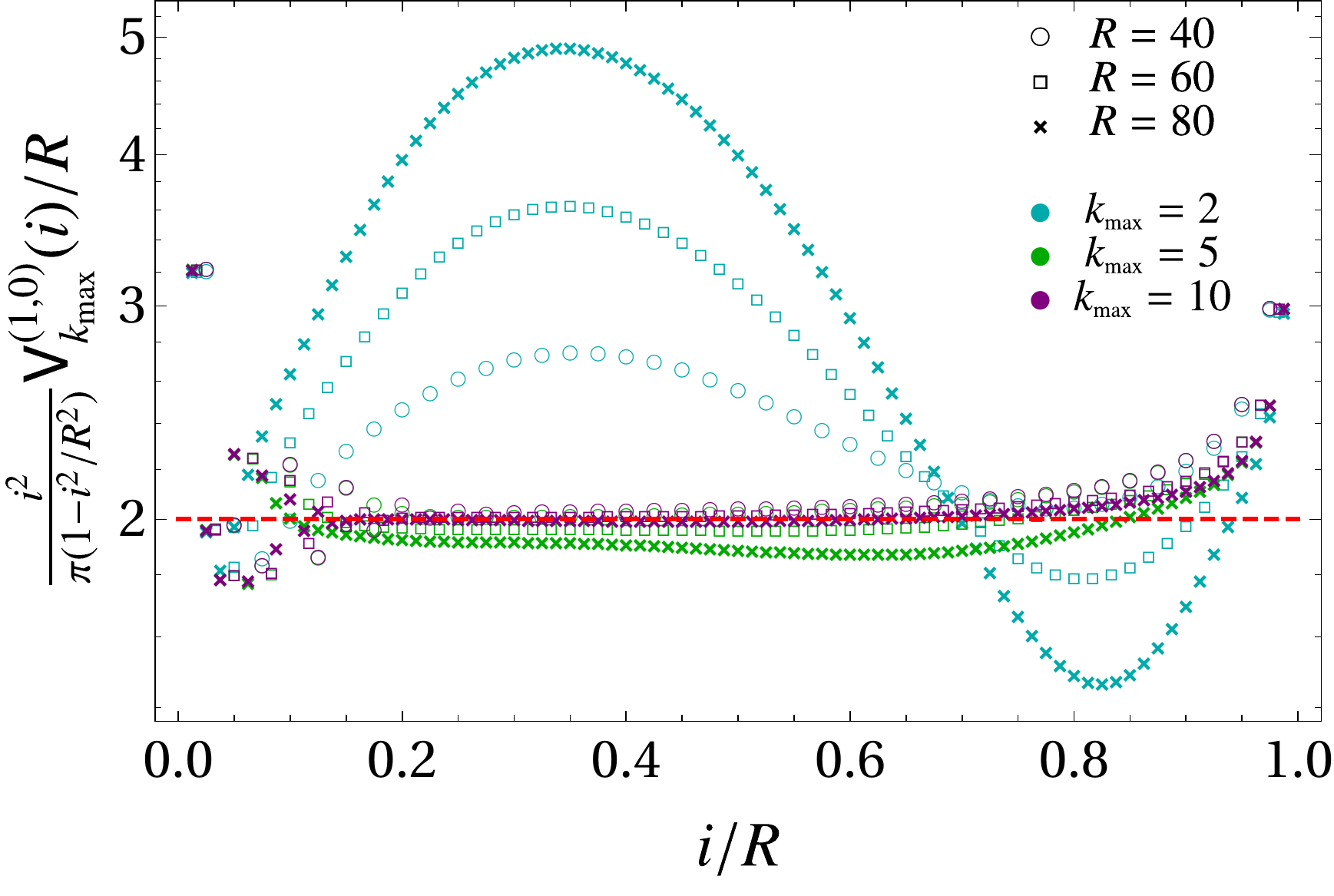}
\end{subfigure}
\hfill
\begin{subfigure}{.33\textwidth}
\hspace{.42cm}
\includegraphics[scale=.338]{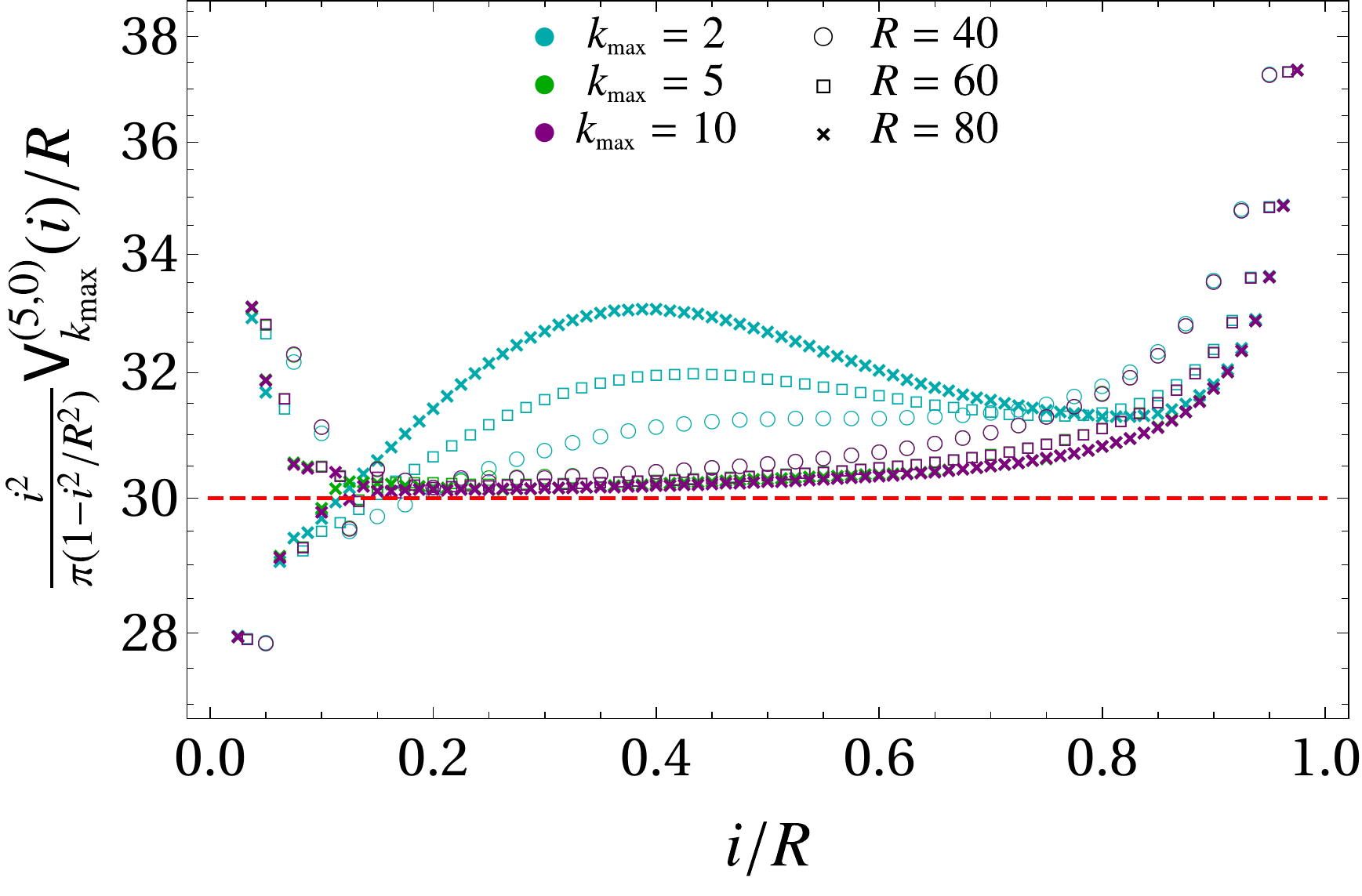}
\end{subfigure}
\hfill
\caption{
\label{fig:V0app}
Combination (\ref{V0-def}) when $\omega = 0$ in the case of $d=3$,
for different values of $k_{\textrm{\tiny{max}}}$.
The dashed red  horizontal lines correspond to $\mu_3(l) = l(l+1)$,
i.e. (\ref{mu-d-def}) when $d=3$.
These panels should be compared with 
the right panels of Fig.\,\ref{beta-V0-horizontal-lines-mode}.
}
\end{figure}

Since very similar results are obtained for $d=2$ and $d=3$,
we report only the latter ones. 
In Fig.\,\ref{fig:T0andV2app} we show numerical data for the combinations 
(\ref{T0-def}) and (\ref{V2-def}),
while in Fig.\,\ref{fig:V0app} the ones for the combination (\ref{V0-def}) are displayed. 
The data corresponding to  $l=0$, $l=2$ and $l=5$ have been reported 
in the left, middle and right panels respectively.  
In each panel, increasing values of the 
summation parameter $k_{\textrm{\tiny{max}}} \in \{ 1,\,2,\,10\}$
are considered. 

The aim of these figures is to show that 
including more diagonals in the summations 
brings the corresponding numerical curve closer 
to the expected curve, predicted by CFT.
This is in agreement with similar analyses made in previous studies
for one dimensional free chains
\cite{Arias:2016nip, Eisler:2017cqi,Eisler:2018ugn, Eisler:2019rnr, DiGiulio:2019cxv}.
For the systems that we have explored, in all the combinations of diagonals 
convergence is attained already when $k_{\textrm{\tiny{max}}}\sim10$ 
for all the values of $l$ considered.
Including more diagonals does not lead to visible improvements. 
In $\mathsf{T}_ {k_{\textrm{\tiny{max}}}}^{(l,0)}$ 
the convergence is faster than the one observed in the other combinations of diagonals; 
indeed already the main diagonal of $T^{(l)}$ 
is already close to the expected CFT result (\ref{beta-def-intro})
(see the top panels of Fig.\,\ref{fig-diagonals-T}).
Instead, for $\mathsf{V}_ {k_{\textrm{\tiny{max}}}}^{(l,2)}$ 
and $\mathsf{V}_{k_{\textrm{\tiny{max}}}}^{(l,0)}$ 
it is evident that $k_{\textrm{\tiny{max}}} > 2$ is needed
to recover the corresponding CFT prediction. 
In particular, while the data of $\mathsf{V}_ {k_{\textrm{\tiny{max}}}}^{(l,2)}$  
display a good agreement with the CFT curve already when $k_{\textrm{\tiny{max}}} = 2$
(see the bottom panels in Fig.\,\ref{fig:T0andV2app}),
the data for $\mathsf{V}_{k_{\textrm{\tiny{max}}}}^{(l,0)}$ in Fig.\,\ref{fig:V0app} 
clearly tell us  that higher values of $k_{\textrm{\tiny{max}}}$
must be considered to recover the expected CFT prediction,
which is characterised by the horizontal dashed red lines obtained from (\ref{mu-d-def}).

\bibliographystyle{nb}

\bibliography{refsEHsphere}

\end{document}